\documentclass[prd,11pt,tightenlines,nofootinbib,superscriptaddress]{revtex4}

\usepackage{amsfonts,amssymb,amsthm,bbm,amsmath}
\usepackage[dvips]{graphicx}
\usepackage{epsfig}

\newcommand{\C}{{\mathbb C}}

\newcommand{\R}{{\mathbb R}}
\newcommand{\Z}{{\mathbb Z}}

\newcommand{\cH}{{\mathcal H}}

\newcommand{\cN}{{\mathcal N}}

\newcommand{\cC}{{\mathcal S}}

\newcommand{\SU}{\mathrm{SU}}
\newcommand{\SL}{\mathrm{SL}}

\newcommand{\be}{\begin{equation}}
\newcommand{\ee}{\end{equation}}
\newcommand{\beq}{\begin{eqnarray}}
\newcommand{\eeq}{\end{eqnarray}}
\newcommand{\bea}{\begin{eqnarray}}
\newcommand{\eea}{\end{eqnarray}}
\newcommand{\nn}{\nonumber}

\newcommand{\bra}{\langle}
\newcommand{\ket}{\rangle}

\newcommand{\ra}{\rangle}

\newcommand{\tr}{{\mathrm Tr}}
\newcommand{\f}{\frac}
\newcommand{\pp}{\partial}
\newcommand{\om}{\omega}
\newcommand{\hsigma}{\hat{\sigma}}

\newcommand\vj{{\vec{\jmath}\,}}

\newcommand{\rd}{\mathrm{d}}

\def\bZ{{\bar{Z}}}

\newcommand{\bz}{\overline{z}}

\begin{document}

\title{{\large Holomorphic Factorization for a Quantum Tetrahedron}}

\author{Laurent Freidel}
\affiliation{Perimeter Institute for Theoretical Physics,
Waterloo N2L 2Y5, Ontario, Canada.}
\author{Kirill Krasnov}
\affiliation{ School of Mathematical Sciences, University of Nottingham, Nottingham NG7 2RD, UK. }
\author{Etera R. Livine}
\affiliation{Laboratoire de Physique, ENS Lyon, CNRS-UMR 5672, 46 All\'ee d'Italie, Lyon 69007, France.}

\date{v2: December 2009}

\begin{abstract}
We provide a holomorphic description of the Hilbert space $\cH_{j_1,\ldots,j_n}$ of $\SU(2)$-invariant tensors
(intertwiners) and establish a holomorphically factorized formula for the decomposition of identity in
$\cH_{j_1,\ldots,j_n}$. Interestingly, the integration kernel that appears in the decomposition
formula turns out to be the $n$-point function of bulk/boundary dualities of string theory.
Our results provide a new interpretation for this quantity as being, in the limit of large
conformal dimensions, the exponential of the K\"ahler potential of the symplectic manifold whose
quantization gives $\cH_{j_1,\ldots,j_n}$. For the case $n$=4, the symplectic manifold in question has
the interpretation of the space of ``shapes'' of a geometric tetrahedron with fixed face areas, and our results
provide a description for the quantum tetrahedron in terms of holomorphic coherent states. We describe how the
holomorphic intertwiners are related to the usual real ones by computing their overlap. The semi-classical analysis
of these overlap coefficients in the case of large spins allows us to obtain an explicit relation between the real
and holomorphic description of the space of shapes of the tetrahedron. Our results are of direct relevance for the
subjects of loop quantum gravity and spin foams, but also add an interesting new twist to the story of the
bulk/boundary correspondence.
\end{abstract}

\maketitle

\section{Introduction}

The main object of interest in the present paper is the space
\be\label{H-intertw}
{\cal{H}}_{\vj} =\left(V_{j_{1}}\otimes \cdots \otimes V_{j_{n}}\right)^{\mathrm{SU(2)}}
\ee
of $\mathrm{SU(2)}$-invariant tensors (intertwiners)
in the tensor product of $n$ irreducible $\mathrm{SU(2)}$ representations
$V^j, {\rm dim}(V^j)\equiv d_j= 2j+1$. Being a vector space with an inner product endowed from that in
representation spaces $V^j$ this space is naturally a Hilbert space. It is finite-dimensional,
with the dimension given by the classical formula:
\be\label{H-dim}
{\rm dim}({\cal{H}}_{\vj})=\frac{2}{\pi}\int_0^\pi d\theta
\sin^2(\theta/2) \,\, \chi^{j_1}(\theta) \ldots \chi^{j_n}(\theta),
\ee
where $\chi^j(\theta)= \sin{((j+1/2)\theta)}/\sin{(\theta/2)}$ are the $\mathrm{SU(2)}$ characters.

The intertwiners (\ref{H-intertw}) figure prominently in many areas of mathematical
physics. They are key players in the theory of angular momentum that has numerous applications
in nuclear and particle physics, atomic and molecular spectroscopy, plasma physics,
quantum chemistry and other disciplines, see e.g. \cite{Varshalovich:1988ye}. A ``q-deformed''
analog of the space ${\cal{H}}_{\vj}$ plays the central role in quantum Chern-Simons (CS) theory
and Wess-Zumino-Witten (WZW) conformal field theory, being nothing else but the Hilbert space of
states of CS theory on a sphere with $n$ marked points, or, equivalently, the
Hilbert space of WZW conformal blocks, see e.g. \cite{Witten:1991mm}. A q-deformed
version of the formula (\ref{H-dim}) is the famous Verlinde formula \cite{Verlinde:1988sn}.
Finally, and this is the main motivation for our interest in (\ref{H-intertw}), intertwiners
play the central role in the quantum geometry or spin foam approach to quantum gravity.
Indeed, in quantum geometry (loop quantum gravity) approach the space of states of geometry is spanned
by the so-called spin network states  based on a graph $\Gamma$.
This space is obtained  by tensoring together the Hilbert spaces $L^2(G)$ of square integrable functions on the group
$G=\mathrm{SU(2)}$ -- one for every edge $e$ of the underlying graph -- while contracting them at vertices $v$ with invariant
tensors to form a gauge-invariant state. Using the Plancherel decomposition the spin network Hilbert space
can therefore be written as: $L^{2}(G_{\Gamma})=\oplus_{j_{e}}{\cal H}_{\Gamma}(j_{e})$, where
${\cal H}_{\Gamma}(j_{e}) = \otimes_{v} {\cal H}_{\vj_{v}}$
is the product of intertwiner spaces one for each vertex $v$ of the graph $\Gamma$.

The Hilbert spaces ${\cal{H}}_{\vj}$ for $n=1,2,3$ are either zero or one-dimensional.
For $n=1$ the space is zero dimensional, for $n=2$ there is a unique (up to rescaling)
invariant tensor only when $j_1=j_2$, and for $n=3$ there is again a unique (up to
normalization) invariant tensor when the ``triangle inequalities''
$j_1+j_2\geq j_3,$ $ j_1+j_3\geq j_2,$ $ j_2+j_3\geq j_1$ are satisfied.
Thus, the first non-trivial case that gives a non trivial
dimension of the Hilbert space of intertwiners is $n=4$. A beautiful geometric interpretation of states from
${\cal{H}}_{j_1,\ldots, j_4}$ has been proposed in \cite{Barbieri}, where it
was shown that the Hilbert space in this case can be obtained via the process of
quantisation of the  space of {\it shapes}  of a geometric tetrahedron in $\R^3$ whose face areas
are fixed to be equal to $j_1,\ldots,j_4$. This space of shapes, to be defined in more details below,
is naturally a phase space (of real dimension two) of finite symplectic volume, and its
(geometric) quantization gives rise to a finite-dimensional Hilbert space ${\cal{H}}_{j_1,\ldots,j_4}$.

In a recent work co-authored by one of us \cite{CF3} the line of thought originating in
\cite{Barbieri} has been further developed. Thus, it was shown that the space of shapes of a tetrahedron
is in fact a K\"ahler manifold of complex dimension one that is conveniently parametrized
by $Z \in \mathbb{C} \backslash \{0,1,\infty\}$. As we shall explain in more details below,
this complex parameter  is just the  cross ratio of the four stereographic coordinates $z_{i}$
labelling the direction of the normals to faces of the tetrahedron. It was also shown in
\cite{CF3} by a direct argument that the two possible viewpoints on ${\cal{H}}_{j_1,\ldots,j_4}$ --
namely that of $\mathrm{SU(2)}$ invariant tensors and that of quantization of the space of shapes
of a geometric tetrahedron -- are equivalent, in line with the general principle of Guillemin and
Sternberg \cite{GS} saying that (geometric) quantization commutes with symplectic reduction.
Moreover, in \cite{CF3} an explicit formula for the decomposition of the identity in
${\cal{H}}_{j_1,\ldots,j_4}$ in terms of certain coherent states was given.

The main aim of this paper is to further develop the holomorphic viewpoint on ${\cal{H}}_{\vj}$
introduced in \cite{CF3}, both for $n=4$ and in more generality. Thus, we give several explicit proofs
of the fact that the Hilbert space ${\cal{H}}_{\vj}$ of intertwiners can be obtained by quantization
of a certain finite volume symplectic manifold $S_{\vj}$, where $S$ stands for {\it shapes}.
The phase space $S_{\vj}$ turns out to be a K\"ahler manifold, with convenient holomorphic coordinates
given by a string of $n-3$ (suitably chosen) cross-ratios
$\{Z_1,\ldots,Z_{n-3}\}$ with $Z_i \in \mathbb{C} \backslash \{0,1,\infty\}$, and it is natural to use the
methods of geometric quantization to get to ${\cal{H}}_{\vj}$. Up to the ``metaplectic correction'' occurring
in geometric quantization of K\"ahler manifolds (see more on this in the main text),
the Hilbert space is constructed, see e.g. \cite{Woodhouse, Tuynman}, as the space of holomorphic functions
$\Psi(z)$ integrable with the measure $\exp(-\Phi(z,\bar{z}))\Omega^k$, where $\Phi(z,\bar{z})$ is the K\"ahler
potential, $\Omega=(1/i)\partial\bar{\partial}\Phi(z,\bar{z})$ is the symplectic form, and $k$ is the
(complex) dimension of the manifold. Alternatively, in the context of K\"ahler geometric quantization,
one can introduce  \cite{Kirwin} the {\it coherent} states $|z\ket$ such that $\Psi(z)=\bra z| \Psi\ket$. 
Then the inner product formula can be rewritten as a formula for the decomposition of the
identity operator in terms of the coherent states:
\be\label{ident-general}
\mathbbm{1} = \int \Omega^k e^{-\Phi(z,\bar{z})} |z\ket \bra z|.
\ee

Our first main result in this paper is a version of formula (\ref{ident-general})
for the identity operator in the Hilbert space
${\cal H}_{\vj}$. The corresponding coherent states shall be denoted by $|\vj, Z\ket \in {\cal H}_{\vj}$,
where $Z$ is a collective notation for the string $Z_1,\ldots,Z_{n-3}$ of cross-ratio coordinates.
We prove that:
\be
\label{unity-H}
\mathbbm{1}_{\vec{\jmath}} =  8\pi^{2}\prod_{i=1}^{n} \frac{{\rd_{j_{i}}}}{2\pi}
\int_{\mathbb{C}}  {\rd^{2}Z}   \,
 \hat{K}_{\vj}(Z,\bar{Z}) \,  |\vj,Z\ket \bra \vj, Z |\,.
\ee
The integration kernel $\hat{K}_{\vj}(Z,\bar{Z})$ here turns out to be just the $n$-point function of
the AdS/CFT duality \cite{Witten}, given by an integral over the 3-dimensional hyperbolic
space of a product of $n$ so-called bulk-to-boundary propagators, see the main text for the
details. A comparison between (\ref{unity-H}) and (\ref{ident-general}) shows that, in the
semi-classical limit of all spins becoming large, the $n$-point function $\hat{K}_{\vj}(Z,\bar{Z})$ must admit an
interpretation of an exponential of the K\"ahler potential on $S_{\vj}$, and we demonstrate by an explicit
computation in what sense this interpretation holds. Thus, as a corollary to our main result (\ref{unity-H})
we obtain a new and rather non-trivial interpretation of the bulk/boundary duality $n$-point
functions.

The formula (\ref{unity-H}) takes a particularly simple form of an
integral over a single cross-ratio coordinate in the case $n=4$ of relevance for
the quantum tetrahedron. The coherent states $|\vj,Z\ket$ are holomorphic functions
of $Z$ that we shall refer to as {\it holomorphic intertwiners}. The resulting holomorphic
description of the Hilbert space of the quantum tetrahedron justifies the title of this paper.

In the second part of the paper we characterize the $n=4$ holomorphic intertwiners $|\vj,Z\ket$
by projecting them onto a more familiar real basis in ${\cal{H}}_{j_1,\ldots,j_4}$.
The real basis $|\vj, k \ket^{ij} $ can be obtained by considering the eigenstates of the operators
$J_{(i)}\cdot J_{(j)}$ representing the scalar product of the area vectors between the faces $i$ and $j$.
We compute these operators as second-order differential operators acting on functions of the
cross-ratio $Z$, and use these results to characterise the pairing (or overlap) between the usual
normalised intertwiners in the channel $(ij)$ : $|\vj, k \ket^{ij} $, and our  holomorphic intertwiner $|\vj, Z\ket$.
Denoting this pairing by $\,{}^{ij}C_{\vj}^{k} \equiv {}^{ij}\bra\vj, k| \vj, Z \ket$, we find it to be
given by the ``shifted'' Jacobi polynomials $P_{n}^{(\alpha,\beta)}$:
\be
{}^{12}C_{\vj}^{k}(Z) = N_{\vj}^{k} \,P_{k- j_{34}}^{(j_{34}-j_{12},\, j_{34} + j_{12})}(1-2Z)
\ee
where $j_{ij}\equiv j_{i}-j_{j}$,
and $N_{\vj}^{k}$ is a normalisation constant to be described below.
This result can be used to express the ${\cal{H}}_{j_1,\ldots,j_4}$ norm of the holomorphic
intertwiner $|\vj,Z\ket$ in a holomorphically factorised form (for any choice of the channel $ij$):
\be
\bra \vj, Z |  \vj, Z \ket = \sum_{k} \rd_{k} |{}^{ij}C_{\vj}^{k}(Z)|^{2},
\ee
which gives another justification for our title. As a by-product of our analysis we also
deduce several non-trivial facts about the holomorphic factorisation of the bulk/boundary
4-point function $K_{\vj}$.

The last step of our analysis is to discuss the asymptotic properties of the (normalized) overlap coefficients
${}^{ij}C_{\vj}^{k}(Z)$ for large spins and the related geometrical interpretation.
This asymptotic analysis allows us to explicitly obtain an extremely non-trivial relation
between the real and holomorphic description of the phase space of shapes of a geometric tetrahedron,
and demonstrates the power of the methods developed in this paper.
As a final and non-trivial consistency check of our analysis we show that
the normalized overlap coefficient is sharply  picked both in $k$ and in $Z$
around a value $k(Z)$ determined by the classical geometry of a tetrahedron.

Our discussion has so far been quite mathematical, so we would now like to switch to a
more heuristic description and explain the significance of our results for the field of quantum
gravity. As we have already mentioned, the $n=4$ intertwiner that we have characterized in
this paper in most details plays a very important role in both the loop quantum gravity and
the spin foam approaches. These intertwiners have so far been characterized using the
real basis $|\vj, k \ket^{ij} $. In particular, the main building blocks of the spin foam
models -- the $(15j)$-symbols and their analogs -- arise as simple pairings of 5 of
such intertwiners (for some choice of the channels $ij$). The main result of this paper is a
{\it holomorphic} description of the space of intertwiners, and, in particular, an explicit
basis in ${\cal{H}}_{j_1,\ldots,j_4}$ given by the holomorphic intertwiners $|\vj,Z\ket$. While
the basis $|\vj, k \ket^{ij} $, being discrete, may be convenient for some purposes, the underlying geometric
interpretation in it is quite hidden. Indeed, recalling the interpretation of the intertwiners
from ${\cal{H}}_{j_1,\ldots,j_4}$ as giving the states of a quantum tetrahedron,
the states $|\vj,k \ket^{ij} $ describe a tetrahedron whose shape is maximally uncertain.
In contrast, the intertwiners $|\vj,Z\ket$, being holomorphic, are coherent states
in that they manage to contain the complete information about the shape of the tetrahedron
coded into the real and imaginary parts of the cross-ratio coordinate $Z$. We give
an explicit description of this in the main text.

Thus, with the holomorphic intertwiners $|\vj,Z\ket$ at our disposal, we can now characterize
the ``quantum geometry'' much more completely than it was possible before. Indeed, we can now
build the spin networks -- states of quantum geometry -- using the holomorphic intertwiners.
The nodes of these spin networks then receive a well-defined geometric interpretation
as corresponding to particular tetrahedral shapes. Similarly, the spin foam model
simplex amplitudes can now be built using the coherent intertwiners, and then the basic
object becomes not the $(15j)$-symbol of previous studies, but the $(10j)$-$(5Z)$-symbol
with a well-defined geometrical interpretation. Where this will lead the subjects of
loop quantum gravity and spin foams remains to be seen, but the very availability of
this new technology opens way to many new developments and, we hope, will give a new
impetus to the field that is already very active after the introduction of the new spin foam
models in \cite{Livine:2007vk,Engle:2007uq,Freidel:2007py,LS2,ELPR}.

The organization of this paper is as follows. In section \ref{shapes} we describe how the phase
space that we would like to quantize arises as a result of the symplectic reduction of a simpler
phase space. Then, starting from an identity decomposition formula for the unconstrained Hilbert 
space, we perform the integration along the directions orthogonal to the constraint surface
by a certain change of variables. The key idea used in this section is that holomorphic
states invariant under the action of ${\rm SU}(2)$ are also ${\rm SL}(2,\C)$-invariant. 
In section \ref{sec:altern} we provide an alternative derivation of the identity
decomposition formula starting from a formula established in
\cite{CF3}, and also utilizing a change of variables argument. 
The two derivations that we give emphasize different geometric
aspects of the problem, and are complementary. Then, in section \ref{Kahlersc} we
analyze the identity decomposition formula in the semi-classical limit of large spins
and show that it takes the form precisely as is expected from the point of view
of geometric quantization. This establishes that the $n$-point function of the
bulk/boundary dualities is the (exponential of the) K\"ahler potential on the
space of shapes. Section \ref{Invarsc} specializes to the case $n=4$
relevant for the quantum tetrahedron and characterizes the holomorphic intertwiners
$|\vj,Z\ket$ by computing their overlap with the standard real intertwiners. This
is done by considering the $\SU(2)$-invariant operators given by the product of two
face normals. Then in section \ref{CFTsc} we use the requirement of hermiticity of these geometric
operators to put some constraints on the inner product kernel $\hat{K}_{\vj}(Z,\bZ)$ given
by the bulk/boundary 4-point function. We show that, remarkably, the hermiticity of these operators
suggests that $\hat{K}_{\vj}(Z,\bZ)$ holomorphically factorizes precisely in the form
required for it to have a CFT 4-point function interpretation. In section \ref{Asymptsc}
we study the asymptotic properties of the holomorphic intertwiners $|\vj,Z\ket$
projected onto the usual real intertwiners $|\vj,k\ket$. We find that the overlap coefficients
are peaked in both $k$ and $Z$ labels, and characterize the relation between the real
and holomorphic descriptions of the space of shapes. We finish with a discussion.

\section{The space of shapes and its holomorphic quantization}
\label{shapes}

\subsection{The space of shapes $S_{\vj}$ as a symplectic quotient}

In this subsection we recall the classical geometry behind the quantization problem we study. We
start with a symplectic manifold obtained as the Cartesian product of $n$ copies of the sphere
with its standard ${\rm SU}(2)$-invariant symplectic structure. Radii of the $n$ spheres
are fixed to be $j_i\in \Z/2, i=1,\ldots, n$ and we denote by $\vj=(j_{1},\cdots, j_{n})$
the $n$-tuple of spins. Thus, the space ${P}_\vj$ we consider is parametrized
by $n$ vectors $j_i N_i$, where $N_i\in \R^3$ are unit vectors $N_i\cdot N_i=1$.
The space ${P}_\vj$ is a $2n$-dimensional symplectic manifold with symplectic form given by the sum
of the sphere symplectic forms.

There is a natural (diagonal) action of the group $G={\rm SU}(2)$ on ${P}_\vj$, which
is generated by the following Hamiltonian:
\be\label{Ham}
H_{\vj}\equiv \sum_i j_i N_i.
\ee
The space of shapes $S_{\vj}$ that we are interested in is obtained by the symplectic reduction,
that is by imposing the constraint $H_{\vj}=0$ and then by considering the space of $G={\rm SU}(2)$-orbits on
the constraint surface. This space can be thought of as that of $n$-faced polygons, with $N_{i}$ being the unit face normals
and $j_{i}$ being the face areas. The following notation for this symplectic reduction is standard:
\be\label{sympl-quot}
S_{\vj}= P_{\vj}/\!/\mathrm{SU(2)}.
\ee
The usual theory of symplectic reductions tells us that the space of shapes is also a symplectic manifold.
The fact of crucial importance for us is that the space $S_{\vj}$ is also a K\"ahler manifold, i.e.
is a complex manifold with a Hermitian metric (satisfying an integrability condition) such that
the metric and the symplectic form arise as the real and imaginary parts of this Hermitian metric.
Indeed, each of the $n$ unit spheres is a K\"ahler manifold. The complex structure on the sphere
is made explicit by the stereographic projection:
\be
(N_{1},N_{2}, N_{3})(z) = \left(   \frac{ z+\bar{z}}{1+|z|^{2}} ,
\frac1{i}\frac{z-\bar{z}}{1+|z|^{2}},\frac{1-|z|^{2}}{1+|z|^{2}} \right),
\ee
The complex structure on the sphere is then the usual complex
structure on the complex $z$-plane, and the symplectic structure is:
\be
\label{Ssymp}
\omega_{j}(z) = \frac{2{j}}{i} \frac{\rd z \wedge \rd {\bar{z}}}{(1+|z|^{2})^{2}}=
 \partial\bar{\partial}\Phi_j(z,\bar{z}) \,\, \frac{\rd z\wedge \rd\bar{z}}{i},
\ee
where $\Phi_j(z,\bar{z})=2j \log(1+|z|^2)$ is the K\"ahler potential.
The complex structure on ${P}_\vj$ is then just the product one, and, crucially, it
turns out to commute with the action of ${\rm SU}(2)$ on ${P}_\vj$. Thus, the
space of shapes $S_{\vj}$ inherits from ${P}_\vj$ a symplectic as well as
a complex structure compatible with the symplectic structure, moreover the
induced metric is positive and is hence a K\"ahler manifold. This is most easily seen via
the Guillemin-Sternberg isomorphism \cite{GS} that expresses the symplectic quotient
(\ref{sympl-quot}) as an unconstrained but complex quotient
of the subset of $P_{\vj}$ consisting of non-coincident points:
\be \label{GS}
S_{\vj} = \{(z_{1},\cdots, z_{n}) \,| \, z_{i}\neq z_{j} \} / \mathrm{SL(2,\mathbb{C})}.
\ee

\subsection{Holomorphic quantization of the unit sphere}

In this subsection, as a preliminary step to the holomorphic quantization of the quotient (\ref{sympl-quot})
we remind the reader how the sphere can be quantized. According to the general spirit
of the geometric quantization, and in the spirit of
(\ref{ident-general}), the holomorphic quantization of the sphere of radius $j$ is achieved
via the ${\rm SU}(2)$ coherent states $|j,z\ket$ that satisfy the completeness relation:
\be\label{id-single}
\mathbbm{1}_j = \frac{\rd_j}{2\pi} \int \frac{\rd^2z}{(1+|z|^2)^{2(j+1)}} |j,z\ket\bra j,z|,
\ee
where $\rd_{j}=2j+1$ is the dimension of the representation $V_{j}$ and $\rd^2z\equiv |dz\wedge d\bar{z}|$.
Note that with this convention $\rd^2 z$ is {\it twice} the canonical area form on the plane.
The coherent states appearing in (\ref{id-single}) are holomorphic, i.e. depend only on $z$ and not on $\bar{z}$.
They are normalized so that $\bra j,z|j,z\ket = (1+|z|^{2})^{2j}$, which, when used in (\ref{id-single})
immediately gives the correct relation ${\rm Tr}(\mathbbm{1}_j)=d_j$.

\subsection{Kinematical and Physical Hilbert spaces}

Let us start with a description of the Hilbert space obtained by quantizing the
unconstrained phase space ${P}_{\vj}$. By the co-adjoint orbits method, this is just the direct
product on $n$ irreducible representation spaces $V^{j_i}$ of ${\rm SU}(2)$. Thus, our ``kinematical'' Hilbert space is:
\be
{\cal H}^{\mathrm{kin}}_{\vj} = V_{j_1} \otimes \ldots \otimes V_{j_n}.
\ee
A holomorphic description of each of these spaces has been given above. However,
for our purposes it is more convenient to use the coherent states description. 
Each space $V_{j}$ can then be described as spanned by holomorphic polynomials
$\psi(z)$ of degree less than $2j$, where $z$ is the usual coordinate on the complex plane.
A state in the tensor product depends on the $n$ variables $(z_1,\ldots,z_n)$
and the inner product is given by
\be\label{inner-n}
\bra\psi_1,\psi_2\ket = \int \prod_{i=1}^n \frac{\rd_{j_i}}{2\pi} \frac{\rd^2z_i}{(1+|z_i|^2)^{2(j_i+1)}}
\overline{\psi_1(z_i)} \psi_2(z_{i}),
\ee
which, after the identification $\Psi(z) \equiv \bra \bar{z}|\Psi\ket $, is just
$n$ copies of the formula (\ref{id-single}) above. As before, $\rd_j=2j+1$ and the convention for the measure is
$d^2z=|dz\wedge d\bar{z}|$. The action of ${\rm SU}(2)$ group elements in this description is given by:
\be
(\hat{T}^j(k^{t})\psi)(z) = (-\beta^* z+\alpha^*)^{2j} \psi( z^{k}),
\ee
where ${}^{t}$ denotes transposition and the action of ${\rm SU}(2)$ on the complex plane is
\be
 k = \left( \begin{array}{cc}\alpha &\beta \\
-\beta^* & \alpha^* \end{array}\right) \in {\rm SU}(2),
\quad\quad
z^k = \frac{\alpha z+\beta}{-\beta^* z+\alpha^*}.
\ee
Since, as is easily checked, $(1+|z^k|^2) = (-\beta^* z+\alpha^*)^{-2}(1+|z|^2)$, and
$d^2z/(1+|z|^2)^2$ is an ${\rm SU}(2)$-invariant measure, the inner product (\ref{inner-n}) is
${\rm SU}(2)$-invariant.

We are interested in computing the inner product of physical, i.e. ${\rm SU}(2)$-invariant
states:
\be
(\hat{T}^{j_1}(k)\otimes \ldots \otimes \hat{T}^{j_n}(k)\psi)(z_1,\ldots,z_n)=
\psi(z_1,\ldots,z_n).
\ee
However, being holomorphic, such states are then automatically invariant under ${\rm SU}_\C(2)={\rm SL}(2,\C)$.
From this we immediately get:
\be\label{transf}
\psi(z_{i}^{g}) =  \prod_{i=1}^n (cz_i+d)^{-2j_i} \psi(z_i),\quad \mathrm{where}
\quad g = \left( \begin{array}{lr}a & b \\
c & d \end{array}\right)\in {\rm SL}(2,\C)
\ee
and the action $z^{g}$ of $\mathrm{SL}(2,\C)$ on the complex plane by rational transformation is given in (\ref{zg}).
Thus, the physical states are completely determined by their values on the moduli space
$\{z_1,\ldots,z_n\}/{\rm SL}(2,\C)$. As we have already mentioned, and as is described at length in \cite{CF3},
this moduli space space is isomorphic to the space of shapes $S_{\vj}$.

\subsection{Moduli Space and Cross-Ratios}

The integral in (\ref{inner-n}) is that over $n$ copies of the complex plane. However,
as we have seen above, on physical states the integrand has very simple transformation
properties under ${\rm SL}(2,\C)$. This suggests that the integral can be computed
by a change of variables where one parametrizes $z_1,\ldots,z_n$ by an element of ${\rm PSL}(2,\C)$ 
together with certain cross-ratios $Z_i, i=4,\ldots, n$.

Indeed, given
the first three complex coordinates $z_1,z_2,z_3$, there exists a unique ${\rm PSL}(2,\C)$
transformation that maps these points to $0,1,\infty$ (and maps the points $z_i, i>3$ to $Z_i$). 
Let us use the inverse of this transformation to parametrize the unconstrained phase
space by an element of ${\rm SL}(2,\C)$ together with $Z_i$. Explicitly, given an  ${\rm SL}(2,\C)$ element
$g$ and $n-3$ cross-ratios $Z_{i}$ we can construct the $n$ points
$\{ 0,1,\infty, Z_i\}^g$ on the complex plane. Explicitly:
\be\label{coord-change}
z_{1}= \frac{b}{d}, \qquad z_{2}=\frac{a+b}{c+d},\qquad z_{3}= \frac{a}{c},\qquad z_{i}=
\frac{ a Z_i+b}{cZ_i+d} \quad i\geq 4.
\ee
This gives us a map
\be
{\rm SL}(2,\C) \times \{Z_4,\ldots,Z_n\} \to \{z_1,\ldots,z_n\},
\qquad g\times Z_j \to z_i(g,Z_{j}).
\ee
which is such that $(z_{1}(g,Z_{j}),\cdots z_{n}(g,Z_{j}) )= \left(0^{g},1^{g},\infty^{g},Z_{i}^{g}\right)$.
This map is $2:1$ since $-g$ and $g$ give the same image. The cross-ratios $Z_i$, together with $g$ (or $a,b,c,d $
satisfying the relation $ad-bc=1$) can be used as (holomorphic) coordinates on
our space $\{z_1,\ldots,z_n\}$. This change of variables is performed in
details in appendix \ref{gmeas} where we find the following relation between the integration measures
\be
\int_{\C^{n}} \prod_{i=1}^{n} \rd^{2} z_{i} \,\, F(z_{i},\overline{z_{i}}) =
{ 8\pi^{2}}
\int_{\C^{n-3}} \! \prod_{i=4}^{n} \rd^{2} Z_{i}
\int_{\mathrm{SL}(2,\C)} \!\!\!\!\!\!\! \rd^{\mathrm{norm}} g  \, \,\,
\frac{F(z_{i}(g,Z_{j}),
\overline{z_{i}(g,Z_{j})})}{|d|^{4} |c+d|^{4} |c|^{4}\prod_{i=4}^{n}| cZ_{i}+d|^{4}}.
\ee
Here $\rd ^{\mathrm{norm}} g$ is the Haar measure on ${\rm SL}(2,\C)$, normalized so that its compact ${\rm SU}(2)$
part measure is just the normalized measure on the unit three-sphere (see appendix \ref{gmeas}). As before, the convention
is that $\rd ^2z=|\rd z\wedge \rd \bar{z}|$.

\subsection{The physical inner product}

Given the transformation property (\ref{transf}) we can describe the functions $\psi(z_i)$
by their values on the moduli space parametrized by $Z_i$. Explicitly:
\be\label{psi-1}
\psi(z_i)= \psi\left(0^{g},1^{g},\infty^{g},Z_{i}^{g}\right) = d^{-2j_1} (c+d)^{-2j_2} c^{-2j_3} 
\prod_{i=4}^n (cZ_i+d)^{-2j_i} \,{\Psi}(Z_i),
\ee
where we have defined a wave functional depending only on the cross ratios as given by the limit
\be\label{psi-Z}
{\Psi}(Z_i)\equiv \lim_{X\to\infty} X^{-2j_3} \psi(0,1,X,Z_i).
\ee
Now, starting from the expression (\ref{inner-n}) for the kinematical inner product, performing
the change of variables from $z_1,\ldots,z_n$ to ${\rm SL}(2,\C)\times \{Z_4,\ldots,Z_n\}$, and
substituting the expression (\ref{psi-1}) for the wave functional we can reduce the inner product of two
physical states to a simple integral over the cross-ratios only. We get
\be\label{inner-phys}
\bra\Psi_1,\Psi_2\ket = 8\pi^2 \prod_{i=1}^n \frac{\rd_j}{2\pi}
\int \prod_{i=4}^n \rd^2 Z_i \,\, \hat{K}_{\vj}(Z_i,\bar{Z}_i) \overline{{\Psi}_1(Z_i)}{\Psi}_2(Z_i),
\ee
where  $\hat{K}_{\vj}$ is given by a group integral
\beq\nonumber
\hat{K}_{\vj}(Z_i,\bar{Z_i})= \int_{\mathrm{SL}(2,\C)}\!\!\!\!\!\!\!\! \rd^{\mathrm{norm}}g \,\,
(|b|^2+|d|^2)^{-2(j_1+1)} (|a+b|^2+|c+d|^2)^{-2(j_2+1)} (|a|^2+|c|^2)^{-2(j_3+1)} \times
\\ \label{ker-phys}
\prod_{i=4}^{n}\left(|aZ_{i}+b|^{2}+|cZ_{i}+d|^{2}\right)^{-2(j_{i}+1)}\, .
\eeq
It is not hard to see that this expression can be obtained from a kernel depending
on $n$ coordinates 
\be\label{ker-2}
K_{\vj}(z_i,\bar{z_i}):= \int_{\mathrm{SL}(2,\C)}\!\!\!\!\!\!\!\! \rd^{\mathrm{norm}}g \,\,
\prod_{i=1}^{n}\left(|cz_{i}+d|^{2}+ |az_{i}+b|^{2}\right)^{-2(j_{i}+1)}
\ee
by taking the limit
\be\label{limK}
\hat{K}_{\vj}(Z_{i},\overline{Z_i}) = \lim_{_{X\to \infty}} |X|^{2\Delta_{{3}}} K_{\vj}(0,1,X,Z_{4},\cdots, Z_{n}).
\ee

The formula (\ref{inner-phys}) for the physical inner product is valid for all $n$ and
admits an illuminating reformulation in terms of the coherent states. Thus, under the identification
$\Psi(Z_i) \equiv \bra \vj, \bar{Z}_i | \Psi \ket$, we get:
\be\label{unity-n}
\mathbbm{1}_{\vec{\jmath}} = 8\pi^2 \prod_{i=1}^n \frac{\rd_j}{2\pi}
\int \prod_{i=4}^n \rd^2 Z_i \,\, \hat{K}_{\vj}(Z_i,\bar{Z}_i) \, |\vj,Z_i\ket\bra \vj,Z_i| \, ,
\ee
which is the formula given in the introduction. 

\subsection{Kernel as the n-point function of the bulk-boundary correspondence}

In this subsection we explicitly relate the kernel $K_{\vj}(z_{i},\overline{z_{i}})$
that we encountered above, see (\ref{ker-2}), to an object familiar from the
bulk-boundary duality of string theory. Indeed, observe that the integrand in
(\ref{ker-2}) is ${\rm SU}(2)$-invariant, so it is enough
to integrate only over the quotient space $H_3={\rm SL}(2,\C)/{\rm SU}(2)$ which is the 3 
dimensional Hyperbolic space or Euclidean AdS space.
This can be achieved by using the Iwasawa decomposition which states
that any matrix of $\mathrm{SL(2,\mathbb{C})}$ can be decomposed as the
product of a unitary matrix $k \in\SU(2)$, a diagonal Hermitian matrix and an upper triangular matrix.
That is, any element $g\in\mathrm{SL(2,\mathbb{C})}$ can be {\it uniquely}
written as
\be\label{Iwasawa}
g = k \left( \begin{array}{cc}\rho^{-\frac12} & 0 \\ 0 & \rho^{\frac12} \end{array} \right)
\left( \begin{array}{cc}1 & -y \\ 0 & 1 \end{array} \right),\quad k \in \mathrm{SU(2)},
\quad \rho \in \mathbb{R}^{+}, \quad y \in \mathbb{C}
\ee
The Haar measure on ${\rm SL}(2,\C)$ written in terms of the coordinates
$k,\rho,y$  reads (see appendix \ref{gmeas}):
\be
\rd^{\mathrm{norm}} g = \rd k\, \frac{\rd \rho} {\rho^{3}} \,\rd y \rd {\bar{y}}
\ee
where $\rd k$ is the normalized $\SU(2)$ Haar measure, and the rest is just the standard
measure on the hyperbolic space $H_3$ whose metric is given by:
\be
ds^{2} = \frac{\rd \rho^{2} + \rd y\rd \bar{y}}{\rho^{2}}.
\ee

Now, expressing  (\ref{ker-2}) in terms of the Iwasawa coordinates one explicitly sees
that the integrand is independent of $k$, so one only has to perform
the integration over the hyperbolic space $H_3$ coordinatized by $\rho,y$.
The first remark is that in these coordinates one immediately recognize that the kernel
is related to the $n$-point function of the bulk-boundary correspondence of string theory \cite{Witten}.
Indeed, the integrand can be easily seen to be a product of the bulk-to-boundary propagators
heavily used in AdS/CFT correspondence of string theory. Thus, let us evaluate the quantity
$\left(|cz+d|^{2}+ |az+d|^{2}\right)^{-2(j+1)}$ on an ${\rm SL}(2,\C)$ group element
appearing in the Iwasawa decomposition (\ref{Iwasawa}). Due to the $\SU(2)$-invariance, it depends
only on the hyperbolic part of the group element $g$, thus on
$$
h=\left( \begin{array}{cc}\rho^{-\frac12} & 0 \\ 0 & \rho^{\frac12} \end{array} \right)
\left( \begin{array}{cc}1 & -y \\ 0 & 1 \end{array} \right)=
\left( \begin{array}{cc}\rho^{-\frac12} & -y\rho^{-1/2} \\ 0 & \rho^{\frac12} \end{array} \right).
$$
Then setting explicitly $a=\rho^{-1/2}, b=-y\rho^{-1/2}, c=0, d=\rho^{1/2}$ and using 
the same convention $\Delta=2(j+1)$ as before,
we see that the integrand is given by the product of the following quantities
\be\label{bulk-boundary}
K_\Delta(h,z)\equiv \frac{\rho^\Delta}{(\rho^2+|z- y|^2)^{\Delta}}.
\ee
This is just the usual expression for the bulk-boundary propagator, see \cite{Witten}, 
where $h$ label a point is the interior of $H_{3}$ while
$z_{i}$ label points in its asymptotic boundary.
Thus, we have shown that the kernel (\ref{ker-2}) can be very compactly written as an
integral over the bulk of $H_{3}$ of a product of $n$ bulk to boundary propagators:
\be\label{n-point}
K_{\vj}(z_i,\overline{z_{i}})=
\int_{H_3} d^3h \, \prod_{i=1}^n K_{\Delta_i}(h,z_i).
\ee
This is the standard definition of the bulk-boundary duality $n$-point function, see \cite{Witten}.

\section{An alternative derivation of the identity decomposition formula}
\label{sec:altern}

In the previous section we have given a simple derivation of the 
decomposition of the identity formula, with the starting point being the
identity formula on the unconstrained Hilbert space. The key idea of
the analysis above was to use the ``analytic continuation'' that implied
that ${\rm SU}(2)$-invariant holomorphic states are also ${\rm SL}(2,\C)$-invariant.
This then allowed us to reduce the integral over $z_i$ to that over cross-ratios.
However, some geometrical aspects remain hidden in this analysis. Thus, our
general derivation made no reference to the symplectic potential on the constraint surface
or metric on the orbits orthogonal to that surface. The aim of this section is
to provide an alternative derivation of the decomposition formula that makes
such geometrical aspects more manifest. 

\subsection{Decomposition of the identity in ${\cal H}_{\vj}$}

The basis of (holomorphic) coherent states $|j,z\ket$ in $V^j$ naturally extends
to a basis in the Hilbert space of intertwiners
$$
{\cal H}_{\vj}
=\left(V_{j_{1}}\otimes \cdots \otimes V_{j_{n}}\right)^{\mathrm{SU(2)}},
$$
by performing the group averaging over $G={\rm SU}(2)$ on a product of coherent states.
This leads to the notion of ``coherent intertwiner'' \cite{Livine:2007vk}
defined as
\be
||\vj, z_{i}\ket \equiv \int_{\mathrm{SU(2)}} \rd g \, T^{j_1}(g)|j_{1},z_{1}\ket\otimes\cdots\otimes
T^{j_n}(g)|j_n,z_n\ket.
\ee
These states are ${\rm SU}(2)$-invariant by construction, and thus are vectors in ${\cal H}_{\vj}$.
Therefore, the operator quantizing the Hamiltonian constraint vanishes on them.
However, the labels of these coherent states  {\it do not} satisfy  the constraint $H_{\vj}(z_i)\equiv \sum_i j_i N(z_i)=0$
(although it can be argued that they are peaked on $H=0$ in the large spin limit)
and thus do not have the interpretation as states in the Hilbert space obtained by quantizing the
space of shapes $S_{\vj}$.

To relate them to some states obtained by quantizing $S_{\vj}$
one can follow the Guillemin-Sternberg prescription \cite{GS} and integrate these states
along the orbits orthogonal to the constraint surface. This was done in \cite{CF3} where the
following result for the projector onto ${\cal H}_{\vj}$ was obtained\footnote{As compared to the
formula (102) in \cite{CF3}, in order to emphasize the holomorphic structure of this formula,
we have stripped out all the dependence on the factors $(1+|z|^2)^2$. It is easy to see that
these factors appearing in the states, the measure and the prefactors all cancel out to give the
formula presented here.}
\be\label{unity-alt}
\mathbbm{1}
_{\vec{\jmath}} = \prod_{i}\frac{\rd_{j_{i}}}{2\pi} \int \prod_{i} d^{2} z_{i} \,
\delta^{(3)}(H_{\vj}(z_{i}))
\mathrm{det}\left(G_{\vj}(z_{i})\right) \,
 K_{\vj}(z_i,\overline{z_{i}}) \,  ||\vj,z_i\ket \bra \vj, z_i||.
\ee
There are two new ingredients in this formula. The first one is the determinant of the 3 by 3
matrix $G_{\vj}$, which is the metric along the $\mathrm{SL(2,\mathbb{C})}$ orbits orthogonal
to the constraint surface. Explicitly
 \be
G_{\vj} ^{ab}(z_{i})= \sum_{i=1}^{n} j_{i} \big(\delta^{ab}- N^{a}(z_{i})N^{b}(z_{i})\big).
\ee
The second ingredient is the $n$-point function $K_{\vj}(z_i,\bar{z}_i)$
entering the measure of integration. It can be defined as the following integral over $\mathrm{SL}(2,\C)$
\be\label{intodo}
K_{\vj}(z_{i},\overline{z_{i}}) =  \int_{\mathrm{SL(2,\mathbb{C})}} \rd^{\mathrm{norm}} g
\prod_{i=1}^{n}\left({\bra z_{i}|g^{\dagger} g |z_{i}\ket}\right)^{-2(j_{i}+1)} \, .
\ee
Here $|z\ket$ is the ${\rm SU}(2)$ coherent state associated with the fundamental spin $1/2$
representation. The normalization of the measure over $\mathrm{SL}(2,\C)$ is given in the appendix
\ref{gmeas}. It is not hard to see that the quantity (\ref{intodo}) coincides with 
(\ref{ker-2}) and so is just the $n$-point function of the bulk-to-boundary dualities \cite{Witten},
as we have reviewed in the previous section.

The integral in (\ref{unity-alt}) is taken over $n$ copies of the complex plane
subject to the closure constraint $\sum_i j_i N(z_i)=0$. Moreover, both the integrand and the
measure are, in fact, ${\rm SU}(2)$-invariant. Thus, this is an integral over our phase space
of shapes $S_{\vj}$. By the Guillemin-Sternberg isomorphism (\ref{GS}) this space is
isomorphic to the unconstrained space $P_{\vj}^{s}$ of $n$ copies of the complex plane
modulo $\mathrm{SL(2,\mathbb{C})}$ transformations. This means that a convenient set
of coordinates on $S_{\vj}$ is given by the $\mathrm{SL(2,\mathbb{C})}$-invariant
cross-ratios. So, the idea is now to express the integral in (\ref{unity-alt}) in term of the cross-ratios.

\subsection{Cross-Ratios and the Holomorphic Intertwiner}

The conformal group ${\rm SL}(2,\C)$ acts on the complex plane by fractional linear
(or M\"obius) transformations
\be\label{zg}
  z\to z^{g}\equiv \frac{az+b}{cz+d}\quad\mathrm{where}\quad
g= \left( \begin{array}{lr}a &b\\
c & d \end{array}\right)\in \mathrm{SL(2,}\mathbb{C})\,.
\ee
The action on the coherent states is also easy to describe:
\be\label{gaction}
 g|z\ket = (cz +d) |z^{g}\ket,
\ee
where  $|z\ket \equiv|1/2,z\ket$ is the coherent state in the fundamental representation
and one obtains a general coherent state simply taking the $2j$'s power of the $|z\ket$ fundamental one.
The $n$-point function $K_{\vj}(z_{i},\overline{z_i})$ is defined as an average over ${\rm SL}(2,\C)$ and
thus transforms covariantly under conformal transformation. Indeed, it is not hard to show that it satisfies
the standard transformation property of a CFT $n$-point function:
\be\label{K-transform}
K_{\vj}(z_i^{g},\overline{z_i^{g}}) = \prod_{i=1}^{n} |cz_i +d|^{2\Delta_{i}} K_{\vj}(z_i,\overline{z_i}),
\quad\mathrm{where}\quad \Delta_{i} = 2(j_{i}+1).
\ee

Given $n$ complex numbers $z_{i}$ we can parametrize
the ${\rm SL}(2,\C)$-invariant set $S_{\vj}$ by
$n-3$ cross-ratios:
\be\label{cross-ratios}
Z_{i}\equiv \frac{z_{i1}z_{23}}{z_{i3}z_{21}}, \quad i = 4,.., n
\qquad
\textrm{with}\quad z_{ij}\equiv z_{i}-z_{j}.
\ee
In other words, there  exists a ${\rm SL}(2,\C)$ transformation $g$  which maps $(z_{1}, z_{2}, z_{3})$ to
$(0, 1, \infty) $ and the  remaining $z_{i}$, $i>3$ to the cross-ratios $Z_{i}$. Explicitly,
this ${\rm SL}(2,\C)$ transformation is given by:
\be\label{abcd}
 \left( \begin{array}{lr}a & b \\
c & d \end{array}\right) =
\frac{1}{\sqrt{z_{23}z_{21}z_{13}}}
\left( \begin{array}{cc} z_{23} & -z_1z_{23} \\
z_{21} & -z_{3}z_{21} \end{array}\right),\quad \mathrm{or} \quad z^{g} = \frac{(z-z_{1}) z_{23}}{(z-z_{3})z_{21}}.
\ee
Thus, we can either use a group element and the cross-ratios $(g,Z_i)$ as well as 
$(z_1, z_2, z_3,\cdots, z_{n})$ as a set of complex coordinates
on the unconstrained phase space ${P}_\vj$. Moreover we can trade the 
$\mathrm{SL}(2,\mathbb{C})$ group element $g$ for the elements $(z_{1},z_{2},z_{3})$,
see, however, below for a subtlety related to the fact that this
parametrization is not one-to-one. Then, using the transformation property (\ref{K-transform})
we can write the $n$-point function $K_{\vj}(z_i,\overline{z_i})$ as
\be\label{KZ}
K_{\vj}(z_{i},\overline{z_{i}}) = |z_{12}|^{ 2\Delta_{{3}} - \Delta }
|z_{23}|^{ \Delta-2\Delta_{{2}} -  2\Delta_{{3}} }  |z_{13}|^{\Delta- 2\Delta_{{1}} -  2\Delta_{{3}} }
\prod_{i=4}^{n} |z_{i3}|^{-2\Delta_{{i}}}\,
\hat{K}_{\vj}(Z_{i},\overline{Z_{i}})\,,
\ee
where $\Delta \equiv \sum_{i=1}^{n} \Delta_{{i}}$. Here ${K}_{\vj}(Z_{i},\overline{Z_{i}})$ 
is defined via (\ref{limK}) and is a function of the cross-ratios only. As such it is  ${\rm SL}(2,\C)$ invariant. 

Now, the function to be integrated in the decomposition of the identity formula (\ref{unity-alt}) is
$K_{\vec{j}\,}(z_i,\overline{z_{i}})\,  ||\vj,z_i\ket \bra \vj, z_i||$. Thus, let us also find
an expression for the coherent intertwiners as functions of the cross-ratios $Z_i$ as well as $z_1, z_2, z_3$.
It is not hard to see from (\ref{gaction}), and was shown explicitly in \cite{CF3}, that since $||\vj,z_{i}\ket$ is a state
invariant under SU(2) it transforms covariantly under $\mathrm{SL(2,\mathbb{C})}$:
\be
 ||\vj,z_{i}^{g}\ket = \prod_{i}(cz_{i} +d)^{\bar{\Delta}_{i}} ||\vj,z_{i}\ket \, ,\quad\mathrm{where}\quad
\bar{\Delta}_{i} = 2 -\Delta_{i} = -2j_{i}
\ee
is the dual conformal dimension. Thus, via a procedure similar to that employed for the $n$-point
function $K_{\vec{j}\,}(z_i,\overline{z_{i}})$, we can express the covariant intertwiner
state in terms of a state depending only on the cross-ratios:
 \be\label{stateZ}
||\vj,z_{i}\ket = z_{12}^{ \bar{\Delta}_{j_{3}}-\bar{\Delta}/2 }
z_{23}^{\bar{\Delta}/2 -\bar{\Delta}_{j_{2}} - \bar{\Delta}_{j_{3}} }
z_{13}^{ \bar{\Delta}/2 -\bar{\Delta}_{j_{1}} -  \bar{\Delta}_{j_{3}}}
\prod_{i=4}^{n} z_{i3}^{-\bar{\Delta}_{j_{i}}} \,\,|\vj, Z_{i}\ket \, ,
\ee
where $\bar{\Delta} = \sum_{i=1}^{n}\bar{\Delta}_{j_{i}}$. We shall refer to the state
$|\vj,Z\ket$ that depends (holomorphically) only on the cross-ratios as the ``holomorphic intertwiner''.
In terms of the coherent intertwiner it is given by
\be\label{Z-states}
|\vj, Z_{i}\ket \equiv \lim_{_{X\to \infty}} (-X)^{\overline{\Delta}_{j_{3}}}  ||\vj, 0,1,X,Z_{4},\cdots, Z_{n} \ket.
\ee

Since the transformation properties of the two factors in our integrand are ``inverse'' of each other,
their product has a very simple description in terms of the cross-ratio coordinates:
\be\label{holo1}
K_{\vj}(z_i,\overline{z_{i}}) \,  ||\vj,z_i\ket \bra \vj, z_i|| =
 \frac{\hat{K}_{\vj}(Z_{i},\overline{Z_{i}}) \,  |\vj,Z_{i}\ket \bra \vj, Z_{i} |}{|z_{12}|^{2(n-2)}
|z_{23}|^{2(4-n)}|z_{13}|^{2(4-n)} \prod_{i=4}^{n}|z_{3i}|^{4}} \, ,
\ee
which is the main result of this subsection.

Of particular interest to us are the two cases $n=3,4$.
In the case $n=3$ there is no cross-ratio coordinate and both
$\hat{K}_{j_1,j_2,j_3}\equiv \hat{K}_{\vj}(Z_{i},\overline{Z_i})|_{n=3}$ and
$|j_1,j_2,j_3\ket\equiv |\vj, Z_{i}\ket|_{n=3}$ are constants (depending only on the
representation labels $\vj$). Thus, in this case, the holomorphic
intertwiner is (up to a normalization denoted by $N_{j_1,j_2,j_3}$ and computed below) just the projector
onto the unique normalised  SU(2) invariant state $|0\ket$. Thus for $n=3$, the
previous formula  (\ref{holo1}) takes the following form:
\be\label{holo3}
K_{\vj}(z_i,\overline{z_{i}}) \,  ||\vj,z_i\ket \bra \vj, z_i|| \Big|_{n=3}=
\frac{N_{j_{1},j_{2},j_{3}}^{2}}{|z_{12}z_{13}z_{23}|^{2}}
\hat{K}_{{j_{1},j_{2},j_{3}}} \,  |0\ket\bra0|.
\ee
In the case $n=4$, which is of main interest to us due to its relation to a quantum tetrahedron, there is a single
cross-ratio parameter $Z$ that possesses a nice geometrical interpretation (it can be
expressed in terms of certain area and angle parameters of the tetrahedron, see \cite{CF3}).
In terms of this cross-ratio the formula (\ref{holo1}) for $n=4$ reads:
\be\label{holo4}
K_{\vj}(z_i,\overline{z_{i}}) \,  ||\vj,z_i\ket \bra \vj, z_i|| \Big|_{n=4}=
\frac{\hat{K}_{\vj}(Z,\overline{Z})}{|z_{12}z_{34}|^{4}}  \,  |\vj,Z \ket \bra \vj, Z  | \,,
\quad \mathrm{where} \quad Z\equiv  \frac{z_{41}z_{23}}{z_{43}z_{21}} \,.
\ee

\subsection{Measure and determinant}

The integral in (\ref{unity-alt}) is over the constraint surface, with the integration measure being
\be
\rd \mu^{(n)}(z_{i})\equiv
\prod_{i=1}^{n}\rd_{j_{i}} \rd^{2} N(z_{i})\, \delta^{(3)}\left(\sum_{i} j_{i} N(z_{i})\right),
\ee
where
\be
\rd^{2}N(z) = \frac{1}{2\pi} \frac{\rd^2z}{(1+|z|^2)^2}
\ee
is the normalized measure on the unit 2-sphere parametrised by $z$.
The factors of $(1+|z|^2)^2$ in the denominator are introduced for
later convenience and are compensated in a formula below. 
Now that we have written the integrand in terms of the coordinates $z_1,z_2,z_3$ and $Z_i, i>3$,
we need to obtain a similar representation for the measure $\rd \mu^{(n)}(z_{i})$. As a first
step towards this goal, we notice that this is an ${\rm SU}(2)$ invariant measure on
the constraint surface that we denote by $P^{(0)}_{\vj}:= H^{-1}_{\vj}(0)$.
Thus, it is given by a product of the Haar measure on ${\rm SU}(2)$ times the symplectic
measure on the quotient manifold $S_{\vj}=P^{(0)}_{\vj}/\mathrm{SU(2)}$. Apart from some
numerical factors (see below), the non-trivial part of this decomposition, i.e.
the symplectic measure is defined as follows. The symplectic structure on $P_{\vj}$ is given by
$\sum_{i} \omega_{j_{i}}(z_{i})$ where $\omega_{j}$ is the sphere symplectic structure (\ref{Ssymp}).
If we denote by $i: P^{(0)}_{\vj}\to P_{\vj}$ the inclusion map and by $\pi: P^{(0)}\to S_{\vj}$ the
projection map, the induced symplectic structure $\Omega_{\vj}$ on $S_{\vj}$ is defined so that
$i^{*} (\sum_{i} \omega_{j_{i}}(z_{i})) = \pi^{*}(\Omega_{\vj})$. We now have the
following formula for the integration measure:
\be\label{measure}
\rd \mu^{(n)}(z_{i}) = 4\pi^2
\left(\prod_{i=1}^{n}\frac{1}{2\pi}\frac{\rd_{j_{i}}}{2j_{i}}\right) \,\, \rd k
\wedge \frac{\Omega^{n-3}_{\vj}}{(n-3)!},
\ee
where $dk$ is the (normalized) Haar measure on $\mathrm{SU(2)}$.

\bigskip

\noindent{\bf Proof:}
In this formula the $j$ dependent prefactor just comes from the discrepancy between the normalisation
of the symplectic measure $2j$ and  the unity decomposition measure $\rd_{j}=2j+1$.
The  numerical prefactor $1/(2\pi)^n$ comes from the relative normalisation of the measure on the sphere relative to the
symplectic measure $ \rd_{j} \rd^{2} N = (1/2\pi)(\rd_{j}/2 j) \omega_{j}$.
The additional factor of $4\pi^{2}$  is the volume of the $\mathrm{SU(2)}$ group
with respect to the unnormalized measure on SU(2) (see appendix \ref{gmeas}).
The fact that  apart from this numerical factor the measure
splits as a product is due to the fact that SU(2) vector fields are orthogonal
with respect to the symplectic measure to vectors tangent to the constraint surface.
Indeed, if $\hat{X}$ is a vector field denoting the action of SU(2),
such that $i_{\hat{X}}\Omega = -dH_{X}$, where $\Omega$ is the symplectic form on
$P_{\vj}$ and $\xi$ is a vector tangent to the constraint surface, then
$\Omega(\xi,\hat{X}) = - \xi i_{\hat{X}} \Omega = \xi (H_{X}) =0.$
This implies that the determinant of the symplectic measure factorises as a product of determinant
for each factor and therefore the measure factorizes. 

\bigskip

To obtain a decomposition of the identity formula in terms of the cross-ratios we now only need to express the
symplectic potential $\Omega_{\vj}$ in term of the cross-ratio coordinates $Z_i$, as well as the coordinates
$z_1,z_2,z_3$. We have explicitly computed this symplectic potential in
the most interesting case for us, that is $n=4$, with the result being~:
\be
\label{Omega}
{\Omega}_{\vj}\Big|_{n=4} =   2 \frac{ {\prod_{i=1}^{4} (2j_{i})}}{ \det(G_{\vj})}
\frac{ |z_{43}z_{21}|^{4}}{\prod_{i=1}^{4} (1+|z_{i}|^{2})^{2}}
\frac{\rd Z \wedge \rd \overline{Z}}{i}\,.
\ee
\noindent{\bf Proof:}
Instead of computing directly the induced symplectic structure it is equivalent but easier
to compute the Poisson bracket of $Z$ with $\overline{Z}$. This is given by
\be
\{Z,\overline{Z}\} = \sum_{i} \frac{i}{2j_{i}} (1+|z_{i}|^{2})^{2} |\partial_{z_{i}}Z|^{2},
\ee
which follows directly from the expression (\ref{Ssymp}) for the symplectic 2-form on the sphere. Now
using the definition of the cross-ratio (\ref{holo4}), one can easily see that
\be
\partial_{z_{i}}Z
 =- \frac{ z_{jk} z_{kl}z_{lj}}{z_{43}^{2}z_{21}^{2}}\, ,
\ee
where $(i,j,k,l) $ stands for an arbitrary (even) permutation of $(1,2,3,4)$.
From this we get the Poisson bracket
\be
\{Z,\overline{Z}\} =\frac{i}{2|z_{43}z_{21}|^{4}} \frac{\prod_{i=1}^{4} (1+|z_{i}|^{2})^{2}}{\prod_{i}(2j_{i})}
16\sum_{i<j<k} j_{i}j_{j}j_{k}
\frac{|z_{ij}|^{2}|z_{jk}|^{2}|z_{ki}|^{2}}{(1+|z_{i}|^{2})^{2}(1+|z_{j}|^{2})^{2}(1+|z_{k}|^{2})^{2}}.
\ee
To finish the computation we need to recognize in this expression the
determinant of the metric $G_{\vj}$. This determinant is computed explicitly in Appendix \ref{Gdet},
where we show that when the closure condition is satisfied $\sum_ij_i N(z_i)=0$,  we have (for any $n$)
\be\label{Gdetfor}
\det\left(G_{\vj}\right)(z_{i}) = 16 \sum_{i<j<k} j_{i}j_{j}j_{k}
\frac{|z_{ij}|^{2}|z_{jk}|^{2}|z_{ki}|^{2}}{(1+|z_{i}|^{2})^{2}(1+|z_{j}|^{2})^{2}(1+|z_{k}|^{2})^{2}}.
\ee
Using this we can rewrite the above formula for the Poisson brackets of $Z,\overline{Z}$ as
\be
\{Z,\overline{Z}\} =
\frac{i}{2|z_{43}z_{21}|^{4}} \frac{\prod_{i=1}^{4} (1+|z_{i}|^{2})^{2}}{\prod_{i}(2j_{i})} \det(G_{\vj}).
\ee
Inverting this Poisson bracket gives the symplectic structure (\ref{Omega}).

\subsection{Holomorphic form of the decomposition of the identity}

We now have all the ingredients to write the identity decomposition in a holomorphically
factorized form. Taking into account the expression (\ref{measure}) for the measure,
as well as the representation (\ref{holo1}), we can now rewrite (\ref{unity-alt}) as the following integral
\beq
\label{unity-alt*}
\mathbbm{1}_{\vec{\jmath}} =  8\pi^2 \prod_{i=1}^{n}\frac{\rd_{j_{i}}}{2\pi}
\int_{S_{\vj}} \frac{\Omega^{n-3}_{\vj}}{(n-3)!}
\left(\frac{\prod_{i=1}^{n}(1+|z_{i}|^{2})^{2}
\mathrm{det}\left(G_{\vj}(z_{i})\right)}
{2\prod_i (2j_i) \, |z_{12}|^{2(n-2)} |z_{23}|^{2(4-n)}|z_{13}|^{2(4-n)}\prod_{i=4}^{n}|z_{3i}|^{4}}\right)
\times \\ \nonumber
\hat{K}_{\vj}(Z_{i},\overline{Z_i}) \,  |\vj,Z_{i}\ket \bra \vj, Z_{i} |
\eeq
where the integral is now over the quotient $S_{\vj}$. Note that we have
dropped the integral over $\SU(2)$ since the integrand is $\SU(2)$-invariant. To transform the result further,
one just has to substitute here an expression for the measure $\Omega^{n-3}_{\vj}$. Above we have
found such an expression for the most interesting cases $n=3,4$, so let
us analyze these cases.

In the case $n=3$ the integral drops out since the quotient space consists of a single point.
The formula (\ref{Gdetfor}) for the determinant implies that the term in parenthesis is equal to
unity. The identity decomposition (\ref{unity-alt*}) in this case thus becomes
\be\label{unity-alt-3}
{1}=
\frac{1}{\pi} \rd_{j_{1}}\rd_{j_{2}}\rd_{j_{3}}
\hat{K}_{{j_{1},j_{2},j_{3}}} \,  N_{j_{1}j_{2}j_{3}}^{2},
\ee
where $N_{j_1,j_2,j_3}$ is the normalization coefficient of the coherent  intertwiner introduced in (\ref{holo3}).

In the case $n=4$ the factors depending on $z_1,z_2,z_3$ in the symplectic 2-form (\ref{Omega}) exactly cancel
those in the parenthesis and the integral (\ref{unity-alt*}) simply becomes
\be
\label{unity-alt-4*}
\mathbbm{1}_{\vec{\jmath}} =  \frac{1}{2\pi^{2}}\prod_{i=1}^{4} {\rd_{j_{i}}}
\int_{\mathbb{C}}  {\rd^{2}Z}   \, \hat{K}_{\vj}(Z,\overline{Z}) \,  |\vj,Z\ket \bra \vj, Z |\,,
\ee
which is our previous result (\ref{unity-n}) specialized to the case $n=4$.

\section{The K\"ahler potential on $S_{\vj}$ and Semi-Classical Limit}
\label{Kahlersc}

The purpose of this section is to study a geometrical interpretation of our main formula
(\ref{unity-n}) in the case of large spins. From the perspective of geometric quantization
of a K\"ahler manifold $S_{\vj}$ we could expect that the integration measure in the
physical inner product formula (\ref{inner-phys}) is given by the exponential of the
K\"ahler potential on $S_{\vj}$. Below we shall see that this is indeed the case
in the limit of large spins.

\subsection{A Useful Representation for the Kernel}

Let us start by remarking that, comparing the expression for the measure on $S_{\vj}$ given
by the formulae (\ref{unity-n}) and (\ref{unity-alt*}), we see that for any $n$
we must have
\be\label{symp-meas}
\frac{\Omega^{n-3}_{\vj}}{(n-3)!} \frac{\mathrm{det}\left(G_{\vj}(z_{i})\right)}{2\prod_i (2j_i)}=
\frac{|z_{12}|^{2(n-2)} |z_{23}|^{2(4-n)}|z_{13}|^{2(4-n)}\prod_{i=4}^{n}|z_{3i}|^{4}}{\prod_{i=1}^{n}(1+|z_{i}|^{2})^{2}}
\prod_{i=4}^n d^2Z_i\,.
\ee
The prefactor on the right-hand-side in this formula, i.e.
\be\label{omega-pfaff}
e^{-2\Phi_\bullet(z_i,\overline{z_i})}\equiv
\frac{|z_{12}|^{2(n-2)} |z_{23}|^{2(4-n)}|z_{13}|^{2(4-n)}\prod_{i=4}^{n}|z_{3i}|^{4}}{\prod_{i=1}^{n}(1+|z_{i}|^{2})^{2}},
\ee
turns out to play an important role in the geometrical description of the constraint surface. Let us give
an expression for the function $\Phi_\bullet(z_i,\overline{z_i})$ in coordinates $g,Z_i$. The change of
coordinates is easily computed using the explicit map (\ref{coord-change}) between the $z_i$'s and the $Z_i$'s,
and we get:
\be\label{Phi-bullet}
e^{\Phi_\bullet(g,Z_i,\overline{Z_i})}=
\left( |b|^{2}+|d|^{2}\right)\left(|a+b|^{2}+|c+d|^{2}\right)\left(|a|^{2} + |c|^{2}\right)
\prod_{i=4}^{n}\left( |aZ_{i}+b|^{2}+|cZ_i+d|^{2}\right).
\ee

Considering the expression (\ref{ker-phys}) for $\hat{K}_{\vj}(Z,\bar{Z})$ in terms of an integral over $g\in\SL(2,\C)$,
we can now give the following useful representation for the kernel
\be\label{ker-rep}
\hat{K}_{\vj}(Z_i,\bar{Z_i})= \int_{\mathrm{SL}(2,\C)}\!\!\!\!\!\!\!\! \rd^{\mathrm{norm}}g \,\,
e^{-\Phi_{\vj}(g,Z_i,\overline{Z_i})-2\Phi_\bullet(g,Z_i,\overline{Z_i})},
\ee
where
\be\label{Phi-Z}
e^{\Phi_{\vj}(g,Z_i,\overline{Z_i})}\equiv
(|b|^2+|d|^2)^{2j_1} (|a+b|^2+|c+d|^2)^{2j_2} (|a|^2+|c|^2)^{2j_3}
\prod_{i=4}^{n}\left(|cZ_{i}+d|^{2}+ |aZ_{i}+b|^{2}\right)^{2j_{i}}\, ,
\ee
and the function $\Phi_\bullet(g,Z_i,\overline{Z_i})$ is given by (\ref{Phi-bullet}) above.
Note that $\Phi_\bullet(g,Z_i,\overline{Z_i})=\Phi_{\vj=1/2}(g,Z_i,\overline{Z_i})$ where all the spins are taken equal to $\f12$.

\subsection{The K\"ahler Potential on $S_{\vj}$}

The purpose of this subsection is to note that the function
$\Phi_{\vj}(g,Z_i,\overline{Z_i})$ that appears in
the exponent of (\ref{ker-rep}) is essentially the K\"ahler potential on the original
unconstrained phase space $P_{\vj}$ but written in coordinates $g,Z_i$, and corrected
by adding to it a holomorphic and an anti-holomorphic function of the coordinates on $P_{\vj}$. Indeed, we have:
\be
\Phi_{\vj}(g,Z_i,\overline{Z_i})= \sum_i 2j_i \log(1+|z_i|^2) + h_{\vj}(g,Z_i)+\overline{h_{\vj}(g,Z_i)},
\ee
where
\be
h_{\vj}(g,Z_i)= 2j_1 \log(d) + 2j_2 \log(c+d)
+ 2j_3 \log(c) + \sum_{i=4}^n 2j_i \log(cZ_i+d)
\ee
is a holomorphic function of $g$ and the cross-ratio coordinates $Z_i$ (and thus the original coordinates $z_i$).
We recognize the standard K\"ahler potential on $n$ copies of the sphere plus
a holomorphic and an anti-holomorphic function of the coordinates
$z_i$. Thus, the function $\Phi_{\vj}(g,Z_i,\overline{Z_i})$ can also be used as the K\"ahler
potential on the unconstrained phase space $P_{\vj}$, for an addition of a holomorphic and
an anti-holomorphic function does not change the symplectic form and thus produces an equivalent potential.

The K\"ahler potential on the constraint surface $S_{\vj}$ is then simply obtained by evaluating
$\sum_i 2j_i \log(1+|z_i|^2)$, or equivalently $\Phi_{\vj}(g,Z_i,\overline{Z_i})$, on this surface.
For a general $n$ and generic values of spins this K\"ahler potential on $S_{\vj}$ is not
easy to characterize in any explicit fashion. However, the described characterization is sufficient
for seeing that the K\"ahler potential gets reproduced in the semi-classical limit of
large spins by the kernel (\ref{ker-phys}). For $n=4$ and all spins being equal,
we evaluate explicitly the K\"ahler potential in section \ref{Asymptsc}.

\subsection{Kernel in the Semi-Classical Limit}

We first note that the formula (\ref{symp-meas}) essentially computes for us the
Pfaffian of the matrix of the symplectic two-form $\Omega_{\vj}$. Recall that the 
Pfaffian of a $2n\times 2n$ antisymmetric matrix is given by
\be
\mathrm{Pf}(\omega)= \frac1{2^{n}n!}\epsilon^{a_{1}\cdots a_{2n}}\omega_{a_{1}a_{2}}\cdots
\omega_{a_{2n-1}a_{2n}}
\ee
and that $\det(\omega)= \mathrm{Pf}(\omega)^{2}$. Thus, from the definition (\ref{omega-pfaff}) of
the function $\Phi_\bullet(z_i,\overline{z_i})$,
we have
\be\label{OG}
\frac{\mathrm{Pf}(\Omega_{\vj})\det(G_{\vj})}{2 \prod_{i=1}^{n}(2j_{i})} = e^{-2\Phi_\bullet(z_i,\overline{z_i})},
\ee
with  $\Phi_\bullet(z_i,\overline{z_i})$ given in terms of the $g,Z_i$ coordinates on the unconstrained
phase space by the equation(\ref{Phi-bullet}).

We can now turn to the main task of this section which is to study the holomorphic factorization
formula (\ref{unity-n}) in the semi-classical limit where all spins are rescaled homogeneously
$j_{i}\to \lambda j_{i}$ with $\lambda \to \infty$. For simplicity, we restrict our
attention to the case of main interest, which is $n=4$, but all arguments remain
essentially unchanged in the general $n$ case. We now use the representation
(\ref{ker-rep}) for the kernel. When all the spin are rescaled we have:
\be\label{ker-asympt}
\hat{K}_{\lambda \vj}(Z,\bar{Z})= \int_{{\rm SL}(2,\C)} 
\rd g \, e^{-\lambda\Phi_{\vj}(g,Z) - 2 \Phi_\bullet(g,Z)}.
\ee
In the limit where $\lambda \to \infty$ the integral is dominated by the points where
the action $\Phi_{\vj}(g,Z)$ reaches its minimum. The analysis performed in \cite{CF3} established
that: (i) the minimum points of the action are the ones satisfying the closure
condition $H_{\vj}(z_{i}(g,Z)) =0$; (ii)  given any $Z \notin \{0,1,\infty\}$ there is a
{\it unique} (up to $g\to -g$) Hermitian $g$ such that the closure condition is satisfied, we denote this solution by
$g(Z)$ and $z_{i}(Z)\equiv z_{i}(g(Z),Z)$; (iii) the Hessian of the action is the metric $G_{\vj}$.
In brief, these results stem from the fact that any orbit on the space
$\{z_{1},\cdots,z_{4}\}$ generated by the transformations in the Hermitian direction of
$\mathrm{SL}(2,\C)$ crosses the constraint surface $H_{\vj}=0$ only once.
Moreover the metric along these orbits is given by $ G_{\vj}$. The fact that the
second derivative of $\Phi_{\vj}$ along the Hermitian direction in $\mathrm{SL}(2,\C)$ is given by
a positive metric $G_{\vj}$ implies that the function is convex and possess a unique 
minimum that lies on the constraint surface.
Therefore, the asymptotic behaviour of the kernel (\ref{ker-asympt}) for large spins $j_i$ is given by
\be
\hat{K}_{\vj}(Z,\bar{Z})\sim \left({2\pi}\right)^{\frac32}
\frac{e^{-\Phi_{\vj}(Z,\overline{Z})}}{e^{2\Phi_\bullet(Z,\overline{Z})}\sqrt{ \det(G_{\vj}(Z))}} ,
\ee
where $\Phi_{\vj}(Z,\overline{Z}) \equiv \Phi_{\vj}(g(Z),Z,\overline{Z})$ and the definition of
$\Phi_\bullet(Z,\overline{Z})$ is similar, and $G_{\vj}(Z)\equiv G_{\vj}(z_{i}(Z))$. We have dropped
the rescaling factor $\lambda$ in order not to clutter the formulae. It is assumed here that all spins
are uniformly large\footnote{In other words, we take the limit $j_{i}\to \infty$ while
$\Delta j / j = O(1/j^{2})$ where $\Delta j$ is the difference between any two spins and $j=1/n \sum_{i=1}^{n}j_{i}$.}.
We can now remark that the term in the denominator is familiar to us. Indeed using (\ref{OG})
we can rewrite it in terms of the symplectic potential:
\be
\hat{K}_{\vj}(Z,\bar{Z})\sim
\sqrt{\frac{(2\pi)^{3}}{2  \prod_{i}(2j_{i})}}\sqrt{\mathrm{Pf}(\Omega_{\vj})} \,
e^{-\Phi_{\vj}(Z,\overline{Z}) -\Phi_\bullet(Z,\overline{Z})}.
\ee
Let us note that $\det (G_{\vj}(Z))$ scales as $j^3$ and that $\mathrm{Pf}(\Omega_{\vj})$ grows
as $j^{n-3}$, i.e as $j$ for the case $n$=4. Using this asymptotic expression for the kernel we can write
an asymptotic decomposition of the identity
\be
\label{asmeta-unity}
\mathbbm{1}_{\vec{\jmath}} \sim  \sqrt{\frac{\prod_{i}(2j_{i})}{{\pi }}}
\int_{\mathbb{C}}  {\rd^{2}Z} \, \sqrt{\mathrm{Pf}(\Omega_{\vj})} \,
e^{-\Phi_{\vj}(Z,\overline{Z}) - \Phi_\bullet(Z,\overline{Z})}  \,
  \,  |\vj,Z\ket \bra \vj, Z |\,.
\ee
This is precisely what is expected from the perspective of geometrical quantization of K\"ahler
manifolds. Indeed, as we have seen in the previous subsection, the quantity $\Phi_{\vj}(Z,\overline{Z})$
is just the K\"ahler potential of the reduced phase space $S_{\vj}$. To give an interpretation
to other terms appearing in this formula let us recall some standard facts about geometric quantization.

In the ``naive''  geometrical quantization scheme which we used
in this paper so far the states $\psi_{0}$ are defined to be
(after a choice of trivialization of the  quantization  bundle\footnote{The quantization bundle is a Hermitian
line bundle over the phase space $P$ with curvature given by $i \omega$, where $\omega$ is the symplectic two-form.
If $P$ is simply connected this quantization bundle is unique.
It exists only if $\hat{\omega}\equiv \omega/2\pi$ is an integral two form, i-e such that
$\int_{S}\hat{\omega} \in \mathbb{N}$ for any closed surface $S$.})
holomorphic  functions (i-e holomorphic $0$-form) on the phase space and the scalar product is given by
\be\label{geom-naive}
||\psi_{0}||^{2}= \int_{P} \mathrm{Pf}(\hat{\omega}) e^{-\Phi} |\psi_{0}|^{2},
\ee
where $\omega$ is the symplectic form on $P$,  $\hat{\omega} \equiv  \omega/ 2\pi $ is  the
integral two form and $\Phi$ is the K\"ahler potential $\bar{\partial} \partial \Phi = i\omega$.
This form of quantization is often called that of Bargmann-Segal in the physics literature,
and geometric quantization  in the mathematics literature,  canonical references are \cite{Woodhouse,Tuynman}.

It is well-known however that a more accurate geometrical quantization includes the so-called ``metaplectic''
or more appropriately
half-form correction. In the geometrical quantization with half-form correction
the states $\psi_{1/2}$ are holomorphic half-forms on the phase space and the scalar product is given by
\be\label{geom-real}
||\psi_{1/2}||^{2} = \sqrt{\mathrm{Pf}\left(\hat{\omega}\right)}(0)
\int_{P} \sqrt{\mathrm{Pf}\left(\hat{\omega}\right)}e^{-\Phi} |\psi_{1/2}|^{2}.
\ee
where $0$ denotes the point at which $\Phi$ reach its minimum.

As an example illustrating the above discussion we mention that the quantization of the sphere given by (\ref{id-single})
can  be written (up to a normalization coefficient) in the ``naive'' K\"ahler form if one takes as
the K\"ahler potential $\Phi_{j}(z)\equiv 2j \ln(1+|z|^{2})$ and as the symplectic structure
$\omega_{j} = \partial \bar{\partial} \Phi_{j}(z) = 2j/(1+|z|^{2})^2$. With this choice the decomposition
of the identity (or the scalar product formula) reads
\be
\mathbbm{1}_j = \frac{\rd_{j}}{2j}  \int \frac{\rd^{2}z}{2\pi}
\mathrm{Pf}(\omega_{j}) e^{-\Phi_{j}(z)} \, |j,z\ket \bra j,z|,
\ee
which coincides with the naive geometrical quantization (\ref{geom-naive}) up to a prefactor
$\rd_j/2j$ that goes to 1 in the large spin limit. However, the correct prescription (\ref{geom-real})
gives (\ref{id-single}) without the need for any prefactors. Indeed, in this case
one chooses the K\"ahler potential to be $\Phi_{j}(z)\equiv (2j+1) \ln(1+|z|^{2})$ (note the shift $2j \to 2j+1$)
and then (\ref{geom-real}) gives precisely (\ref{id-single}).

Comparing (\ref{geom-real}) and (\ref{asmeta-unity}) we see that the K\"ahler potential
$\Phi_{\vj}(Z,\overline{Z})$ on the space of shapes $S_{\vj}$ is correctly reproduced.
Moreover, our result (\ref{asmeta-unity}) reproduces
not just the ``naive'' K\"ahler potential $\Phi_{\vj}$, but even the metaplectic corrected one
given in this case by $\Phi_{\vj}+\Phi_{\bullet}$, which also amounts to the shift $2j_{i} \to 2j_{i}+1$.

It is worth emphasizing that the integration kernel in (\ref{inner-phys}) is only
equal to (minus) the exponential of the K\"ahler potential on $S_{\vj}$ in the limit
of large spins. For generic spins the two quantities differ by  quantum
corrections. Thus, for generic values of spins, the quantization of $S_{\vj}$ provided by the
identity decomposition formula (\ref{unity-n}), even though equivalent (in the sense that there is
an isomorphism of the resulting Hilbert spaces) to the geometric
quantization (\ref{geom-real}), is in details {\it different} from it. The isomorphism
between the two Hilbert spaces is quite non-trivial, involves quantum corrections and
is only unitary asymptotically (for large spins), see \cite{BHall} for more details
on this point.

In this section we have only discussed the case $n=4$, but it is easy to see that
it generalizes without any difficulty to the arbitrary $n$ case. The only novelty in the general
case is a different numerical prefactor (i.e., the right-hand-side should be multiplied by
$1/(2\pi)^{n-4}$) and the replacement of $Z$ by $Z_{i}$'s.

We would like to finish this section by pointing out that our results imply that
the $n$-point function of the bulk-boundary correspondence of string theory
has the interpretation of the (exponential of the) K\"ahler potential on
the space of shapes $S_j$. This is surprising, at least to the present
authors, and appears to be a new result. It would be of interest to provide
some more direct argument for why this is the case,
but we leave this interesting question to further work.

\section{Invariant Operators on the Intertwiners Space}
\label{Invarsc}

In this section we will focus exclusively on the case $n=4$ relevant for
the quantum tetrahedron, even though some of our results can easily be extended to the
general case.

Now that we have defined the holomorphic intertwiners  $|\vj,Z\ra$ and shown that they provide
(\ref{unity-n}) an over-complete basis in the Hilbert space of intertwiners
${\cal H}_{\vj} = \left(V_{j_{1}}\otimes\cdots \otimes V_{j_{4}}\right)^{\SU(2)}$,
we would like to describe and study  the action  of $\mathrm{SU}(2)$-invariant operators on this basis.
It is well known that a generating set of such operators is given by
\be
 J_{ij} \equiv (J_{(i)} + J_{(j)})^{2} = J_{(i)}^{2} + J_{(j)}^{2} + 2 \vec{J}_{(i)}\cdot\vec{J}_{(j)},
\ee
where the $\mathfrak{su}(2)$ generators $J^a_{(i)}$, $a=1..3$, act on the $i$-th vector space $V_{j_{i}}$, with
the action of $J^2_{(i)}$ on $V_{j_{i}}$ (and thus ${\cal H}_{\vj}$) being diagonal with eigenvalue $j_{i}(j_{i}+1)$.
The operators $J_{ij}$ are then positive Hermitian operators acting on the intertwiners space
${\cal H}_{\vj}$. Diagonalising these operators one gets an orthonormal basis in ${\cal H}_{\vj}$
denoted $|\vj,k\ket^{ij}$, where the subscript ${ij}$ labels which ``channel'' one works with ,
and $k$  is the ``intermediate'' spin in the corresponding channel, running from $|j_i-j_j|$ to $(j_i+j_j)$.
These intertwiners are defined by the conditions
\be
J_{ij} |\vj,k\ket^{ij} = k(k+1)  |\vj,k\ket^{ij},\qquad
{}^{ij}\bra \vj,k'|\vj, k\ket^{ij} = \frac{\delta_{k,k'}}{\rd_{k}}.
\ee

Our goal in this section is to express the operators $J_{ij}$ as second-order
differential operators in the complex variable $Z$ and then describe the eigenstates as holomorphic functions of $Z$.
From this we will deduce the overlap function between the real and holomorphic intertwiners:
\be
{}^{ij}C_{\vj}^{k}(Z) \equiv  {}^{ij}\bra \vj, k | \vj, {Z}  \ket.
\ee
We find that these eigenvectors are essentially hypergeometric polynomials.
As a by-product of our results we obtain a check of the phased Gaussian ansatz
for coherent states on the quantum tetrahedron \cite{Rovelli:2006fw}, \cite{Livine:2007vk},
as well as an integral formula for the $\{6j\}$-symbol in terms of $Z$.

\subsection{Scalar Product Operators and their Eigenfunctions}

In this subsection we obtain and study $J_{ij}$ as second-order differential operators in the complex variable
$Z$. The action of $J_{ij}$ on coherent states was computed in \cite{CF3}. One finds that it acts as a holomorphic
differential operator in the variables $z_i,z_j$ on which the coherent intertwiner $||\vj,z_i\ra$ depends. Explicitly,
\be
J_{ij}\,=\,
-z_{ij}^2\pp_i\pp_j + 2z_{ij}(j_i\pp_j-j_j\pp_i)+ (j_{i}+j_{j})(j_{i}+j_{j}+1).
\ee
It is easy to check that this second-order differential operator has the correct spectrum by
computing its action on polynomials in $z_{ij}=z_i-z_j$. Thus, we have:
\be
J_{ij}\,\left(z_{ij}\right)^l\,=\,
\left[
(j_i+j_j-l)(j_i+j_j-l+1)
\right]\,\left(z_{ij}\right)^l.
\ee
The identification $k=(j_i+j_j-l)\ge 0$ shows that such polynomials are indeed eigenvectors of
eigenvalue $k(k+1)$. However, the state $z_{ij}^{l}$ is not an $\SU(2)$-invariant vector. In order
to find eigenvectors in the space of invariant vectors (intertwiners), we need to express this
operator as acting on functions of $Z$. This is an exercise in a change of variables. Indeed,
recall that the holomorphic intertwiners $|\vj,Z\ra$ are related to the coherent intertwiners $||\vj,z_i\ra$
by a non trivial pre-factor (\ref{stateZ}):
\be\label{prefact}
||\vj,z_1,\ldots,z_4\ket = P_{\vj}(z_1,\ldots,z_4) \,\,|\vj, Z\ket,\qquad
P_{\vj}(z_1,\ldots,z_4)\,=\, z_{12}^{j_{1}+j_{2}-j_{34}}
z_{23}^{j_{34}-j_{12}}
z_{31}^{j_{34}+j_{12}}
z_{43}^{ 2j_{4}},
\ee
where we have denoted $ j_{ij}\equiv j_{i}-j_{j}$. One can now commute the differential operator
$J_{ij}$ through the pre-factor  $P_{\vj}(z_1,\ldots,z_4)$, and then translate the partial derivatives $\pp_i,\pp_j$
with respect to the complex labels $z_i$ and $z_j$ into the derivative $\pp_Z$ with respect to the cross-ratio
$Z$. We denote the resulting operator by $\Delta_{ij}$:
\be
P_\vj(z_1,\ldots,z_4)\,\Delta_{ij}|\vj, Z\ket\,\equiv
J_{ij}\,P_{\vj}(z_1,\ldots,z_4)\,|\vj, Z\ket.
\ee

For definiteness, we now focus on the operator $\Delta_{12}$. As we shall see, this operator happens
to be the simplest one in the sense that it is precisely of the hypergeometric form, while
to relate the other operators to the hypergeometric-form ones one needs to do some extra work,
see below. In order to convert the $\pp_1,\pp_2$ derivatives we use the following identities
valid for an arbitrary function $\phi(Z)$:
\be
\pp_1\phi \,=\, \f{z_{42}}{z_{21}z_{41}}Z\pp_Z\phi,\quad
\pp_2\phi \,=\, \f{z_{31}}{z_{23}z_{21}}Z\pp_Z\phi,\quad
\pp_2\pp_1\phi \,=\, \f{1}{z_{21}^2}\left[Z(1-Z)\pp_Z^2\phi+(1-2Z)\pp_Z\phi\right].
\ee
After straightforward but somewhat lengthy algebra one gets
\beq\label{Delta}
\Delta_{12}&=&
 Z(Z-1)\pp_Z^2\,+
\left(2(j_{34}+1)Z-{(1+j_{34}-j_{12})}\right)\pp_Z + j_{34}(j_{34}+1).
\eeq
One recognizes a second order differential operator of the hypergeometric type.
It is then well-known that the hypergeometric function $F(a,b;c|Z)$ gives one
of the two linearly-independent solutions of $\Delta_{(a,b;c)} \phi=0$, where
\be\label{hyper-type}
\Delta_{(a,b;c)} \equiv  Z(Z-1)\pp_Z^2\,+
\left( (a+b+1)Z- c \right)\pp_Z + ab.
\ee
The $\SU(2)$-invariant operator $\Delta_{12}-k(k+1)$ is then a hypergeometric operator $\Delta_{(a,b;c)}$ with parameters
\be
a= -k+ j_{34},\quad b= k + j_{34}+1,\quad c=  j_{34}-j_{12} +1.
\ee

However, since the operator $\Delta_{(a,b;c)}$ is second-order, there are two linearly-independent
solutions. The question is then which linear combinations of them corresponds to the eigenvectors
of $\Delta_{12}$ that we are after. The answer to this is as follows. When $k$ respects the bounds
$\mathrm{max}(|j_{12}|,|j_{34}|)\leq k\leq \mathrm{min}(j_{1}+j_{2},j_{3}+j_{4})$
one of the solution of the arising hypergeometric equation is polynomial.
This is the eigenfunction we are looking for, and it is given by
\be\label{Jac}
\hat{P}_{k-j_{34}}^{(j_{34}-j_{12},\,j_{34}+j_{12})}(Z)=\frac{(k-j_{12})!}{(k-j_{34})!(j_{34}-j_{12})!}
F(-k+j_{34},k+j_{34}+1; j_{34}-j_{12} +1;Z ).
\ee
Here $P_{n}^{(a,b)}$ denotes the Jacobi polynomial and $\hat{P}_{n}^{(a,b)}(Z) \equiv {P}_{n}^{(a,b)}(1-2Z)$ is
the {\it shifted } Jacobi polynomial (see appendix \ref{app:Jac} for some useful facts about the shifted polynomials).
This expression is valid if $j_{34} \geq |j_{12}|$, and we can assume that this inequality is satisfied without
any loss of generality. Indeed, we can always assume that $j_{12}\geq 0$ since otherwise we can exchange the
role of $1$ and $2$ as is implied by the equality
\be\label{ex1}
\hat{P}_{k-j_{34}}^{(j_{34}-j_{12},\,j_{34}+j_{12})}(Z)= (-1)^{k-j_{34}}
\hat{P}_{k-j_{34}}^{(j_{34}+j_{12},\,j_{34}-j_{12})}(1-Z).
\ee
We can also assume that $j_{34}\geq j_{12}$ since otherwise we can exchange $(12)$ with $(34)$
using to the exchange identity
\be\label{ex2}
\hat{P}_{k-j_{34}}^{(j_{34}-j_{12},\,j_{34}+j_{12})}(Z) =
\frac{(k-j_{12})!(k+j_{12})!}{(k-j_{34})!(k+j_{34})!} (-Z)^{j_{12}-j_{34}}
\hat{P}_{k-j_{12}}^{(j_{12}-j_{34},\,j_{12}+j_{34})}(Z),
\ee
which is valid if $ j_{12}\geq j_{34}$.

A special case where all formulae simplify considerably
is when all representations are equal, $j_1=j_2=j_3=j_4=j$, which correspond to a tetrahedron
with all faces having the same area. Then our eigenfunctions reduce to the shifted Legendre polynomial:
\be
\hat{P}^{(0,0)}_{i}(Z)=P_{i}(1-2Z)
\,=\,
\sum_{l=0}^{i}
\left(\begin{array}{c}i\\l\end{array}\right)^2
\,(-Z)^{i-l}(1-Z)^l.
\ee
Note that in this case the eigenvectors actually do not depend on the spin $j$. The dependence on $j$ will
nevertheless reappear in the normalization of these states.

By construction the polynomials $\hat{P}_{k-j_{34}}^{(j_{34}-j_{12}, \,j_{34}+j_{12})}$ are eigenstates
of the operator $\Delta_{12}$. These eigenvectors give, up to normalization, matrix elements of the
change of basis between the holomorphic intertwiner $|\vj, Z\ket$ and the usual orthonormal
intertwiners $| \vj, k \ket^{12}$ that diagonalize $\Delta_{12}$. More precisely, if we define the
overlap\footnote{From now on when we work in the channel $12$ drop the superscript $12$ to avoid notation cluttering.}
\be
C_{\vj}^{k}(Z) \equiv  \bra  \vj, k | \vj, Z \ket\, ,
\ee
the above discussion shows that it is proportional to the Jacobi polynomial:
\be\label{C-Jac}
C_{\vj}^{k}(Z) = {N_{\vj}^{k}} \, \hat{P}_{k-j_{34}}^{(j_{34}-j_{12},\, j_{34}+j_{12})}(Z),
\ee
where the non-trivial normalization coefficient is given by the integral
\be\label{normP}
\left(N_{\vj}^{k}\right)^{-2} = \frac{1}{2\pi^{2}} \prod_{i=1}^{4}\rd_{j_{i}}
\int \rd^{2} Z  \, \hat{K}_{\vj}(Z,\overline{Z}) \left| \hat{P}_{k-j_{34}}^{(j_{34}-j_{12}, \,j_{34}+j_{12})}(Z)\right|^{2}.
\ee
Our task is now to determine these normalization coefficients.

\subsection{3-Point Function and Normalization of the 4-Point Intertwiner}

Instead of computing the integral (\ref{normP}) directly, which is quite non-trivial,
we will take an alternative route and express the normalized intertwiner $|\vj, k\ket$ as a combination of
certain Clebsch-Gordan coefficients. To this end, we start by reminding the reader some information about
the coherent intertwiner for $n$=3.

In this trivalent case the space of intertwiners is one dimensional and
we have denoted in section \ref{shapes} the unique normalized $\bra 0|0\ket=1$ intertwiner by $|0\ket$. As the analysis
of that section shows, the coherent intertwiner $||\vj,z_i\ket$ is proportional to $|0\ket$,
with the proportionality coefficient being the normalization factor $N_{j_1,j_2,j_3}$ times
a $z_i$-dependent pre-factor. Explicitly:
\be\label{3-coherent}
||\vj,z_1,z_2,z_3\ket = C_{j_1,j_2,j_3}(z_1,z_2,z_3)|0\ket = N_{j_{1},j_{2},j_{3}}
\,\,z_{12}^{j_{1}+j_{2}-j_{3}}z_{23}^{-j_{1}+j_{2}+j_{3}}z_{13}^{j_{1}-j_{2}+j_{3}} |0\ket.
\ee
In (\ref{unity-alt-3}) we have related the normalization coefficient $N_{j_{1},j_{2},j_{3}}$ to the
3-point function $\hat{K}_{j_1,j_2,j_3}$. The 3-point function can be computed explicitly,
see (\ref{K3}). One gets the result~:
\be\label{3norm}
N_{j_{1},j_{2},j_{3}}^{2} = \frac{[2j_{1}]![2j_{2}]![2j_{3}]!}
{[j_{1}+j_{2}+j_{3}+1]! [-j_{1}+j_{2}+j_{3}]! [j_{1}-j_{2}+j_{3}]! [j_{1}+j_{2}-j_{3}]! }.
\ee

Now given the trivalent Clebsch-Gordan map we can construct the normalized $4$-valent intertwiner
-- an eigenstate of $J_{12}$ -- by gluing two 3-valent intertwiners. Indeed, as described in \cite{CF3},
there is a gluing map $g: V_{k} \otimes V_{k} \to \C $ that can be represented in terms of coherent states as
\be
g =\rd_{k}\int \frac{\rd^{2} N(z)}{  (1+|z|^{2})^{2k}} \bar{z}^{2k} | -1/ \bz \ket \otimes  | z\ket.
\ee
In terms of this gluing map the normalized intertwiner is given by
 \be
C^{k}_{\vj}(z_{i}) = \rd_{k}
\int   \frac{C_{j_{1},j_{2},k}(z_{1},z_{2}, -1/ \bz) C_{k, j_{3}, j_{4}}(z,z_{3},z_{4})}{(1+|z|^{2})^{2k}}
\bar{z}^{2k} \, \rd^{2}N(z)\,.
\ee

At first sight this integral seems quite cumbersome.
However, we know that the integral being holomorphic and $\SU(2)$-invariant
is entirely determined by its values at the special points $(0,1,\infty,Z)$.
Thus, the overlap between the holomorphic $|\vj,Z\ket$ and the real
$|\vj, k\ket$ intertwiners can be extracted as the limit
$C^{k}_{\vj}(Z) =\lim_{X\to \infty} (-X)^{-2j_{3}} C^{k}_{\vj}(0,1,X,Z)$.
Using the explicit expression (\ref{3-coherent}) for the ${\rm SU}(2)$-invariant 3-point function
we get the following  integral representation for the overlap:
\be \nonumber
C_{\vj}^{k}(Z) =  (-1)^{s-2k}
\rd_{i}  N_{j_{1},j_{2},k} N_{k, j_{3},j_{4}}\,
I_{\vj}^{k}(Z),
\ee
where $s=j_1+j_2+j_3+j_4$ and
\be
I_{\vj}^{k}(Z)\equiv \int \rd^{2}N(z)
\frac{(\bar{z} +1)^{k-j_{12}}(z-Z)^{k-j_{34}}}{(1+|z|^{2})^{2k}}.
\ee
In order to compute the integral we perform the following change of variables:
\be
\sqrt{u}e^{i\phi} = \frac{ z}{\sqrt{1+|z|^{2}}},\quad \sqrt{1-u} = \frac{ 1}{\sqrt{1+|z|^{2}}},
\quad \rd u \frac{\rd \phi}{2\pi} = \rd^{2}N(z),
\ee
then expand all the terms and perform the integration over $\phi$. We are left with
\be
I_{\vj}^{k}(Z) =\sum_{n} \frac{(k-j_{12})!(k-j_{34})!}{(k-j_{12}-n)! (k-j_{34}-n)! (n!)^{2}}
\left(\int_{0}^{1}u^{n}(1-u)^{2k-n} \rd u\right) (-Z)^{k-j_{34}-n }.
\ee
Now making use of the standard formula
\be
\int_{0}^{1} u^{n}(1-u)^{m} =\frac{n! m!}{(n+m+1)!}
\ee
we recognized that the integral in question is proportional to the hypergeometric function
\beq
I_{\vj}^{k}(Z) &=&\frac{(k-j_{12})! (k+j_{34})!}{(2k+1)! (j_{34}-j_{12})!} F(-k+j_{34},k+j_{34}+1; j_{34}-j_{12} +1;Z ).
\eeq
Thus we get the final expression in terms of the Jacobi polynomial (\ref{C-Jac})
with the normalization coefficient given by
\be\nonumber
N_{\vj}^{k} = (-1)^{s-2k}
\sqrt{\frac{(2j_{1})!(2j_{2})!(2j_{3})!(2j_{4})!(k+j_{34})! (k-j_{34})!}
{(j_{1}+j_{2}+k+1)!(j_{3}+j_{4}+k+1)! (j_{1}+j_{2}-k)! (j_{3}+j_{4}-k)!(k+j_{12})! (k-j_{12})! }}.
\ee
By writing the factorials in this formula as $\Gamma$-functions we can
extend the definition of the normalization coefficient $N_{\vj}^{k}$ beyond its initial domain of validity
$j_{34}\leq k\leq \mathrm{min}(j_{1}+j_{2},j_{3}+j_{4})$ (recall that we are under the assumption $j_{34}\geq |j_{12}|$).
Since $1/\Gamma(0)=0$ one sees however that $N_{\vj}^{k}=0$ if $k= j_{1}+j_{2}+1$ or $k= j_{3}+j_{4}+1$.
This implies that the Jacobi polynomial corresponding to this value is not normalisable with respect
to our norm $\int d^{2}Z K_{\vj}$, and so this particular Jacobi polynomial is not part of the Hilbert space.
To explore the other boundary  $k=j_{34}-1$ one first needs to rewrite the overlap in terms of the
hypergeometric function $C_{\vj}^{k} = \tilde{N}_{\vj}^{k} F$ and notice again that the normalization
coefficient $\tilde{N}_{\vj}^{k}$ vanishes at the boundary $k=j_{34}-1$, as long as $j_{34}>j_{12}$.

In the case all $j_{i}$'s are equal to a given spin $j$, the expression (\ref{C-Jac}) simplifies to
\be\label{C-equiarea}
C_{\vj}^{k}(Z) = \frac{(-1)^{2k}}{2j+k+1} \frac{(2j)!(2j)!}{(2j+k)! ( 2j-k)! }\, \hat{P}_{k}(Z).
\ee
We will need this expression in section \ref{Asymptsc}.

\subsection{Other Channels}

In the previous two subsections we have studied the channel $12$ and the associated operator $J_{12}$
whose eigenstates $|\vj, k\ket^{12}$ provided a real basis in the 4-valent intertwiners Hilbert space.
This choice of the channel is somewhat distinguished by the fact that, with our choice (\ref{holo4}) for
the cross-ratio coordinate $Z$, the second-order holomorphic operator $\Delta_{12}$ turned out to be
precisely of the hypergeometric type (\ref{hyper-type}) so that the eigenstates -- the real intertwiners --
are just the Jacobi polynomials (\ref{Jac}).

It is also interesting and important to compute the other channel operators and their eigenstates.
Indeed, consider for example the channel $23$. There is similarly an operator $J_{23}$ and the
basis in $\cH_{j_1,j_2,j_3,j_4}$ given by its eigenstates $|\vj, k\ket^{23}$. The overlap
${}^{23}\bra \vj, k| \vj, l\ket^{12}$ is the $6j$-symbol, and this is why the other basis
in the Hilbert space is of interest. We can similarly find the holomorphic representation of
$J_{23}$ by commuting it with the prefactor (\ref{prefact}). With our choice (\ref{holo4}) of the cross-ratio,
however, the resulting holomorphic operator is not exactly of the type (\ref{hyper-type}).
Indeed, the computation is completely similar to the one performed in the $12$ channel.
We use:
\be
\pp_3\phi \,=\, \f{z_{24}}{z_{23}z_{43}}Z\pp_Z\phi,\quad
\pp_2\phi \,=\, \f{z_{31}}{z_{23}z_{21}}Z\pp_Z\phi,\quad
\pp_3\pp_2\phi(Z)
\,=\, \f{1}{z_{23}^2}\left[Z^2(Z-1)\pp_Z^2\phi+Z^2\pp_Z\phi\right].
\ee
to obtain
\beq\label{Delta-23}
\Delta_{23}&=&
 Z^2(1-Z)\pp_Z^2+
\left[(j_{1}+j_{2}-j_{34}+2j_{4}-1)Z-2(j_1+j_4)\right]Z\pp_Z\nn\\
&&-2j_{4}(j_{1}+j_{2}-j_{34})Z +(j_1+j_4)(j_1+j_4+1).
\eeq
Thus, this operators is not exactly of the hypergeometric type. However,
as we shall explain below, its eigenfunctions are also related to Jacobi
polynomials via a simple transformation of the cross-ratio coordinate.

Let us also mention that one can similarly compute $\Delta_{13}$ with the result being:
\beq
\Delta_{13}&=&Z(Z-1)^2\pp_Z^2
+\left[(j_1+j_2-j_{34}+2j_4-1)Z+j_{34}-j_{12}+1\right](1-Z)\pp_Z\nn\\
&&+2j_4 (j_1+j_2-j_{34})(Z-1)+ (j_2+j_4)(j_2+j_4+1),
\eeq
which is also not of the hypergeometric type. It is then easy to check that:
\be\label{Delta-rel}
\Delta_{12}+\Delta_{13}+\Delta_{23}=j_1(j_1+1)+j_2(j_2+1)+j_3(j_3+1)+j_4(j_4+1),
\ee
which is an expected relation that follows from the definition of $J_{ij}$ and
the condition $\vec{J}_{(1)}+\vec{J}_{(2)}+\vec{J}_{(3)}+\vec{J}_{(4)}=0$ that is just the
requirement of $\SU(2)$-invariance of the intertwiner space. Thus, in view of
(\ref{Delta-rel}) it is sufficient to compute only two scalar product operators,
say $\Delta_{12}$ and $\Delta_{13}$, in order to get the expression of all scalar
product operators. Indeed these operators are symmetric, $\Delta_{12}=\Delta_{21}$,
and operators with opposite labels are equal, $\Delta_{12}=\Delta_{34}$.

Moreover, as we shall now explain, it is in fact sufficient to compute only one
of these operators, for the two other inequivalent operators, as well as their
eigenstates can be obtained by considering the action of the group of
permutations acting on $z_1,\ldots,z_4$. Thus,  we denote by $\sigma_{ij}$ the permutation
that exchanges the variables $(j_i,z_i)$ and $(j_j,z_j)$ in the functional $P(z_1,\ldots,z_4)$
defined in (\ref{prefact}) and in the state $||\vj,z_1,\ldots,z_4\ra$. The action of these permutation
can then be extended to the holomorphic intertwiners by
\be
\hat{\sigma}_{ij}| \vj, Z\ket \equiv P_{\vj}(z_1,\ldots,z_4)^{-1} \Big(
\sigma_{ij} P_{\vj}(z_1,\ldots,z_4) |\vj,Z\ra \Big).
\ee
Since $P_{\vj}(z_1,\ldots,z_4) $ and $\sigma_{ij}P_{\vj}(z_1,\ldots,z_4)$ have the same transformation
properties under conformal transformations, i.e., $P_{\vj}\,(z_{i}^{g}) = \prod_{i}(cz_{i}+d)^{-2j_{i}} P_{\vj}(z_{i})$,
it follows that $\sigma_{ij}$ is well defined as an operator acting purely on the cross-ratio $Z$.
The action of all 24 different permutations on the coherent intertwiner $|\vj,Z\ket$ is given in appendix
\ref{perm}. For instance, we have
\beq\nn
\hat{\sigma}_{12}|\vj, Z\ket &=&(-1)^{s-2j_3} \, |\vj_{12},1-Z\ket,\\ \label{perm-action}
\hat{\sigma}_{23}|\vj, Z\ket &=&(-1)^{2j_2} \, \left(1-Z\right)^{2j_{4}} \left|\vj_{23}, \frac{Z}{Z-1} \right\ket,\\ \nn
\hat{\sigma}_{13}|\vj, Z\ket &=& (-1)^{s}\, Z^{2j_{4}} \left|\vj_{13}, \frac1{Z} \right\ket.
\eeq
where $s=j_{1}+j_{2}+j_{3}+j_{4}$ and  $\vj_{ij}\equiv \hat{\sigma}_{ij}(\vj)$.

Using (\ref{perm-action}) we can now understand why the computed above operators
$\Delta_{13},\Delta_{23}$ turned out to be not of the hypergeometric type. Indeed,
these operators can be obtained from $\Delta_{12}$ by conjugation with $\hat{\sigma}_{ij}$. We have:
$\hat{\sigma}_{23} \Delta_{12}\hat{\sigma}_{23} = \Delta_{13}, \hat{\sigma}_{13} \Delta_{12}\hat{\sigma}_{13} = \Delta_{23}$.
Let us consider the channel $23$ in more details. Thus, if we denote by $\hat{\Delta}_{23}$ the hypergeometric
operator obtained from $\Delta_{12}$ (\ref{Delta}) by exchanging  $j_{1}$ with $j_{3}$, then the operator
$\Delta_{23}$ is related to this hypergeometric operator by a change of variables $Z \to Z^{-1}$
followed by the conjugation with $Z^{2j_{4}}$, see (\ref{perm-action}). In other words, we have:
\be
(\Delta_{23}F ) (Z^{-1}) = \left(Z^{-2j_{4}} \tilde{\Delta}_{23} Z^{2j_{4}}\right) F(Z^{-1}).
\ee
Let us see that this is indeed the procedure that gives (\ref{Delta-23}). To this end,
let us first write $\Delta_{23}$ in terms of the variable $X\equiv Z^{-1}$. One gets:
\beq\nn
\tilde{\Delta}_{23}&=&
 X(X-1)\pp_X^2+
\left[2(j_1+j_4+1)X -{(j_{1}+j_{2}-j_{34}+2j_{4}+1)}\right]\pp_X\\ \nn
&& -\frac{2j_{4}(j_{1}+j_{2}-j_{34})}{X}+(j_1+j_4)(j_1+j_4+1).
\eeq
Now conjugating this operator with $X^{-2j_{4}}$ one gets
\be\nn
X^{2j_{4}}\tilde{\Delta}_{23}X^{-2j_{4}}=
X(X-1)\pp_X^2+
\left[2(j_1-j_4+1)X -{(j_{1}-j_{4}+j_{2}-j_{3}+1)}\right]\pp_X +(j_1-j_4)(j_1-j_4+1).
\ee
which is the hypergeometric operator obtained from $\Delta_{12}$ by exchanging $j_{1}$ and
$j_{3}$ as expected.

An interesting application of the above discussion is as follows. Let us consider the overlap
${}^{23}C_{\vj}^{k}(Z)$ of the holomorphic intertwiner $|\vj,Z\ket$ with the basis $|\vj,k\ket^{23}$
diagonalizing $J_{23}$ (and thus $\Delta_{23}$). We can then express these overlap coefficients in terms
of the ones in the $12$ channel given by the Jacobi polynomials:
\be\label{C-1-Z}
{}^{23}C_{\vj}^{k}(Z) =
(-1)^{s}\, Z^{2j_{4}} \, {}^{12}C_{\vj}^{k}(Z^{-1}).
\ee
This formula allows to get an interesting expression for the usual $6j$-symbol of $\SU(2)$.
Indeed, the $6j$-symbol is given by the overlap between basis states in two different channels.
For example, we can consider
\be
{}^{23} \bra  \vj, l | \vj, k \ket^{12} = \left\{
\begin{array}{ccc}
j_1 & j_2 & k \\
j_3 & j_4 & l
\end{array}
\right\}\, .
\ee
Now inserting into this formula the holomorphic identity decomposition (\ref{unity-alt-4*}),
and using (\ref{C-1-Z}) we get:
\be
\left\{
\begin{array}{ccc}
j_1 & j_2 & k \\
j_3 & j_4 & l
\end{array}
\right\} \, =\,
\frac{1}{2\pi^2} \prod_{i=1}^4 \rd_i \int d^2Z\, \hat{K}_\vj(Z,\bZ)\,
(-1)^{s}\, \bZ^{2j_{4}} \, {}^{12}C_{\vj}^{l}(1/\bZ) \, {}^{12}C_{\vj}^{k}(Z).
\ee
Here the coefficients ${}^{12}C_{\vj}^k(Z)$ are essentially the Jacobi polynomials, see
(\ref{C-Jac}).

\section{On the bulk/boundary 4-point function}
\label{CFTsc}

In this section we derive some non-trivial properties of the 4-point function $\hat{K}_{\vj}(Z,\bZ)$
and compare them with what is known in the literature about this object. Here we shall use some
of the facts about the action of permutation group derived in the previous section.

\subsection{Hermiticity and Measure}

The operator $\Delta_{12}$ studied in the previous section is a positive Hermitian operator
with respect to the inner product on $\cH_{j_1,j_2,j_3,j_4}$ defined by the kernel $\hat{K}_{\vj}(Z,\bZ)$.
Indeed, it descends from $(J_{(1)}+J_{(2)})^{2}$, which is obviously Hermitian. However, since it is given by
a hypergeometric-type operator $\Delta_{12} = \Delta_{a,b,c}$ where
\be\label{Delta-abc}
\Delta_{a,b,c}=Z(Z-1)\pp_Z^2+((a+b+1)Z-c)\pp_Z-ab
\ee
with the hypergeometric parameters
\be\label{hyper-param-12}
a= j_{34},\quad b= j_{34}+1,\quad c=j_{34}-j_{12}+1,
\ee
the hermiticity of $\Delta_{12}$ implies the equality
\be
\bra \phi|\Delta_{12}|\phi \ket=
\int d^2Z \, K_{\vj}(Z,\bZ)\,\overline{\phi(Z)}\,\Delta\phi(Z)\,=\,
\int d^2Z \, K_{\vj}(Z,\bZ)\,\overline{\Delta\phi(Z)}\,\phi(Z)
\ee
for all holomorphic functions $\phi$. In turn, integrating by parts, this leads to a constraint
on the kernel $\hat{K}_{\vj}(Z,\bar{Z})$. Thus, one finds that $\hat{K}$ satisfies a ``balanced"
hypergeometric equation in both $Z$ and $\bZ$:
\be
\label{equationonK}
\Delta'_{a,b,c}\,\hat{K}_{\vj}(Z,\bar{Z}) = \overline{\Delta'_{a,b,c}}\,\hat{K}_{\vj}(Z,\bar{Z}),
\ee
where $\Delta'$ is the transpose  of $\Delta_{a,b,c}$, which is an operator
of the same type, i-e $\Delta'_{a,b,c}=\Delta_{a',b',c'}$ with:
\be
a'=1-a = 1-j_{34}, \quad b'=1-b=-j_{34}, \quad c'=2-c=1-j_{34}+j_{12}.
\ee

This equation strongly suggests that $\hat{K}(Z,\bar{Z})$ must be given by an expansion in terms of
joint equal eigenvalue eigenstates of $\Delta'$ and $\overline{\Delta'}$. Moreover, because of the positivity
of $\Delta_{12}$ these common eigenvalue should be positive. The space of positive eigenvalue $\lambda(\lambda+1)$
eigenstates of $\Delta'_{a,b,c}$ is (real) two-dimensional. Let us now discuss the eigenfunctions.
When $j_{12}\geq j_{34}$ a convenient set of linearly independent solutions of
$\Delta'_{a,b,c}\phi=\lambda(\lambda-1)\phi$, which is the same as the
set of linearly independent solutions of $\Delta_{-j_{34}+\lambda,1-j_{34}-\lambda,1-j_{34}+j_{12}}\phi=0$
is given by\footnote{Because for integral $\lambda$ all coefficient entering
the hypergeometric function are integers we cannot take a basis of two hypergeometric functions with the same
argument as it would be usual for generic hypergeometric functions.
This is why we use here a mixed basis with arguments $Z$ and $1/(1-Z)$.
The existence of such a basis follows from the Kummer's relations. For the equation
$\Delta_{a,b,c}\phi(z)=0$ we take the set $F(a,b;c;z)$ and
$z^{1-c}(z-1)^{c-a-1}F(a-c+1,1-b;a-b+1;1/(1-z))$.}
\beq\nonumber
F(-j_{34} +\lambda, 1-j_{34}-\lambda;1-j_{34}+j_{12}; Z), \\ \label{lin-indep}
\mathrm{and} \quad
Z^{j_{34}-j_{12}}(Z-1)^{j_{12}-\lambda}  F(- j_{12}+\lambda,j_{34}+\lambda; 2 \lambda; 1/(1-Z)).
\eeq
The problem is to find which linear combination of these solutions arises
in the holomorphic decomposition of $K_{\vj}$.

To see this, we note that, as reviewed in appendix \ref{app-n-point}, the kernel
$\hat{K}_{\vj}(Z,\bar{Z})$ admits a representation as a double series expansion around $Z=0$.
Thus, in particular it is regular at $Z=0$. However, it must also be regular at $Z=\infty$.
Indeed, the 4-point function $K_{\vj}(z_{i})$ transforms covariantly under the permutations $\sigma_{ij}$ of
$(j_{i}, z_{i})$. This translates into non-trivial identities for the kernel $\hat{K}_{\vj}(Z,\bar{Z})$,
similar to those derived in the previous section for the holomorphic intertwiner. For instance,
we have:
\be\label{K-perm}
\hat{K}_{j_1,j_2,j_3,j_4}(Z,\bZ) = \hat{K}_{j_2,j_1,j_3,j_4}(1-Z,1-\bZ) =
|1-Z|^{-2\Delta_4} \hat{K}_{j_2,j_3,j_1,j_4}\left(\f1{1-Z},\f1{1-\bZ}\right),
\ee
where $\Delta_{i}= 2(j_{i}+1)$. These should be compared
with (\ref{perm-action}). The last equality plus our regularity condition at $Z=0$ shows that
$\hat{K}_{\vj}(Z,\bZ)$ vanishes in the limit $Z\to \infty$ as $|Z|^{-2\Delta_{4}}$, and so is regular.
This immediately implies that the expansion of $\hat{K}_{\vj}(Z,\bZ)$ should be in terms of functions regular
at infinity. Only the second set of solutions in (\ref{lin-indep}) is regular at infinity, so
we must expect an expansion
\be\label{K-factor}
\hat{K}_{\vj}(Z,\bar{Z}) =\sum_{\lambda} a_{\lambda}
|Z^{j_{34}-j_{12}}(Z-1)^{j_{12}-\lambda}  F(- j_{12}+\lambda,j_{34}+\lambda; 2 \lambda; 1/(1-Z))|^{2}\, ,
\ee
where the sum is taken over $\lambda \geq j_{12}$ so that the whole expansion is regular at infinity.

Thus, the requirement of hermiticity of the $J_{ij}$ operators (of which we have considered
only one, but the others lead to the same conclusion) strongly suggests that the kernel
$\hat{K}_{\vj}(Z,\bar{Z})$ {\it holomorphically factorizes} as indicated in (\ref{K-factor}).
However, this holomorphic factorization that might seem surprising from the point of view
taken in this article, is not at all surprising and in fact very much desired from the
point of view of bulk/boundary dualities, that would like to interpret $\hat{K}_{\vj}(Z,\bar{Z})$
(or various closely related objects, see e.g. \cite{Arutyunov:2000ku}) as the 4-point
function of some CFT. If this interpretation is valid, then (\ref{K-factor}) is not
surprising, and follows directly from the defining properties of the CFT as we now review.

\subsection{Holomorphic factorization in CFT}

In this section we would like to understand to what extent the
the quantity $\hat{K}_{\vj}(Z,\bar{Z})$ can be interpreted as
a CFT 4-point function of operators $\phi_{\Delta_{i}}$ of conformal dimensions $\Delta_1,\ldots,\Delta_4$
inserted at points $0,1,\infty,Z$. We base our discussion here on \cite{Dolan:2003hv}.

The first consideration of interest for us is to look at the limit of the 4-point
function when the points $z_1\to z_2$. The cross-ratio $Z$ that we have been working
with is not a convenient coordinate to study this limit, as $Z\sim 1/z_{21}\to\infty$.
Thus, let us introduce a different cross-ratio:
\be\label{U}
U=\frac{z_{12}z_{34}}{z_{13}z_{24}}, \qquad U = \frac{1}{1-Z},
\ee
which goes to zero in the limit $z_1\to z_2$.

Let us now recall some standard facts about 4-point functions in conformal field theory. A very
important role in CFT is played by the so-called operator product expansion. This
interprets the 4-point function as a sum over contributions of (primary and their descendants)
operators of conformal dimension $\Delta$ and spin $l$ to the operator product expansion of $\phi_1(z_1)$
and $\phi_2(z_2)$, where $\phi_{1,2}$ are the so-called primary operators
of given conformal dimension $\Delta_{1,2}$. To make this more precise, let us recall that,
as we have witnessed in this paper already on many occasions, the CFT 4-point function
transforms covariantly under conformal transformations and thus can in general be expressed as:
\be\label{mes-4-point}
\bra \phi_1(z_1)\phi_2(z_2)\phi_3(z_3)\phi_4(z_4)\ket
= \frac{1}{|z_{12}|^{\Delta_1+\Delta_2}|z_{34}|^{\Delta_3+\Delta_4}}
\left(\frac{|z_{24}|}{|z_{14}|}\right)^{\Delta_{12}}
\left(\frac{|z_{14}|}{|z_{13}|}\right)^{\Delta_{34}} F(|U|^2,|V|^2),
\ee
where $\Delta_{ij}=\Delta_i-\Delta_j$, the cross-ratio $U$
is as introduced above (\ref{U}), $V=1-U$, and $F(|U|^2,|V|^2)$ is the 4-point
function as a function of the conformal invariants. The formula (\ref{mes-4-point})
is as given in \cite{Dolan:2003hv}. With appropriate modifications it is valid in any dimensions,
but only in 2 dimensions the two cross-ratios $U,V$ are simply related as $V=1-U$. For later use
we note that the 4-point function (\ref{mes-4-point}) is given as a function of the cross-ratio $Z$ by:
\be\label{4-point-limit}
\lim_{X\to\infty} |X|^{2\Delta_3} \bra \phi_1(0)\phi_2(1)\phi_3(X)\phi_4(Z)\ket =
|1-Z|^{\Delta_{12}}|Z|^{\Delta_{34}-\Delta_{12}} F(|U|^2,|V|^2),
\ee
where $U={1}/{1-Z}$, $V= {Z}/{Z-1}$ are understood to be  functions of $Z$.

For any CFT, the function $F(|U|^2,|V|^2)$ can be
decomposed into a sum of contributions of primary operators with given $\Delta,l$,
where $\Delta$ is a conformal dimension and $l$ is the angular momenta.
so we have:
\be\label{partial-waves}
F(|U|^2,|V|^2)=\sum_{\Delta,l} a_{\Delta,l} \,G_\Delta^{(l)}(|U|^2,|V|^2),
\ee
where the sum is taken over the range of $\Delta,l$ which constitutes the {\it spectrum} of the CFT.
CFT's for which this range is discrete and finite are called rational. But in general
the sum here is an integral.
This expansion correspond to the operator product expansion
$O_{\Delta_{1}}(z_{1},\bar{z}_{1})O_{\Delta_{2}}(z_{2},\bar{z}_{2}) \sim \sum_{\Delta,l} a_{\Delta,l}\, O_{\Delta, l }(z,\bar{z})$.

The requirement of {\it unitarity} states that all the
coefficients $a_{\Delta,l}$ are positive. The representation (\ref{partial-waves})
is also known as the partial wave decomposition, see \cite{Dolan:2003hv} for more
details on this, in particular for analogous formulae in higher dimensions.

The partial waves $G_\Delta^{(l)}(|U|^2,|V|^2)$ can (in any dimension) be characterized completely.
The situation is especially simple in two dimensions. These are functions satisfying the
following eigenvalue differential equation:
\be
L^2 G_\Delta^{(l)}(|U|^2,|V|^2) = - C_{\Delta,l} G_\Delta^{(l)}(|U|^2,|V|^2), \qquad
C_{\Delta,l} = \Delta(\Delta-2) + l^2,
\ee
or its appropriate generalization to higher dimensions. Here $L^2=(1/2){\rm Tr}(L^2)$
is the ``quadratic Casimir'' for the operator $L=L_{1}+L_{2}$ given by the sum of the
generators of the conformal group acting on $\phi_1,\phi_2$. In two dimensions
the operator $L^2$ factorizes into a sum of a holomorphic and an anti-holomorphic
one with respect to $U,\bar{U}$, with each of these being (related to those) of the
hypergeometric type. The solution that behaves as:
\be
G_\Delta^{(l)}(|U|^2,|V|^2)\sim U^{\lambda_1} \bar{U}^{\lambda_2}, \qquad U\to 0,
\ee
where
\be
\lambda_1=\frac{1}{2}(\Delta+l), \qquad \lambda_2=\frac{1}{2}(\Delta-l)
\ee
is given by:
\beq
G_\Delta^{(l)}(|U|^2,|V|^2)&=& U^{\lambda_1} \bar{U}^{\lambda_2}
F(\lambda_1-\frac{1}{2}\Delta_{12},\lambda_1+\frac{1}{2}\Delta_{34},2\lambda_1;U)
F(\lambda_2-\frac{1}{2}\Delta_{12},\lambda_2+\frac{1}{2}\Delta_{34},2\lambda_2;\bar{U}) \nn\\
&& + U\leftrightarrow \bar{U}.
\eeq
Here $F(a,b,c;z)$ is the usual hypergeometric function. See e.g. \cite{Dolan:2003hv}
for a demonstration of all this facts, as well as for a generalization to higher
dimensions.

We can now compare these standard CFT facts with the formula (\ref{K-factor}) we have
been led to by the requirement of hermiticity in the previous subsection. Recalling
(\ref{4-point-limit}) and rewriting everything in terms of the $Z$ cross-ratio,
we see that the 4-point function factorization formula takes precisely the form
(\ref{K-factor}), where in (\ref{K-factor}) the sum is restricted to intermediate
states with zero angular momentum $l=0$ and thus $\lambda_1=\lambda_2=\lambda$. Thus, we learn
that what the requirement of hermiticity suggests for the kernel $\hat{K}_{\vj}(Z,\bZ)$ is in fact
the standard CFT 4-point function partial wave decomposition formula.

To summarize, we see that the kernel function $\hat{K}_{\vj}(Z,\bZ)$ has all the properties of a
CFT 4-point function. It behaves covariantly under the conformal transformations,
of which at the level of the cross-ratio coordinate $Z$ only the permutations
(\ref{K-perm}) remain.
It must moreover admit the holomorphic
factorization of the standard CFT form.
Quite remarkable as we have seen this holomorphic factorisation follows from the requirement
of hermiticity of invariant operators.
However, the underlying CFT
is unknown. Moreover, it is even not obvious that there is an underlying {\it unitary}
CFT, i.e., the one where the partial wave decomposition (\ref{K-factor}) gives rise to
positive coefficients $a_\lambda$. One could argue that this can be determined by explicitly
computing the function $K(Z,\bZ)$ for a given set of conformal dimensions $\Delta_1,\ldots,\Delta_4$
and then expanding the result into partial waves. However, the outcome of this certainly depends on the
conformal dimensions chosen. It may be expected that for those integral dimensions of interest to us
here $\Delta=2(j+1)$ the situation must be simpler. However, the spectrum that should be expected in
(\ref{K-factor}) does not seem to be known in the literature. What has been worked out in the applications
to AdS/CFT correspondence of string theory is the first few terms of the expansion (\ref{K-factor}) (in the case
of four dimensions) for some simple integral conformal dimensions, and these have been shown to match the
boundary CFT predictions. However, no general result seems to be known even in the
simplest case of two dimensions.

We finish this section by pointing out that from the perspective taken in this paper,
the positivity of $a_{\lambda}$ (and thus the existence of some underlying unitary CFT) is
highly plausible since the kernel $\hat{K}_{\vj}(Z,\bZ)$ is a strictly positive functional with the
interpretation of an exponent of the K\"ahler potential on an appropriate moduli space,
see section \ref{Kahlersc}. It is also tempting to think that the factorizability of a CFT
$n$-point function in any dimension could be interpretable as following from the hermiticity of
certain invariant operators, but we leave an attempt to prove all this to future work.

\section{Asymptotic properties of intertwiners}
\label{Asymptsc}

The purpose of this section is to describe the large spins asymptotic properties of
the overlap coefficients $C^k_{\vj}(Z)=\bra \vj, k|\vj, Z\ket$ characterizing
the holomorphic intertwiners in the usual real basis. Recalling that this quantity
consists of the normalization coefficient times the shifted Jacobi polynomial,
we need to develop the asymptotic understanding of both of these pieces. As a
warm-up, let us analyze the case $n=3$, where there is only the normalization
coefficient to consider.

\subsection{A Geometric Interpretation of the $n=3$ Intertwiner}

Recall, that for $n=3$ the coherent intertwiner $||\vj,z_i\ket$, up to a simple
$z$-dependent prefactor, see (\ref{3-coherent}), is the normalization coefficient
$N_{j_1,j_2,j_3}$ times the unique normalized intertwiner $|0\ket$. In this subsection
we would like to develop a geometric interpretation for the normalization
coefficient. This can be achieved by considering the limit of large spins.

The normalization coefficient $N_{j_{1},j_{2},j_{3}}$ is given by
(\ref{3norm}) as a ratio of factorials. In order to evaluate it in the limit of large spins
we use the Stirling formula
\be
n! = \sqrt{2\pi n}\, n^{n}e^{-n} \phi(n)
\ee
where $\phi(n) = 1 + 1/12n +O(1/n^{2})$.
Up to terms of order $O(1/j^{2})$ one finds
\beq\label{3-asympt}
N_{j_{1}j_{2}j_{3}}^{-2}
\approx \sqrt{4\pi}
A(j_{i}) \frac{\Sigma+1}{\sqrt{j_{1} j_{2} j_{3}}}
\left(\frac{(j_{1}+j_{2})^{2}-j_{3}^{2}}{4j_{1} j_{2}}\right)^{2J_{3}}
\left(\frac{(j_{3}+j_{1})^{2}-j_{2}^{2}}{4j_{3} j_{1}}\right)^{2J_{2}}
\left(\frac{(j_{2}+j_{3})^{2}-j_{1}^{2}}{4j_{2} j_{3}}\right)^{2J_{1}}
\eeq
where $\Sigma= j_{1}+j_{2}+j_{3}$ and $ 2J_{i} = \Sigma - 2j_{i}$.
In this formula $$A(j_{i})=\sqrt{[j_{1}+j_{2}+j_{3}] [-j_{1}+j_{2}+j_{3}] [j_{1}-j_{2}+j_{3}] [j_{1}+j_{2}-j_{3}]}/4$$
denotes the area of the triangle of edge length $j_{i}$.

As such this formula is not particularly illuminating, but a nice geometric interpretation can
be proposed by noting that the quantities in brackets are invariant under a simultaneous
rescaling of all the spins. So, they can be rewritten in terms of unit length vectors.
Thus, let us introduce three vectors $j_{1}N_{1},j_{2}N_{2},j_{3}N_{3}$, where $N_{i}^{2}=1$
are unit vectors, with the condition that $j_i N_i$ satisfy the closure  constraint
$j_{1}N_{1}+ j_{2}N_{2}+j_{3}N_{3}=0$. Then these are the three edge vectors of a triangle with edge
length $j_{i}$. One can now check that
\be\label{N12geo}
\left| N_{1} - {N_{2}} \right|^{2}
= \frac{ (j_{1}+j_{2}-j_{3})(j_1+j_2+j_3)} {j_{1}j_{2}},
\ee
which are essentially the factors appearing in the asymptotics (\ref{3-asympt}) of the normalization coefficient.
We can moreover see that the  prefactor entering (\ref{3-asympt}) have a nice geometrical interpretation as those arising
from the determinant of the metric on orbits orthogonal to the constraint surface. Indeed, in case
$n=3$ this determinant, see (\ref{detGN}), is given simply by
\be
\det G_{\vj}=\f{j_1j_2j_3}{4}|N_1-N_2|^2|N_2-N_3|^2|N_3-N_1|^2 = \frac{4A^{2}(j_{i})\Sigma^{2}}{j_{1}j_{2}j_{3}}.
\ee
Thus, approximating $\Sigma+1\sim \Sigma$ at leading order in (\ref{3-asympt}),
the asymptotic evaluation of the normalization coefficient reads
\be
N_{j_{1}j_{2}j_{3}}^{-2}\approx \sqrt{\pi \det(G_{\vj})}
\left(\frac{|N_{1}-N_{2}|}{2}\right)^{4J_{3}}
\left(\frac{|N_{3}-N_{1}|}{2}\right)^{4J_{2}}\left(\frac{|N_{2}-N_{3}|}{2}\right)^{4J_{1}}.
\ee

\subsection{Asymptotics of the Jacobi Polynomial}

Let us now switch to the case of real interest -- that of $n=4$. In this subsection we study the
asymptotics of the shifted Jacobi polynomials. Thus, we are interested in the limit
$j_{i}\to \lambda j_{i}, \lambda \to \infty$. In order to avoid unnecessary cluttering of notation we do not
make the parameter $\lambda$ explicit, with the understanding that the evaluation is performed in the limit
of uniformly large spins.

As shown in appendix \ref{app:Jac}, a convenient integral representation for the shifted Jacobi polynomials
is given by a contour integral:
\be\label{con-integral}
\hat{P}_{k-j_{34}}^{(j_{34}-j_{12},j_{34}+j_{12})}(Z)
= Z^{j_{12}-j_{34}} (1-Z)^{-j_{12}-j_{34}} I(Z),
\qquad I(Z)\equiv \frac{1}{2i\pi}\oint \frac{\rd \omega}{\omega}   e^{-  S_{Z}(\omega)},
\ee
where the contour is around the origin and should avoid the other singularities at $\omega=-Z$ and $\om = 1-Z$.
The  $Z$ dependent action is given by
\be\label{Jac-action}
S_{Z}(\omega) =  (k-j_{34})\ln \omega- (k-j_{12})\ln (Z+\omega) - (k+ j_{12})\ln(1-Z-\omega).
\ee
Note that we have included in the action only the terms that are proportional to the large
parameter $\lambda$, and this is the reason for leaving $1/\om$ outside of the exponent
in (\ref{con-integral}).  In the uniformly large spin limit this integral is evaluated by the steepest
descent method by deforming the integration contour so that it passes through the stationary
point\footnote{The steepest descent method used here may be not completely standard for some readers, so
we briefly review the basic idea. Consider the problem of computing the integral $\int dz g(z) \exp{-\lambda S(z)}$
along some contour in the complex $z$-plane. Here $S(z), g(z)$ are some holomorphic functions of $z$.
Let us, for simplicity, assume that the stationary point $z_0: \partial_z S(z)|_{z=z_0}=0$ is unique and does not
coincide with any of the singularities of the integrand so that the contour can be deformed to pass
through $z_0$. In the limit $\lambda\to\infty$ the integral can be evaluated as follows. Around the
stationary point the ``action'' $S(z)$ admits an expansion $S(z)=S(z_0)+(1/2)S^{(2)}(z_0)(z-z_0)^2+\ldots$.
The second derivative $S^{(2)}(z_0)$ at the stationary point is a complex number, which we
parametrize as $Ae^{i\phi}$. Let us also introduce polar coordinates on the $z$-plane via
$z=z_0+re^{i\theta}$. Then the action near the stationary point behaves as
$S(r,\theta)=S(z_0)+(1/2)Ar^2 e^{i(\phi+2\theta)}$. Therefore, the paths of steepest
descent to the stationary point is $\theta=-\phi/2$. Along this path the integration
measure is $dz=e^{-i\phi/2}dr$ and the resulting real integral can be evaluated
using the usual steepest descent method with the result being
$e^{-i\phi/2} g(z_0) e^{-\lambda S(z_0)} \sqrt{2\pi/(\lambda A)}$. However,
this can be written very compactly as $g(z_0)e^{-\lambda S(z_0)} \sqrt{2\pi/(\lambda S^{(2)}(z_0))}$.}.
The stationary phase equation $\partial_{\omega}S_{Z}(\omega) =0$
leads to a quadratic equation
$$
(k+j_{34}) \omega^{2} + \omega\left((j_{34}+j_{12})(Z-1) + (j_{34}-j_{12}) Z\right) + (j_{34}-k) Z(Z-1)=0
$$
whose solutions are $\omega_{\pm}(Z)\equiv  L(Z) \pm \sqrt{Q(Z)}$,
where $L,Q$ are a linear and quadratic function of $Z$ given by the following expressions:
\beq
L(Z) &\equiv& -\frac{(j_{34}+j_{12})(Z-1) + (j_{34}-j_{12}) Z}{2 (k+j_{34})}\\
Q(Z) &\equiv& \frac{(j_{34}+j_{12})^{2}(Z-1)^{2} + (j_{34}-j_{12})^{2} Z
^{2} +2(2k^{2}-j_{34}^{2}-j_{12}^{2})Z(Z-1)}{4(k+j_{34})^{2}}.
\eeq
Recall now that for non-degenerate tetrahedra $Z$ is complex.
This means that in general $\omega_{\pm}$ are complex numbers and the corresponding
on-shell actions $S_{\pm}(Z) \equiv S_{Z}(\omega_{\pm}(Z))$ are also complex.
In the semi-classical limit of uniformly large spin, only the
root possessing the smallest real value of $S_{Z}$ dominates, the other one being exponentially suppressed.
Without loss of generality we can assume that this root is $\omega_{+}$. Then we get
\be
I(Z) \sim\frac{1}{i\sqrt{2\pi S^{(2)}_{+}(Z)}} \frac{e^{-  S_{+}(Z)}}{\omega_{+}(Z)},
\ee
where  $ S^{(2)}_{\pm}(Z) \equiv \partial_{\omega}^{2}S_{Z}(\omega_{\pm}(Z))$.

\subsection{The Equi-Area Case: Peakedness with respect to $k$}

\begin{figure}[ht]
\begin{center}
\includegraphics[width=5.5cm,angle=-90]{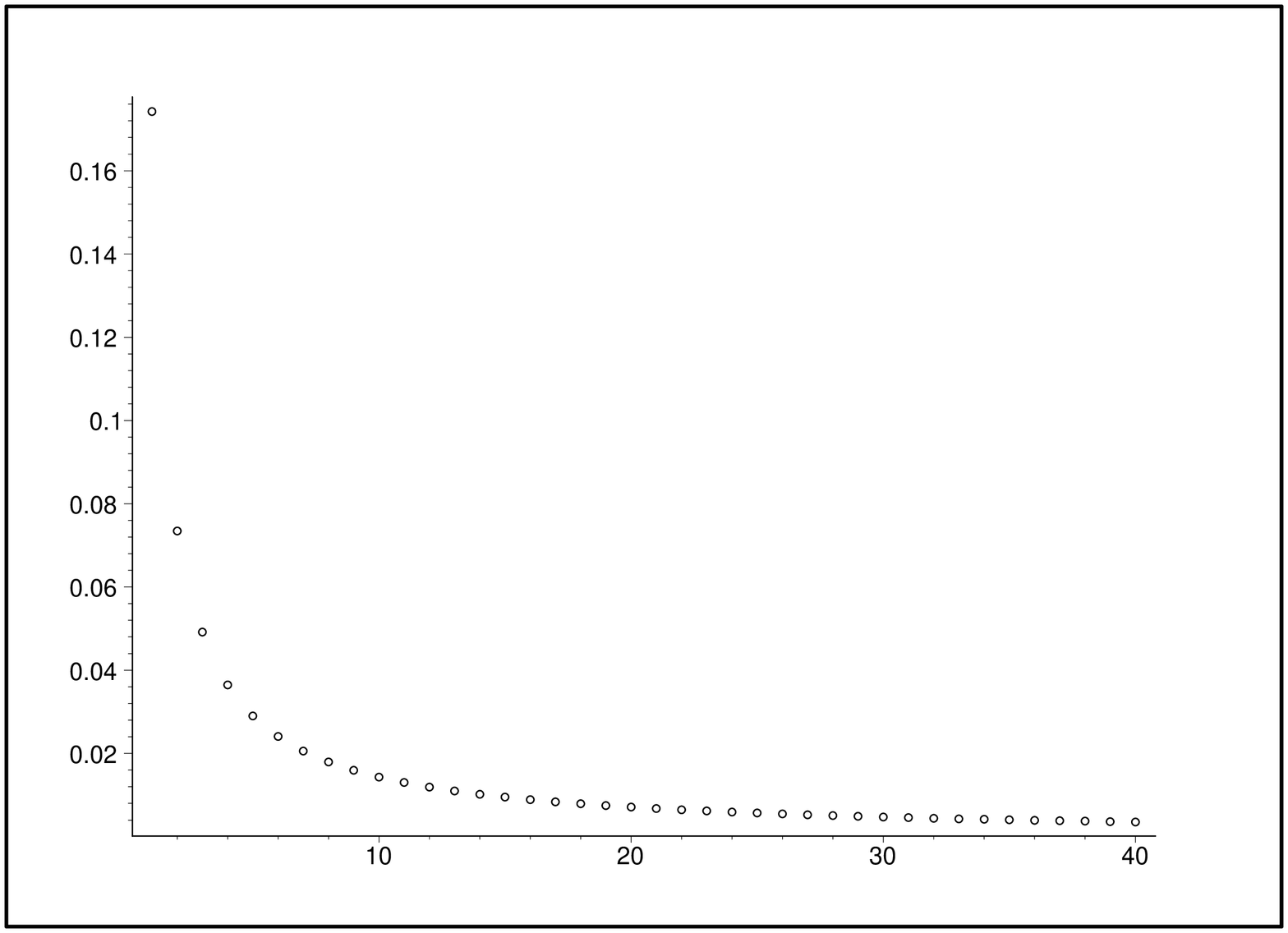}
\includegraphics[width=5.5cm,angle=-90]{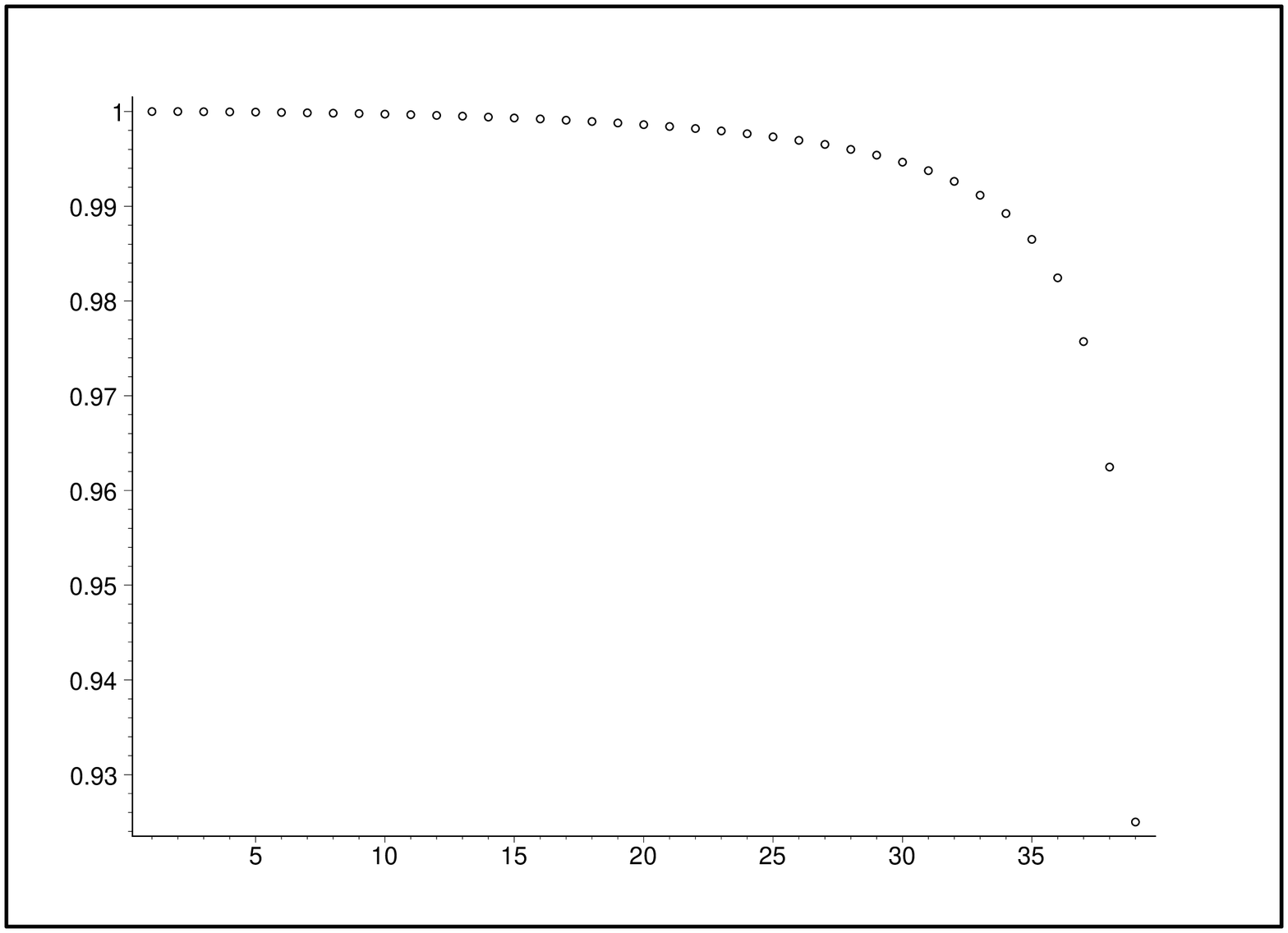}
\caption{On the left, we plot (as a function of $k$) the modulus of the (normalized) difference between the (shifted) Legendre
polynomial $\hat{P}_k(Z)$ and its approximation (\ref{Legendreapprox}) for the value of the cross-ratio
$Z=\exp(i\pi/3)$ corresponding to the equilateral tetrahedron. We see that the asymptotic formula
is already good at 2\% from $k=8$ and at 1\% from $k=15$. On the right, we've plotted (also as a function of $k$) the ratio
between the binomial coefficient $(2j)!^2/(2j-k)!(2j+k)!$ and its approximation (\ref{Binomialapprox})
for $j=20$. We see that the approximation is excellent as long as $k$ doesn't get too close to its maximal value $2j$.}
\label{LegendreApprox}
\end{center}
\end{figure}

In the ``equi-area'' case where all four representations are equal $j_i=j,\,\forall i=1,2,3,4$,
all equations simplify considerably. Thus, the action (\ref{Jac-action}) reduces to:
\be
S(\om)=k\ln\f{\om}{(Z+\om)(1-Z-\om)}.
\ee
The two roots are given by  $\omega_{\pm} = \pm\sqrt{Z(Z-1)}$.
In order to compute the on-shell action it is convenient to introduce a complex
angle $\Theta(Z)$ such that $Z = \cosh^{2} \Theta(Z)$. Then the two roots are given by:
\be
\omega_{\pm} = \pm\sinh\Theta\cosh \Theta=\pm\f12\sinh 2\Theta.
\ee
Changing $\Theta\rightarrow -\Theta$ (which does not affect $Z$) simply exchanges the two roots $\om_+\leftrightarrow\om_-$.
Then it is easy to check that
\beq
(Z+\omega_\pm) = \cosh \Theta \,e^{\pm \Theta},\quad
(1-Z-\omega_\pm) =\mp \sinh \Theta \,e^{\pm \Theta},
\eeq
and so the on-shell action is given by:
\be
S_{\pm}(Z) = \mp 2k \Theta(Z) + ik\pi(2l+1), \qquad l\in \Z.
\ee
The Hessian of $S_{Z}$ at the stationary points can also be computed and we find
\be
S^{(2)}_{\pm}(Z) = \mp\frac{4k }{ \sinh 2\Theta e^{\pm 2 \Theta} }.
\ee
Let us now assume, for definiteness, that the real part of $\Theta$ is positive.
Then the stationary point that dominates is $\om_+$ and we have:
\be
\hat{P}_{k}(Z) \approx (-1)^{k} \sqrt{\frac{1}{ 2\pi  k \sinh 2\Theta(Z)}} e^{(2k+1)\Theta(Z)}.
\label{Legendreapprox}
\ee
One should keep in mind that both the cross-ratio $Z$ and the angle parameter $\Theta(Z)$ are
generically complex, since a real cross-ratio would correspond to a degenerate flat tetrahedron.
Thus, there is both an exponential and oscillating behaviour of the integral. We have compared
the asymptotic formula (\ref{Legendreapprox}) for the Jacobi polynomial to the exact quantity
numerically in Fig.~\ref{LegendreApprox}.

Let us now discuss the numerical prefactor coming from the normalization coefficient.
As we have seen earlier, in the case $j_{i}=j$, the 4-valent overlap coefficient is given by (\ref{C-equiarea}).
Using the Stirling formula, it is easy to give the leading order behaviour of the pre-factor:
\beq
\f{(2j)!^2}{(2j-k)!(2j+k)!}\sim
\f{e^{-4j\Lambda(x)}}{\sqrt{1-x^2}}, \quad
\Lambda(x) \equiv
(1/2)(1-x)\ln(1-x)+(1/2)(1+x)\ln(1+x),
\label{Binomialapprox}
\eeq
where $x=k/2j$. Note that $\Lambda(x)\approx x^2$ for small $x$, so we have a Gaussian
peaked near $x=0$. The ratio between the approximation (\ref{Binomialapprox}) and the
exact quantity is plotted in Fig.~\ref{LegendreApprox}.

Putting everything together we get the semi-classical estimate:
\be\label{C-semi}
C^k_j(Z)\,\sim\,
\f{(-1)^k e^{\Theta(Z)}}{(2j)^{3/2}(1+x)}\sqrt{\frac{1}{2\pi x (1-x^2)\sinh 2 \Theta(Z)}}
\, e^{-4j(\Lambda(x)-x\Theta(Z)) }\, .
\ee
We see that the exponent in (\ref{C-semi}) describes a Gaussian peaked at $x_c$
such that:
\be
\frac{d\Lambda(x)}{dx}\Big|_{x=x_c} = {\rm Re}(\Theta(Z)),
\ee
times an oscillating exponent $\exp{[4i\,j x_c\,{\rm Im}(\Theta(Z))]}$. It is not hard to find $x_c$.
We have for the first derivative $\Lambda^{(1)}(x)
= (1/2)\log[(1+x)/(1-x)]$ and thus $x_{c}(Z)= \tanh \mathrm{Re}(\Theta(Z))$.
In other words
\be\label{kdefthet}
k_{c}(Z)= 2j \tanh \mathrm{Re}(\Theta(Z)).
\ee
It is natural to
expect that the corresponding value $k_c$ should  be the classical value associated with the tetrahedron
in question, that is
\be\label{k-theta}
k_{c}^{2}= j^{2}_{1}+j_{2}^{2} + 2j_{1}j_{2} \cos \theta_{12},
\ee
where $\theta_{12}$  is the dihedral angle between the faces 1 and 2. It is also natural to expect that the phase factor
$2 {\rm Im}(\Theta(Z))$ should be the conjugate variable to $k$, which is the angle
$\varphi_{c}$ between the edges $(12)$ and $(34)$ of the classical tetrahedra determined by $Z$.
Putting the real and imaginary parts together, we should therefore expect the following
relation between the cross-ratio parameter $\Theta(Z)$ and the geometrical variables
\be\label{kphiZ}
e^{2\Theta}(Z) = \frac{2j+k_{c}}{2j-k_{c}} e^{i \varphi_{c}}.
\ee
This formula is relates the two very different descriptions of the phase
space of shapes of a classical tetrahedron -- the real one in terms of the $k,\phi$
parameters and the complex one in terms of the cross-ratio coordinate $Z$. As
is clear from this formula, the relation between the two descriptions is very
non-trivial. Here we have only identified the simplest case of this relation
when all areas are equal, leaving the general case to future studies. Below
we shall check this geometrical interpretation in the case of the equilateral tetrahedron.

Let us now explicitly write the exponent of (\ref{C-semi}) as a Gaussian peaked at $x=x_c$
times some prefactor. The imaginary part of the quantity in the exponent is just
$2k{\rm Im}(\Theta(Z))=k\phi_c$ according to our real parametrization (\ref{kphiZ}).
For the real part we have $\Lambda(x)-x\mathrm{Re}(\Theta(Z))=\Lambda(x_c)-\Lambda^{(1)}(x_c) x_c +
(1/2) \Lambda^{(2)}(x_c)(x-x_c)^2 +\ldots$. The second derivative is $\Lambda^{(2)}=1/(1-x^2)$,
while $\Lambda-\Lambda^{(1)} x = (1/2)\log(1-x^2)$. Thus, going back to the parameter $k$, we see that
the most essential part of the asymptotics (\ref{C-semi}) written in terms of the coordinates $k_c,\phi_c$,
see (\ref{kphiZ}), is given by the following Gaussian:
\be
C^k_j(Z)\,\propto\,
\frac{1}{\left(1-\f{k_{c}^{2}}{4j^{2}}\right)^{2j} }\,\,
\exp{\left( -2j \f{(k-k_{c})^2}{(4j^{2}-k^{2}_{c})}+ i k_{c}\varphi_{c}\right)}.
\ee

An important feature of this state is the fact that its width
\be\label{width}
\sigma= (4j^{2}-k^{2}_{c})/2j = \frac{2j}{\cosh^2 \mathrm{Re}(\Theta(Z))}
\ee
depends not only on $j$ but also on the classical value $k_{c}(Z)$.
This is in qualitative agreement with the analysis performed in \cite{Rovelli:2006fw}.
In this work a Gaussian ansatz for the semi-classical states was postulated and the width was calculated
by asking that it is independent of the channel used. This led to an expression of the width in terms of
matrix elements of the Hessian of the  Regge action. It would be interesting to check that this is indeed
the case to provide an additional justification for the hypothesis made in \cite{Rovelli:2006fw}
as well as to relate our explicit parametrization to the Regge action.

\begin{figure}
\begin{center}
\includegraphics[width=5.5cm,angle=-90]{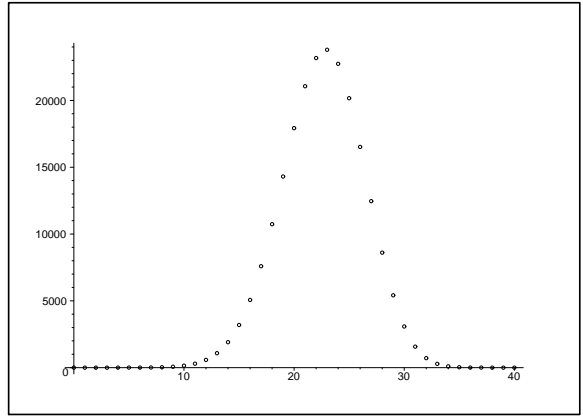}
\caption{We plot the modulus of the equi-area case state $C^k_\vj(Z)$ (for $j=20$) as a function of the spin label $k$,
for the value of the cross-ratio $Z=\exp(i \pi/3)$ that corresponds to the equilateral tetrahedron. It is obvious that
the distribution looks Gaussian. We also see that the maximum is reached for $k_c=2j/\sqrt{3}\sim 23.09$, which
agrees with our asymptotic analysis.}
\label{GaussianPlot-fixed}
\end{center}
\end{figure}

\begin{figure}[h]
\begin{center}
\includegraphics[width=3.7cm,angle=-90]{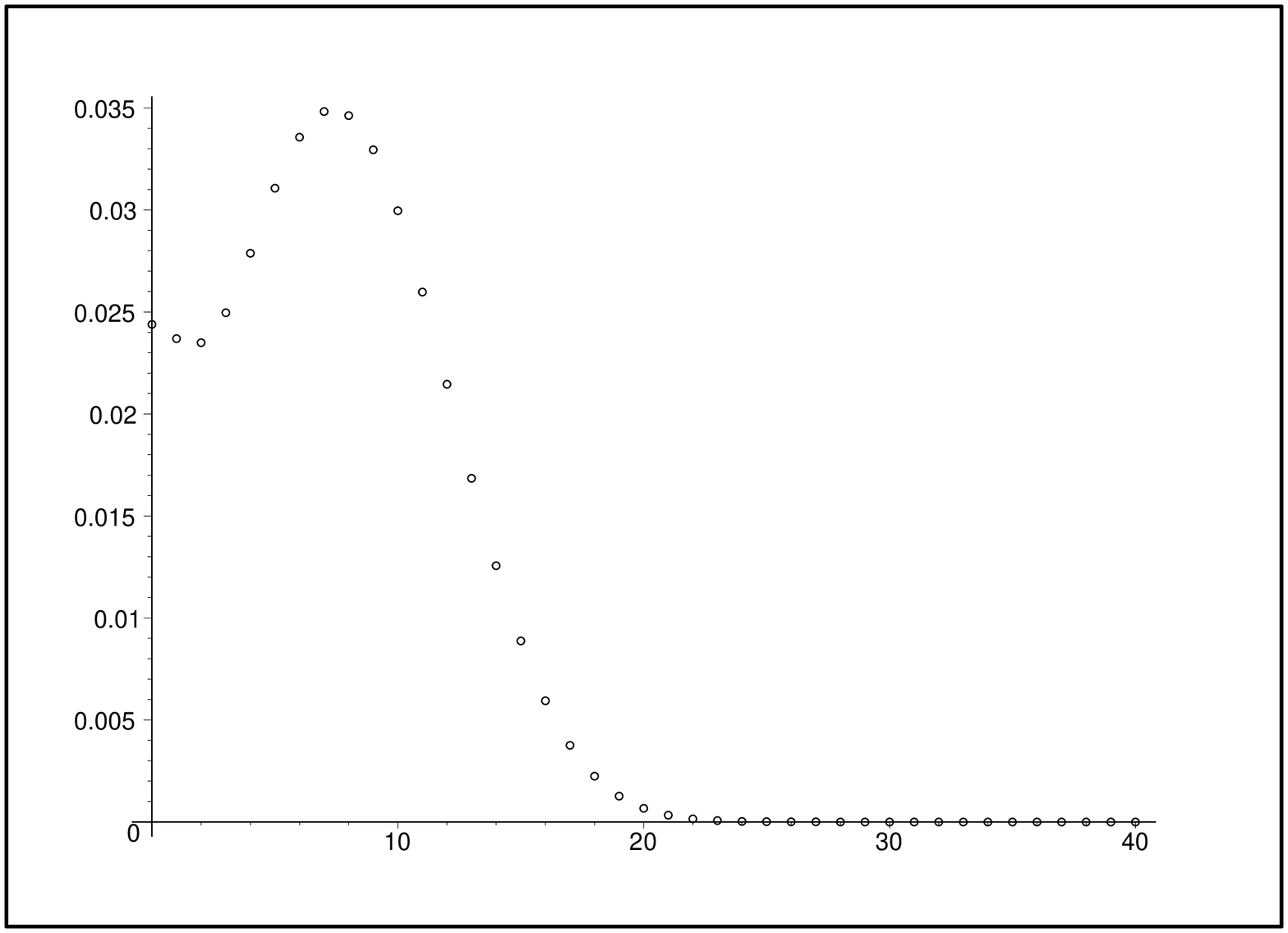}
\includegraphics[width=3.7cm,angle=-90]{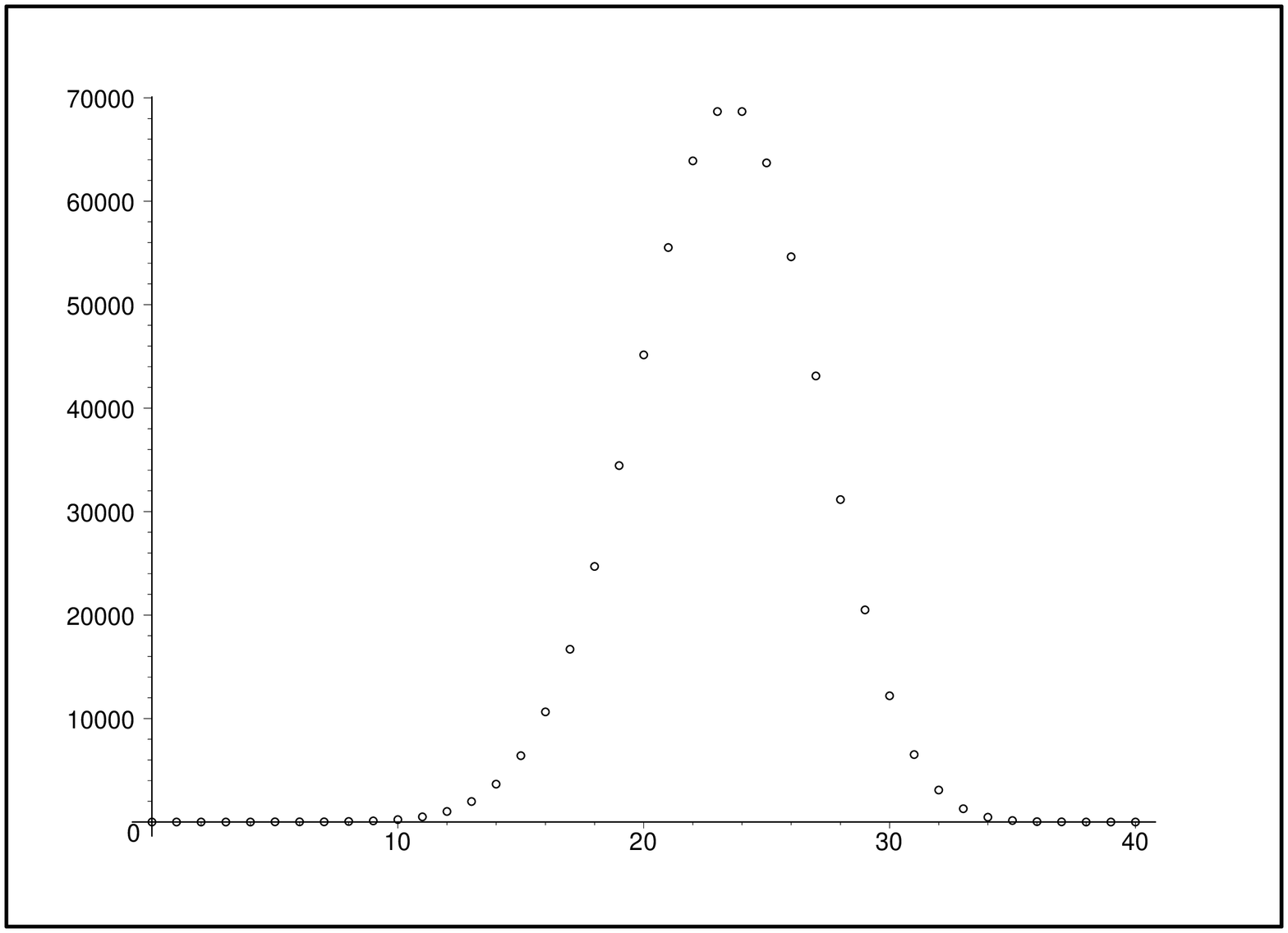}
\includegraphics[width=3.7cm,angle=-90]{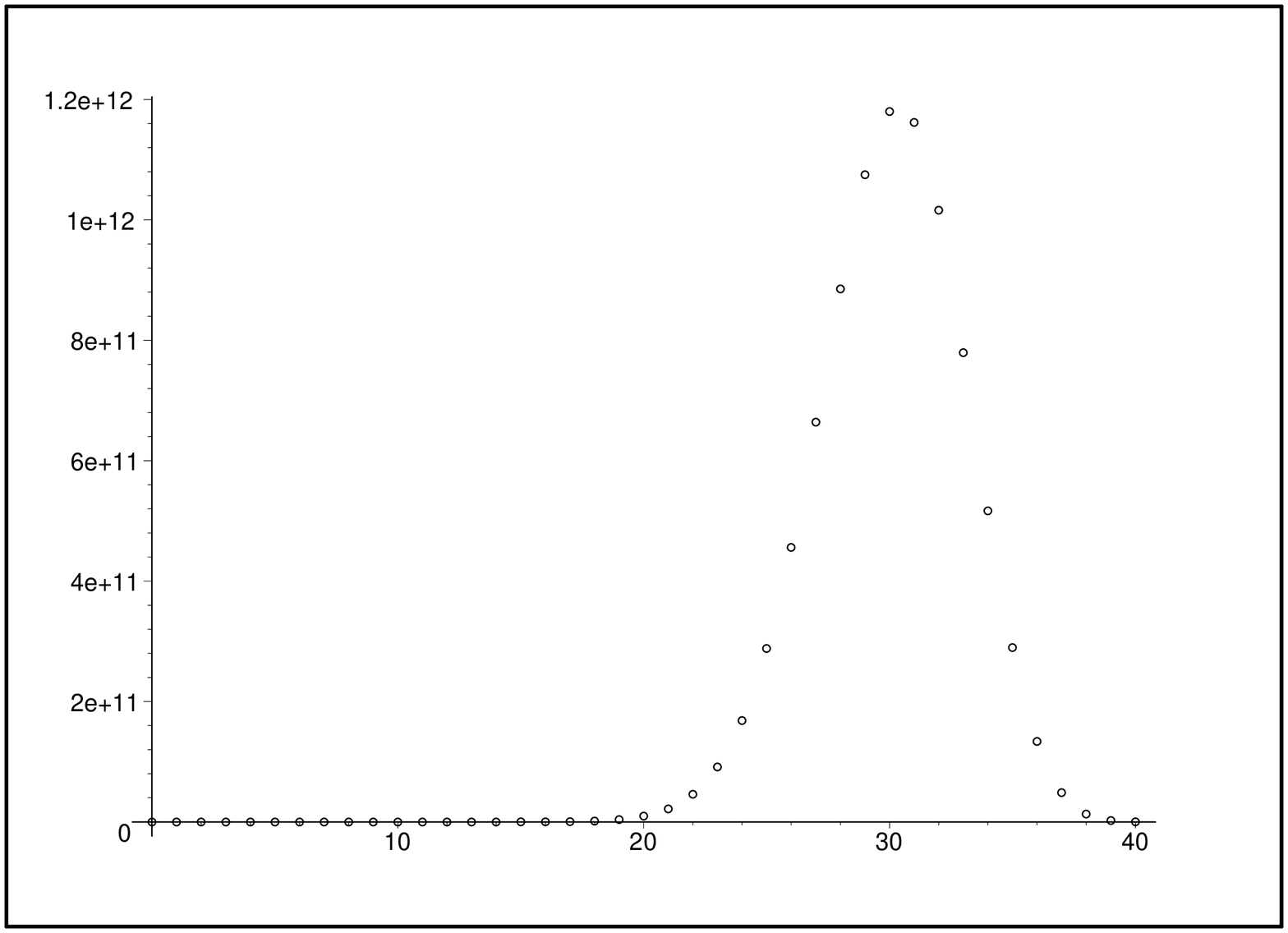}
\caption{We have plotted the modulus of the $j=20$ equi-area state $C^k_\vj(Z)$ for increasing cross-ratios
$Z=0.1i,0.8i,1.7i$. We see the Gaussian distribution progressively moving its peak from $0$ to $2j$. This illustrates
how changing the value of $Z$ affects the semi-classical geometry of the tetrahedron.}
\label{GaussianPlots}
\end{center}
\end{figure}

The simplest example in which we can check everything is the regular equilateral tetrahedron, which
corresponds to the value $Z=\exp(i\pi/3)$. In this case, the complex angle is easily computed:
$$
\Theta=\ln\left(\f{1+\sqrt{3}}{\sqrt{2}}\right)+i\f\pi4.
$$
This gives the position of the peak, the angle  and the deviation:
$$
k_{c}= 2j \tanh \mathrm{Re}(\Theta)=\f{2j}{\sqrt{3}}, \qquad
\varphi_{c}= \frac{\pi}2,\quad
\sigma=2j/\cosh^2\mathrm{Re}(\Theta)=4j/3.
$$
This fits perfectly the expected classical values for an equilateral tetrahedron
and the deviation $\sigma$ agrees with that proposed in \cite{Rovelli:2006fw}. A plot of
the modulus of the quantity $C^k_\vj(Z)$ (for $j=20$ and $Z=\exp(i\pi/3)$) as a function of the spin
label $k$ is given in Fig.~\ref{GaussianPlot-fixed} and confirms the above semi-classical analysis.
Similar plots for other values of $Z$ are given in Fig.~\ref{GaussianPlots} and show how the position
of the peak depends on the cross-ratio coordinate $Z$.

%

\subsection{The Equi-Area Case: Peakedness with respect to $Z$}

In the previous subsection we have considered the overlap function
$C^k_{\vj}(Z)=\bra \vj, k|\vj, Z\ket$ and saw that, in the semi-classical limit of large
spins, it is essentially a Gaussian in the $k$-variable peaked around
some classical value $k_c$ parametrized by $Z$ via (\ref{kphiZ}). In this
subsection we are going to continue our study of the equi-area case, and
consider the peakedness properties of the same overlap function but now
viewed as a function of the cross-ratio coordinate. To do this, it is important
to obtain an expression for the K\"ahler potential $\Phi_{\vj}(Z,\bZ)$ on the constraint
surface parametrized by $Z$, for, as we shall see below, it is the
wave-function $e^{-(1/2)\Phi_{\vj}(Z,\bZ)} C^k_{\vj}(Z)$ that is
peaked in $Z$.

Thus, let us first obtain an explicit expression for the equi-area
integration kernel $\hat{K}(Z,\bZ)$ that, as we know from the analysis of the
section \ref{Kahlersc} is essentially the (exponential of the) K\"ahler potential
on the constraint surface. A representation for $\hat{K}(Z,\bZ)$
as an integral over the orbits orthogonal to the constraint surface was given earlier
in (\ref{ker-asympt}), and the following discussion established that in the semi-classical
limit of large spins the K\"ahler potential on the constraint surface is essentially given
by the function $\Phi_{\vj}(z_i,\bz_i)$, given by e.g. (\ref{Phi-Z}), evaluated on the constraint surface. In the
equi-area case this function can be computed explicitly. Thus, let us start with the function
(\ref{omega-pfaff}), which in the case $n=4$ is given by:
\be
e^{-\Phi_\bullet(z_i,\bz_i)}\Big|_{n=4} = \frac{|z_{12}z_{34}|^{2}}{\prod_{i=1}^{4} (1+|z_{i}|^2)}
= \left|\frac{N_1-N_2}{2}\right|^2 \left|\frac{N_3-N_4}{2}\right|^2,
\ee
where we have used (\ref{N12}) to write the second equality. Using $|N_1-N_2|^2=2(1-\cos\theta_{12})$,
where $\theta_{12}$ is the dihedral angle between the faces 1 and 2, as well as the fact
that in the equi-area case on the constraint surface we have $\cos\theta_{12}=\cos\theta_{34}$, and
recalling the relation (\ref{k-theta}) between the parameter $k_c$ and $\theta_{12}$, we can write the above formula as:
\be
e^{-\Phi_\bullet(z_i,\bz_i)}\Big|_{\sum_{i=1}^4 j_i N_i=0} \equiv e^{-\Phi_\bullet(Z,\bZ)}= (1-x_c^2)^2,
\ee
where, as before $x_c=k_c/2j$. Now, the semi-classical K\"ahler potential $\Phi_{\vj}(Z,\bZ)$
in the equi-area case is equal to $2j\Phi_\bullet(Z,\bZ)$,
see (\ref{Phi-Z}) and (\ref{Phi-bullet}). Let us write an expression for this K\"ahler potential
in terms of the parameter $\Theta(Z)$. The quantity $(1-x_c^2)$ was computed in (\ref{width}) and
we get the following simple expression:
\be\label{Kahler}
\Phi_{j}(Z,\bar{Z}) = 8j \ln\left[ \cosh \mathrm{Re}(\Theta)\right].
\ee

Now, given the K\"ahler potential, we can compute the corresponding symplectic form
\beq\label{Omega-Theta}
\Omega_{\vj} &\equiv& \f1{i}  \partial_{Z}\partial_{\bar{Z}}\Phi_{j}\,\rd Z\wedge \rd \bar{Z}
= - \frac{2j}{i}\frac{\rd \Theta \wedge \rd \bar{\Theta}}{ \cosh^{2}\mathrm{Re}(\Theta)}\\ \nn
&=& 4j \frac{\rd \mathrm{Re}(\Theta) \wedge \rd \mathrm{Im}(\Theta)}{ \cosh^{2}\mathrm{Re}(\Theta)}
 = \rd k_{c}(Z) \wedge \rd \varphi_{c}(Z),
\eeq
where in the last equality we have used the differential of (\ref{kdefthet}) and
the definition of $\varphi_{c}=2 \mathrm{Im}(\Theta)$. This demonstrates that $k_{c}$ and
$\varphi_{c}$ are canonically conjugate variables, as anticipated in the previous subsection.

We would now like to compute the inner product $\bra\vj, k|\vj, l\ket$ of two real intertwiners
as an integral over the cross-ratio coordinate $Z$. In the semi-classical approximation of
large spins the integration kernel is found explicitly in (\ref{asmeta-unity}), so we
are interested in computing (in the equi-area case):
\be
\bra\vj, k|\vj, l\ket \sim
\frac{(2j)^2}{\sqrt{\pi}} \int \rd^2Z \sqrt{{\rm Pf}(\Omega_\vj)}
e^{-(2j+1)\Phi_\bullet(Z,\bZ)} \overline{C^k_{j}(Z)}C^l_{j}(Z).
\ee
To analyze this integral it is very convenient to switch to $\Theta$ coordinate instead of $Z$.
The change of variables is easy to work out. Indeed, we have $\rd Z=\sinh(2\Theta) \rd\Theta$, and the
Pfaffian of the symplectic form is available from (\ref{Omega-Theta}). Note that we get a
factor of $|\sinh 2\Theta|^2$ from the change of integration measure, as well as a factor
of $|\sinh 2\Theta|^{-1}$ from the Pfaffian. We also need to discuss the range of
integration of $\Theta$. Rewriting the definition $Z=\cosh^2(\Theta)$ in terms of the real
$\mathrm{R}\equiv\mathrm{Re}(\Theta)$ and imaginary $\mathrm{I}\equiv \mathrm{Im}(\Theta)$
parts of $\Theta$ we get:
\be
Z=\frac{1}{2}+\frac12\cosh(2\mathrm{R})\cos(2\mathrm{I})+ \frac{i}{2}\sinh(2\mathrm{R})\sin(2\mathrm{I}).
\ee
It is then clear that the range $\mathrm{R}\in(-\infty,\infty),2\mathrm{I}\in[0,2\pi]$ covers the
whole $Z$-plane {\it twice} (note that in this parametrization there are two cuts in the $Z$-plane
along the real axes starting each at $\pm1$ and going to infinity). After this change of variables
we get:
\be
\bra\vj, k|\vj, l\ket \sim
 (2j)^{2} \sqrt{\f{4j}{\pi}} \int  \f{\rd^{2}\Theta}{2} \,\f{|\sinh 2\Theta|}{\cosh \mathrm{Re}(\Theta)}
e^{-(2j+1)\Phi_\bullet(\Theta,\overline{\Theta})} \overline{C^k_{j}(\Theta)}C^l_{j}(\Theta),
\ee
where, as we have computed above,
$\Phi_\bullet(\Theta,\overline{\Theta})=4\ln\left[ \cosh \mathrm{Re}(\Theta)\right]$,
the integration over $\Theta$ is carried over the specified above domain, and an additional factor
of $1/2$ was introduced because the original domain is now covered twice. We can now
substitute into this expression the semi-classical expression (\ref{C-semi}) to get:
\be\label{semi-CC}
\bra\vj, k|\vj, l\ket \sim \sqrt{\frac{j}{\pi}} \int \rd^{2} \Theta \,
\overline{\hat{C}^k_{j}(\Theta)}\hat{C}^l_{j}(\Theta),
\ee
where we have introduced the new states:
\be
\hat{C}_{j}^{k}(\Theta)
\equiv
2j \sqrt{\f{\sinh 2\Theta}{\cosh \mathrm{Re}(\Theta)}}
\,  e^{-(j+1/2) \Phi_\bullet(\Theta,\bar{\Theta})}\, {C}_{\vj}^{k}(\Theta)
\,\sim\,
(-1)^k\frac{{\cN}(x,\Theta)}{\sqrt{2\pi k}} e^{-4j{\cal{S}}(x,\Theta)},
\ee
where we have introduced, still with $x=k/2j$:
\be
\cN(x,\Theta)\equiv
\f{e^{-\f12\Phi_\bullet(\Theta,\overline{\Theta}) +\mathrm{R} +i\mathrm{I}}}{\left(1+x\right)\sqrt{\left(1-x^{2}\right)}}
\f{1 }{ \sqrt{\cosh  \mathrm{R} }}, \qquad
{\cal{S}}(x,\Theta)\equiv \Lambda(x) - x(\mathrm{R} +i\mathrm{I}) + \ln\cosh \mathrm{R},
\ee
and wrote everything in terms of the real $\mathrm{R}\equiv\mathrm{Re}(\Theta)$ and
imaginary parts $\mathrm{I}\equiv \mathrm{Im}(\Theta)$.

Taking into account the fact that with our conventions $\rd^2\Theta = 2\rd\mathrm{R} \rd\mathrm{I}$,
we can easily perform the integral over $\mathrm{I}$:
$$
\int_0^\pi 2\rd \mathrm{I}\,e^{2iI(l-k)}
\,=\,
2\pi\delta_{k,l}.
$$
Thus the integral over $\mathrm{I}$ just imposes $k=l$. The integral over the real
part $\mathrm{R}$ is more interesting. It can again be computed using the steepest descent method. The
value $R_c$ that dominates the integral is given by
\be
\tanh \mathrm{R}_c =x.
\ee
Remarkably, this is the same relation, see (\ref{kdefthet}) that we have obtained in the previous
subsection by extremizing the integrand with respect to $x$. One also sees that the value of
$\mathrm{Re}(\cC)$ at the minimum is simply equal to $0$ which reflects the fact that the states are normalized.
To compute the integral over $\mathrm{R}$ in (\ref{semi-CC}) it remains to compute the Hessian at the
critical point, which is given by:
\be
\left.{\pp^2_{R}\mathrm{Re}(\cC)}\right|_{\mathrm{R}_c}=\f{1}{\cosh^2\mathrm{R}_c}.
\ee
At the critical point the normalization coefficient $\cN(x,\Theta)$ also simplifies
considerably since
\be
\f{e^{-\f12\Phi(\mathrm{R}_{c}) +\mathrm{R}_{c} }}{\left(1+x\right)\sqrt{\left(1-x^{2}\right)}} =1,
\quad \mathrm{hence}\quad \left|\cN(x,\mathrm{R}_{c})\right|\,=\,
\f{1 }{ \sqrt{\cosh  \mathrm{R}_{c} }   }.
\ee
Then putting all the pieces together, the steepest descent evaluation of the remaining integral over $\mathrm{R}$ gives:
\be
\bra\vj,k|\vj, l\ket=
\sqrt{\f{j}{{\pi }}}\, (2\pi \delta_{k,l})\,
\frac{\left|\cN(x,\mathrm{R}_{c})\right|^{2}}{2\pi k}
\sqrt{ \f{2\pi\cosh^2\mathrm{R}_{c}}{8j}}
=\f{\delta_{k,l}}{2k} \, .
\ee
This allows us to recover explicitly the expected orthonormality of the states $|\vj, k\ket$ at leading
order in $k$ (the exact normalization factor would $1/(2k+1)$ instead of $1/2k$). This provides a highly non trivial consistency
check of all our asymptotic evaluations, i.e., of the  asymptotic formulae for the kernel $\hat{K}_\vj$ and the states $C^k_\vj(Z)$.

\section{Discussion}

In this paper we have introduced and studied in detail a holomorphic basis for the Hilbert space $\cH_{j_1,\ldots,j_n}$ of
$\SU(2)$ intertwiners. We have considered the general $n$ case, but gave more details for the
4-valent intertwiners that can be interpreted as quantum states of a ``quantum tetrahedron''. Our main result
is the formula (\ref{inner-phys}) for the inner product in $\cH_{j_1,\ldots,j_n}$ in terms of a holomorphic
integral over the space of ``shapes'' parametrized by the cross-ratio coordinates $Z_i$. In the ``tetrahedral''
$n=4$ case there is a single cross-ratio $Z$. Somewhat unexpectedly, we have found that the integration kernel
$\hat{K}(Z_i,\bZ_i)$ is given by the $n$-point function of the bulk/boundary dualities of string theory,
and this fact allowed to give to $\hat{K}(Z_i,\bZ_i)$ an interpretation that related them to the K\"ahler
potential on the space of ``shapes''. The new viewpoint on the $n$-point functions $\hat{K}(Z_i,\bZ_i)$
as being a kernel in the inner product formula for $\cH_{j_1,\ldots,j_n}$ also led us to the expectation
that the $n$-point functions should satisfy the holomorphic factorization formula (\ref{K-factor})
of the type expected of an $n$-point function of a 2-dimensional conformal field theory. It would
be of interest to develop this line of thought further by proving (\ref{K-factor}), computing 
the coefficients $a_\lambda$ and thus finding the spectrum of this
CFT, as well as possibly even identifying the underlying conformal field theory, if there is one.
It would also be of considerable interest to see if a group-theoretic interpretation similar to the one
developed here for 2-dimensional $n$-point function also exists for d-dimensional
objects, with the case of most interest for applications in string theory being of course $d=4$.

In spite of $n$-point functions of bulk/boundary correspondence of string theory showing a
somewhat unexpected appearance, our results are most relevant to the subjects of loop quantum gravity and
spin foam models. It is here that we believe the new techniques developed in this paper can lead
to new advances, impossible without the new coherent states we introduced. Indeed, we have shown
that the $n$=4 holomorphic intertwiners $|\vj,Z\ket$ parametrized by a single cross-ratio variable $Z$ are true
coherent states in that they form an over-complete basis of the Hilbert space of intertwiners and are semi-classical
states peaked on the geometry of a classical tetrahedron. We have also shown how the new holomorphic intertwiners
are related to the standard ``spin basis" $|\vj,k\ket$ of intertwiners that are usually used in loop quantum gravity and
spin foam models, and found that the change of basis coefficients are given by Jacobi polynomials.

With the mathematics of the new holomorphic intertwiners hopefully being clarified by this work, the next step
is to apply the techniques developed here to loop quantum gravity and spin foam models. In the canonical framework of
loop quantum gravity, spin network states of quantum geometry are labeled by a graph as well as by $\SU(2)$ representations
on the graph's edges $e$ and intertwiners on its vertices $v$. 
We can now put holomorphic intertwiners at the vertices of the graph,
which introduces the new spin networks labeled by representations $j_e$ and cross-ratios $Z_v$. Since each holomorphic
intertwiner can be associated to a classical tetrahedron, we could truly interpret these new spin network states
as discrete geometries. In particular, geometrical observables such as the volume can be expected to be peaked on their
classical values. This fact should be of great help when looking at the dynamics of the spin network states and
when studying how they are coarse-grained and refined. We leave these interesting developments to future research.

The holomorphic intertwiners are also of direct relevance to spin foam models. 
Indeed, spin foam amplitudes encode the dynamics
of spin network states. The coherent intertwiners, introduced in 
\cite{Livine:2007vk}, already led to significant improvement on
the understanding of the semi-classical behavior of spin foams 
\cite{Freidel:2007py,LS2,CF1,CF2,Winston}. It is clear that the new
holomorphic intertwiners developed in this paper (building upon 
\cite{CF3}) should lead to further progress in the same direction.
Indeed, the basic spin foam building blocks are the $\{15j\}$-symbols, 
which is an $\SU(2)$ invariant labeled by 10 representations
and 5 intertwiners. With holomorphic intertwiners at hand we can now 
define a $\{10j,5Z\}$-symbol. It is of great interest to understand
the large spin behaviour of this new invariant, for it can be expected 
that this asymptotics contains a great deal of information about
the classical geometry of a 4-simplex. It is also important to consider 
that case of the twice larger gauge group
$\textrm{Spin}(4)=\SU(2)_L\times\SU(2)_R$ and construct the corresponding 
$\{10j_L,5Z_L\}\{10j_R,5Z_R\}$-symbols. These, with
appropriate constraints between the labels $j_L,j_R,Z_L,Z_R$ as in 
\cite{Engle:2007uq,Freidel:2007py,LS2,ELPR}, should correspond
to semi-classical 4-simplices in 4d gravity. Another very important task 
is to express the (area-)Regge action in terms of the
spins and the cross-ratios $Z$. We leave all these exciting problems to 
future research and hope that the analysis performed in
the present paper will provide a solid first step in the direction outlined.

\section*{Acknowledgements} The authors would like to thank Florian Conrady for comments on the manuscript. Research at Perimeter Institute is supported by the Government of Canada through Industry Canada and by the Province of Ontario through the Ministry of Research \& Innovation. KK was supported by an EPSRC Advanced Fellowship. EL was partly supported by a ANR ``Programme Blanc" grant.

\begin{appendix}

\section{$G$ determinant}
\label{Gdet}

Here we give a proof of the formula for the determinant of the metric $G_{\vj}$ on orbits orthogonal
to the constraint surface. Let us recall that $G_{\vj}(z_{i})$ is a 3 by 3 metric
\be
G_{\vj} ^{ab}(z_{i})= \sum_{i=1}^{n} j_{i} (\delta^{ab}- N^{a}(z_{i})N^{b}(z_{i})).
\ee
We derive the following expression for the determinant of this matrix:
 \be
 \det\left(G_{\vj}\right)(z_{i}) = 16 \sum_{i<j<k} j_{i}j_{j}j_{k}
\frac{|z_{ij}|^{2}|z_{jk}|^{2}|z_{ki}|^{2}}{(1+|z_{i}|^{2})^{2}(1+|z_{j}|^{2})^{2}(1+|z_{k}|^{2})^{2}}
+ G_{\vj}\,(H_{\vj}, H_{\vj})
 \ee
where $H_{\vj}= \sum_ij_i N(z_i)$ is the closure vector
and the sum is over all ordered triples belonging to $\{1,\cdots,n\}$.
When the closure condition is satisfied the last term vanishes and we recover the
evaluation of the determinant that we use in the main text.

\vspace{8 pt}
\noindent
{\bf Proof:}
$G$ is a 3$\times$3 matrix and so we can compute its determinant in terms of traces of its powers:
\be
\det G= \f16\left[(\tr G )^3-3(\tr G)(\tr G^2) +2\tr G^3\right].
\ee
We compute:
\bea
\tr G &=& 2\sum_i j_i, \nonumber\\
\tr G^2 &=& \sum_{i,k} j_ij_k\left[1+(N_i\cdot N_k)^2\right],\nonumber \\
\tr G^3&=&\sum_{i,k,l}j_ij_kj_l \left[3(N_i\cdot N_k)^2-(N_i\cdot N_k)(N_k\cdot N_l)(N_l\cdot N_i)\right],\nonumber
\eea
where we have introduced the convention $N_i\equiv N(z_i)$. We can simplify the formula and get:
\be
3\det G=\sum_{i,k,l}j_ij_kj_l\left[1-(N_i\cdot N_k)(N_k\cdot N_l)(N_l\cdot N_i)\right].
\ee
Using the trivial formula for the scalar product of two unit vectors:
$$
N_i\cdot N_k=1-\f12 \left| N_i -N_k \right|^{2},
$$
we expand the previous formula and obtain:
\beq
3\det G_{\vj}&=&
\f18\sum_{i,k,l}j_ij_kj_l | N_i -N_k |^{2}| N_k -N_l |^{2}| N_l -N_i |^{2} \nn\\
&+&\f32\sum_{i,k,l}j_ij_kj_l | N_i -N_k |^{2}
-\f34\sum_{i,k,l}j_ij_kj_l  | N_i -N_k |^{2}| N_k -N_l |^{2}. \nn
\eeq
The first term here is the one we would like to keep. It doesn't vanish if and only if the indices $i,k,l$
are all different. We can simplify the two remaining terms by noticing the symmetry under the exchange of
the indices $i,k,l$ and using the scalar product formula the other way round, $|N_i -N_k|^ 2=2(1-N_i\cdot N_k)$.
We then obtain a much simpler formula (notice that this formula is valid for any number of vectors $N_i$):
\beq
\det G_{\vj}&=&
\f1{24}\sum_{i,k,l}j_ij_kj_l | N_i -N_k |^{2}| N_k -N_l |^{2}| N_l -N_i |^{2} +
\sum_l j_l(H_{\vj}\cdot H_{\vj}-(H_{\vj}\cdot N_l)^2)
\nn\\
&=&\label{detGN}
\f1{4}\sum_{i<k<l}j_ij_kj_l | N_i -N_k |^{2}| N_k -N_l |^{2}| N_l -N_i |^{2}
+ G_{\vj}(H_{\vj},H_{\vj} ),
\eeq
where we have introduced the closure vector $H_{\vj}\equiv\sum_i j_i N_i$.
Finally, we use the following identity
\be\label{N12}
\frac14 \left| N_1 -N_2 \right|^{2}= \frac{|z_{12}|^{2}}{(1+|z_{1}|^{2})(1+|z_{2}|^{2})},
\ee
where $z_{12}=z_1-z_2$, and derive the desired result. $\Box$
\vspace{8 pt}
\noindent

\section{Measures on ${\rm SL}(2,\C)$}
\label{gmeas}

Here we obtain two expressions for the group-invariant measure on ${\rm SL}(2,\C)$ that
are used in the main text.

We start from the following expression for the Haar measure on $\mathrm{SL}(2,\C)$:
\be
\rd g = \rd a\wedge \rd b\wedge \rd c\wedge \rd d \,  \wedge
\rd \bar{a}\wedge \rd \bar{b}\wedge \rd \bar{c}\wedge \rd \bar{d} \,\, \delta^{(2)}(ad-bc-1).
\ee
Resolving the delta function we can express this measure as the  the product of the holomorphic and
anti-holomorphic pieces, $\rd g =\rd^{\mathrm{hol}} g\wedge \overline{\rd^{\mathrm{hol}} g}$
with the holomorphic piece given by, e.g.:
\be
\rd ^{\mathrm{hol}} g = - \frac{\rd a\wedge \rd b \wedge \rd c}{a}.
\ee
Let us now introduce the following parametrization:
\be
z_{1}= \frac{b}{d}, \quad z_{2}=\frac{a+b}{c+d},\quad z_{3}= \frac{a}{c}.
\ee
It is easy to compute
\be\nonumber
\rd z_1\wedge \rd z_2\wedge \rd z_3 = \frac{(ad-bc)}{( c(c+d)d)^{2}}
\left( -a \rd b \wedge \rd c\wedge \rd d +  b \rd a \wedge \rd c\wedge \rd d -  c \rd a \wedge \rd b \wedge \rd c
+  d \rd a \wedge \rd b \wedge \rd c \right)
\ee
If one imposes the determinant condition $ad-bc=1$ one gets
\be
\rd z_1\wedge \rd z_2\wedge \rd z_3  =\frac{ 2 \rd ^{\mathrm{hol}} g}{( c(c+d)d)^{2}}
\ee
Note that since $z_{21} z_{13}z_{23}= ( c(c+d)d)^{-2}$ we can equivalently rewrite this relation as
\be
\rd ^{\mathrm{hol}} g = \frac{\rd z_1\wedge \rd z_2\wedge \rd z_3}{2z_{21}z_{23}z_{13}}.
\ee

We can now  compute the measure $\wedge_{i=1}^n dz_i$ in terms of the new coordinates
$a,b,c,d, Z_i$. The idea is to first decompose it in terms of $z_{1}, z_{2},z_{3}$ and $Z_{i}$
and then use the previous relations to express it in terms of $a,b,c,d$ and $Z_{i}$.
We have:
\beq\label{mes-1}
\bigwedge_{i=1}^n \rd z_i &=& \rd z_{1}\wedge \rd z_{2}\wedge \rd z_{3} \bigwedge_{i} \frac{\rd Z_i}{(cZ_i+d)^2} \\
&=&   2\frac{\rd^{\mathrm{hol}} g \wedge_{i=4}^{n} \rd Z_i}{ d^{2} (c+d)^{2} c^{2}\prod_{i}(cZ_i+d)^2}.
\eeq
Thus, the total measure is:
\be
\prod_{i=1}^{n} \rd^{2} z_{i} = 4
\frac{\rd g \,\, \prod_{i=4}^{n} \rd^{2}Z_{i}}{|d|^{4} |c+d|^{4} |c|^{4}\prod_{i=4}^{n}| cZ_{i}+d|^{4}},
\ee
where our convention is that $\rd ^2z=|\rd z\wedge \rd \bar{z}|$.
Below we show that the relationship between the Haar measure on the group $\rd g$ and the
normalized one for which the volume of SU(2) subgroup is normalized is given by
$\rd g = (2 \pi)^{2} \rd^{\mathrm{norm}}g$, and thus the total measure is
\be
\prod_{i=1}^{n} \rd^{2} z_{i} = (4\pi)^{2}
\frac{\rd^{\mathrm{norm}} g \,\, \prod_{i=4}^{n} \rd^{2}Z_{i}}{|d|^{4} |c+d|^{4} |c|^{4}\prod_{i=4}^{n}| cZ_{i}+d|^{4}}.
\ee
The last subtlety comes from the fact that the map $ (g,Z_{i}) \to (z_{i}(g,Z_{i}))$ is $2:1$ thus
we have to insert an extra factor $1/2$ when integrating and this leads us to the relation we were looking for:
\be
\int_{\C^{n}} \prod_{i=1}^{n} \rd^{2} z_{i} \,\, F(z_{i},\overline{z_{i}}) =
{ 8\pi^{2}}
\int_{\C^{n-3}} \! \prod_{i=4}^{n} \rd^{2} Z_{i}
\int_{\mathrm{SL}(2,\C)} \!\!\!\!\!\!\! \rd^{\mathrm{norm}} g  \, \,\,
\frac{F(Z_{i}^{g}, \overline{Z_{i}^{g}})}{|d|^{4} |c+d|^{4} |c|^{4}\prod_{i=4}^{n}| cZ_{i}+d|^{4}}.
\ee

We now compute the Haar measure in terms of the  Iwasawa
decomposition ${\rm SL}(2,\C)=KAN$.  From (\ref{Iwasawa}) we have the following explicit
parametrization:
\beq
a= \cos(\theta)e^{i\phi} \rho^{-1/2}, \qquad b = -\cos(\theta)e^{i\phi} \rho^{-1/2} n +
\sin(\theta) e^{i\psi} \rho^{1/2}, \\ \nonumber
c= - \sin(\theta)e^{-i\psi} \rho^{-1/2}, \qquad
d = \sin(\theta)e^{-i\psi} \rho^{-1/2} n + \cos(\theta)e^{-i\phi}\rho^{1/2},
\eeq
where the range of the angular coordinates on the ${\rm SU}(2)$ part is
$\theta\in[0,\pi/2]$ and $\phi,\psi\in[0,2\pi]$. A simple explicit computation
gives:
\be
\rd a\wedge \rd \bar{a}\wedge \rd c\wedge \rd \bar{c}=\frac{\rd \rho}{\rho^3}\wedge 2\sin(\theta)\cos(\theta)
\rd \theta\wedge \rd \phi\wedge \rd \psi.
\ee
Thus, we only need to compute the $\rd n\wedge \rd \bar{n}$ part of $\rd d\wedge \rd\bar{d}$, which is
just $\sin^2(\theta)\rho^{-1}\rd n\wedge \rd \bar{n}$. Multiplying it all together and dividing by
$|c|^2$ we get:
\be
\rd g = \frac{\rd \rho}{\rho^3}\wedge \rd n\wedge \rd \bar{n} \wedge 2\sin(\theta)\cos(\theta)
\rd \theta\wedge \rd \phi\wedge \rd \psi.
\ee

Now, the ${\rm SU}(2)$ measure that appears here is not a normalized one - the corresponding
volume of ${\rm SU}(2)$ that it gives is $(2\pi)^2$. Thus, the normalized measure is given by:
\be
\rd^{\mathrm{norm}} g = \frac{\rd \rho}{\rho^3}\wedge \rd n\wedge \rd \bar{n} \wedge \frac{1}{4\pi^2} \sin(2\theta)
\rd \theta\wedge \rd \phi\wedge \rd \psi.
\ee

\section{Permutations and cross-ratios}
\label{perm}

In order to define the cross-ratio we need to chose an order among the $z_{i}$.
A permutation of $z_{1},z_{2},z_{3},z_{4}$ changes this order and  naturally acts on the cross-ratios.
The other cross ratios one obtains this way are given by
\be
Z = \frac{z_{41}z_{23}}{z_{43}z_{21}},
\quad 1-Z  = \frac{z_{42}z_{31}}{z_{43}z_{21}}=\hsigma_{12}(Z),
\quad \frac{Z}{Z-1}=  \frac{z_{41}z_{32}}{z_{42}z_{31}}=\hsigma_{23}(Z),
\ee
as well as their inverses
\be
\frac1{Z} = \hsigma_{13}(Z),\quad \frac1{1-Z}= \hsigma_{12}\hsigma_{23}(Z), \quad \frac{Z-1}Z = \hsigma_{23}\hsigma_{12}(Z).
\ee
The reason why there is only $6$ different cross-ratios while the number of permutations is 24, is that
the initial cross-ratio $Z$ is fixed by $4$ permutations.
These are generated by the identity and  the three following permutations exchanging a pair of indices
\be\label{app-refl}
 \hsigma_{23} \hsigma_{14},\quad  \hsigma_{13}\hsigma_{24},\quad\hsigma_{12}\hsigma_{34}.
\ee
The whole permutation group (even the reflections (\ref{app-refl})) acts non-trivially on the prefactors
(\ref{prefact}) and thus on the holomorphic intertwiners $|\vj, Z\ket$.
The action of the non-trivial permutations is
\beq
\hsigma_{12}|\vj, Z\ket &=&(-1)^{s-2j_3} \, |\vj_{12},1-Z\ket,\\
\hsigma_{23}|\vj, Z\ket &=&(-1)^{2j_2} \, \left(1-Z\right)^{2j_{4}} \left|\vj_{23}, \frac{Z}{Z-1} \right\ket,\\
\hsigma_{13}|\vj, Z\ket &=& (-1)^{s}\, Z^{2j_{4}} \left|\vj_{13}, \frac1{Z} \right\ket,\\
\hsigma_{23}\hsigma_{12}|\vj, Z\ket &=& (-1)^{2s-2j_3}\, (1-Z)^{2j_{4}} \left|\vj_{231}, \frac1{1-Z}  \right\ket,\\
\hsigma_{12}\hsigma_{23}|\vj, Z\ket &=& (-1)^{s+2j_2}\, (-Z)^{2j_{4}} \left|\vj_{312},\frac{Z-1}{Z} \right\ket.
\eeq
where $s=j_{1}+j_{2}+j_{3}+j_{4}$, $\vj_{abcd} = (j_{a},j_{b},j_{c},j_{d})$
and $\vj_{abc}=\vj_{abc4}$ etc.

\section{Bulk-boundary dualities $n$-point function}
\label{app-n-point}

In this section we are interested in reviewing some
properties of the $n$-point function (\ref{n-point})
\be
\label{n-point2}
K_{\vec{j}}(z_i,\overline{z_i})=
\int_{\R_+}\f{d\rho}{\rho^3}\int_\C d^2n \,
\prod_{i=1}^n\frac{\rho^{\Delta_{i}}}{(\rho^2+|z_{i}- n|^2)^{\Delta_{i}}},
\ee
where we have introduced  the conformal dimensions: $\Delta_i\equiv2(j_i+1)$.
Because this function is covariant under $\mathrm{SL(2,\mathbb{C}})$,
it can be expressed in terms of a function
$\hat{K}_{\vj}(Z_{i},\overline{Z_i}) =\lim_{X\to \infty} |X|^{2\Delta_{3}}K_{\vj}(0,1,X,Z_{i})$
that only depends on the cross-ratios $Z_i$.

The quantities (\ref{n-point2}) can be computed using the Feynman
parameter technique proposed in this context already by Symanzik \cite{Symanzik:1972wj},
see also \cite{Arutyunov:2000ku},\cite{Krasnov:2005fu}. The technique is based on the following formuli:
\be
\f{1}{z^\lambda}=\f{1}{\Gamma(\lambda)}\int_0^\infty dt\,t^{\lambda-1}e^{-tz},
\ee
\be
\int_0^\infty\f{d\rho}{\rho^{3}}\,\rho^{\sum_i\Delta_i}e^{-\sum_i t_i\rho^2}=
\frac{1}{2} S^{1-\sum_i\Delta_i/2}\,\Gamma(\sum_i\Delta_i/2-1),
\ee
\be
\int_{\C}\rd^2n \, e^{-\sum_i t_i|z_i-n|^2}=\f{2\pi}{S}e^{-\frac{1}{S}\sum_{i,j}t_it_j|z_i-z_j|^2},
\ee
where $S=\sum_i t_i$.
In the last integral our measure convention is that $\rd^{2}n =|dn \wedge \rd \bar{n}|$.
The formuli given are an adaptation of the general ones reviewed in e.g. \cite{Krasnov:2005fu} to
$d=2$. Using these we find
\beq\label{intt}
K_{\vec{j}}(z_i,\overline{z_i})
= \pi\f{\Gamma\left(\Sigma -1\right)}{\prod_{i=1}^n \Gamma(\Delta_{i})}
\int_{\R_+^n}\prod_{i=1}^n \rd t_{i} \, t_i^{\Delta_{i}-1}
\, S^{-\Sigma}e^{-\f1{S}\sum_{i,j} t_it_j |z_{ij}|^2},
\eeq
where $z_{ij}\equiv (z_i-z_j)$ and $\Sigma= \sum_{i}(j_{i}+1)$.
The crucial observation of \cite{Symanzik:1972wj} is that the quantity $S$ in (\ref{intt}) can be modified
without changing the integral to $S=\sum_{i} a_{i}t_{i}$, where again $a_{i}\geq 0$ are arbitrary positive
coefficients not all zero. Indeed, supposing that $S$ is of this form, we make the following change of variables:
\be
u_i\equiv \f1{\sqrt{S}}t_i,\quad \mathrm{or} \quad t_i=u_i\left(\sum_k a_{k}u_k\right),
\ee
where we have used $U\equiv \sum_i a_{i}u_i= \sqrt{S}$.
The determinant of the arising $n\times n$ Jacobian matrix is easy to compute:
\be
\f{\pp t_j}{\pp u_i}=\delta_{ij}U+a_{i}u_j,\qquad
\det\left(\f{\pp t_j}{\pp u_i}\right)=2U^n.
\ee
This leads to the simple expression in terms of the norms of the complex numbers $z_{ij}\equiv (z_i-z_j)$:
\beq
K_{\vec{j}}(z_i,\overline{z_i})
=2\pi\f{\Gamma\left(\Sigma -1\right)}{\prod_{i=1}^n \Gamma(\Delta_{i})}
\int\prod_{i=1}^n \rd u_{i}\, u_i^{\Delta_{i}-1}
\, e^{-\sum_{i<j} u_iu_j |z_{ij}|^2}.
\eeq
The result is independent of $a_{i}$, and so we can choose these numbers arbitrarily already at
the level of (\ref{intt}).

Let us now specialize to the most interesting for us case $n=4$. We are interested in
computing the limit (\ref{limK}). We first note that we can
reabsorb the multiplicative factor $|X|^{2\Delta_{3}}$  by a rescaling of the Feynman parameter
$t_{3} \to t_{3}/|X|^{2}$. We then evaluate (\ref{intt}) with with $z_{1}=0, z_{2}=1, z_3=X\to\infty$
and the choice $a_{i}=\delta_{3i}$ that corresponds to $S=t_3$. This gives:
\be\nonumber
\hat{K}_{\vec{j}}(Z,\overline{Z})
= \f{ \pi\Gamma\left(\Sigma -1\right)}{\prod_{i=1}^n \Gamma(\Delta_{i})}
\int\prod_{i=1}^n \frac{\rd t_{i}}{t_i}\,
t_1^{\Delta_{1}} t_{2}^{\Delta_{2}} t_{3}^{\Delta_{3}-\Sigma} t_{4}^{\Delta_{4}}
\, e^{-\frac{t_1t_2}{t_3}-(t_1+t_2+t_4)-\frac{t_1t_4}{t_3}|Z|^{2}-\frac{t_2t_4}{t_3}|1-Z|^{2}}.
\ee
We now perform a change of variables $\tilde{t}_{3} = t_{1}t_{2}/t_{3}$, and write the result
omitting the tilde on the new variable $t_3$. We get:
\beq\nonumber
\hat{K}_{\vec{j}}(Z,\overline{Z})
= \f{ \pi\Gamma\left(\Sigma -1\right)}{\prod_{i=1}^n \Gamma(\Delta_{i})}
\int\prod_{i=1}^n \frac{\rd t_{i}}{t_i}
t_1^{\Delta_{1}+\Delta_{3}-\Sigma} t_{2}^{\Delta_{2}+\Delta_{3}-\Sigma} t_{3}^{\Sigma-\Delta_{3}} t_{4}^{\Delta_{4}}
e^{-(t_{1}+t_{2}+t_{3}+t_{4})-\frac{t_{3}t_{4}}{t_{2}}|Z|^{2}-\f{t_{3}t_{4}}{t_{1}}|1-Z|^{2}}.
\eeq

Now the case $n=3$ can be recovered by putting $\Delta_{4}=0$. In this case the integral over $t_{4}$
is trivial and we are left with
\be\label{K3}
\hat{K}_{j_1,j_2,j_3}
= \pi\f{\Gamma\left(\Sigma -1\right) \Gamma(\Delta_{1}+\Delta_{2}-\Delta_{3})
\Gamma(\Delta_{1}-\Delta_{2}+\Delta_{3})
\Gamma(-\Delta_{1}+\Delta_{2}+\Delta_{3})}{ \Gamma(\Delta_{1}) \Gamma(\Delta_{2})\Gamma(\Delta_{3})},
\ee
where $\Sigma=\Delta_{1}+\Delta_{2}+\Delta_{3}$.

Returning to the case $n=4$ we can, following \cite{Arutyunov:2000ku},
use the Barnes-Mellin expansion of the exponential
\be
e^{-\lambda} =\frac1{2i\pi} \int \rd s\, \Gamma(-s) \lambda^{s},
\ee
with the integration contour running along the imaginary axes. Using this integral representation
for the two exponentials containing the cross-ratio, and performing all Feynman parameter integrations,
we get the Barnes-Mellin representation for the 4-point function
\beq\label{K4}
\hat{K}_{\vec{j}}(Z,\overline{Z})\nonumber
= \f{ \pi\Gamma\left(\Sigma -1\right)}{\prod_{i=1}^n \Gamma(\Delta_{i})}
\int \frac{dt ds}{(2\pi i)^2} && \Big[
\Gamma(-s)\Gamma(-t)\Gamma(\Delta_{1}+\Delta_{3}-\Sigma-t) \Gamma(\Delta_{2}+\Delta_{3}-\Sigma-s)\\
&& \Gamma(\Sigma- \Delta_{3}+s+t)\Gamma(\Delta_{4}+s+t) |Z|^{2 s} |1-Z|^{2t} \Big].
\eeq
The double integral here is of the Barnes-Mellin type with the integration contour running
to the left of the imaginary axes. However, one must be careful when evaluating the integrals
in terms of the pole contributions as $\Delta_{i}$ are integers and so there
are double poles.

Unfortunately, the integral expression (\ref{K4}) is not very suitable for producing a series
expansion in $Z$ as it contains powers of $|1-Z|^2$. It can be converted to a more suitable
integral representation by assuming $\Delta_{1}+\Delta_{3}-\Sigma\geq 0$ and using the Barnes lemma:
\beq\nonumber
 \int \frac{\rd t}{2i\pi}\!\!\!\!\!\!\! & &\Gamma(a+t) \Gamma(b+t) \Gamma(c-t)\Gamma(-t) X^{t}\\
&=&
\frac{\Gamma(a+c)\Gamma(b+c)\Gamma(a)
\Gamma(b)}{\Gamma(a+b+c)}
F(a, b;a+b+c| 1-X)\\ \nonumber
&=&\Gamma(a+c)\Gamma(b+c) \sum_{n}\frac{\Gamma(a+n)\Gamma(b+n)}{\Gamma(a+b+c+n) n!} (1-X)^{n}
\eeq
which is valid as long as $a,b,a+c,b+c$ are not negative integers.
Using this we can perform the $t$-integration and get a converging
(for $|Z|<1$) double power series expansion:
\beq\nonumber
K_{\vj}(Z) = \f{ \pi\Gamma\left(\Sigma -1\right)}{\prod_{i=1}^n \Gamma(\Delta_{i})}
\sum_{n=0}^\infty \frac{(1-|1-Z|^2)^n}{n!} \Big\{
\int \frac{\rd s }{2\pi i} \Big[ \Gamma(-s) \Gamma(\Delta_{2}+\Delta_{3}-\Sigma-s)\\
\Gamma(\Delta_{1}+s)\Gamma(\Delta_{1}+\Delta_{3}+ \Delta_{4}-\Sigma+s)
\frac{ \Gamma(\Sigma- \Delta_{3}+s+n)\Gamma(\Delta_{4}+s+n)}{\Gamma(\Delta_{1}+\Delta_{4}+ 2s+n)}\,
 |Z|^{2 s} \Big]\Big\}.
\eeq
The remaining Barnes-Mellin integral receives contributions from double poles, and an
explicit formula for collecting these, as well as explicit expressions for the arising
double power series for particular values of spins are given in \cite{Arutyunov:2000ku}.

\section {The shifted Jacobi Polynomial}
\label{app:Jac}

Jacobi Polynomials $P_{n}^{(a,b)}(x)$ can be defined in terms of the Rodrigues formula
\be
P_{n}^{(a,b)}(x)
= \frac{(-1)^{n}}{2^n n!} (1-x)^{-a}(1+x)^{-b}\partial_{x}^{n}\left((1-x)^{a+n}(1+x)^{b+n}\right).
\ee
Here we will give some useful formuli for the shifted Jacobi polynomial
$\hat{P}_{n}^{(a,b)}(Z) \equiv P_{n}^{(a,b)}(1-2Z)$. We define $1-x =2Z$, $1+x= 2(1-Z)$, and then
$-2 \partial_{x}=\partial_{Z}$. The ``shifted'' Rodriguez formula now reads
\be
\hat{P}_{n}^{(a,b)}(Z)
=  \frac{1}{n!} Z^{-a}(1-Z)^{-b}\partial_{Z}^{n}\left(Z^{a+n}(1-Z)^{b+n}\right).
\ee
We can therefore express the shifted Jacobi polynomial as the following contour integral
\be\label{app-integr-repr}
 \hat{P}_{n}^{(a,b)}(Z) =\oint \frac{\rd w}{2i\pi} \frac{(Z +w)^{n+a}(1-Z-w)^{n+b}}{Z^{a}(1-Z)^{b}} \frac{1}{\omega^{n+1}},
\ee
where the contour of integration is around $\omega=0$ in the positive direction and the
contour should  avoid the other poles at $ \omega = -Z $ and $\omega = 1-Z$.

The shifted Jacobi polynomials satisfy a recurrence relation given by
\bea\label{app-jac}
(1-2Z) \hat{P}^{(a,b)}_{n} &=&
\frac{2 (n + 1) (n + a + b + 1)}{(1 + a + b + 2 n) (2 + a + b + 2 n)} \hat{P}^{(a,b)}_{n+1}
\\
&+& \frac{(b^2 - a^2)}{(2n +a + b ) (2n + a + b + 2 )} \hat{P}^{(a,b)}_{n}
+ \frac{2 (n + a) (n + b)}{ (2n+a + b ) (2n+  a + b + 1)}
\hat{P}^{(a,b)}_{n-1}, \nonumber
\eea
where all polynomials are evaluated at $(1-2Z)$. Now, using the fact that the action of
the hypergeometric operator ${\cal{O}}_n\,\equiv\,Z(1-Z)\pp_Z^2+[(a+1)-(a+b+2)Z]\pp_Z+n(n+a+b+1)$
on the Jacobi polynomial is diagonal:
\be
{\cal{O}}_nP_n^{(a,b)} =0,\quad
{\cal{O}}_nP_{n-1}^{(a,b)} = [2n+a+b]\,P_{n-1}^{(a,b)},\quad
{\cal{O}}_nP_{n+1}^{(a,b)} = -[2n+a+b+2]\,P_{n+1}^{(a,b)},
\ee
we can apply the hypergeometric operator to equation (\ref{app-jac}), and obtain a first order differential
recursion relation:
\bea
 Z(1-Z) \partial_{Z} \hat{P}_{n}^{(a,b)} &=&
\frac{n (n + 1) (n + a + b + 1)}{(2n + a + b + 1) (2n + a + b + 2 )} \hat{P}^{(a,b)}_{n+1}
\\
&+& \frac{n(b - a)(n +a+b +1)}{(2n +a + b ) (2n + a + b + 2 )}
\hat{P}^{(a,b)}_{n} - \frac{(n + a) (n + b) (n+a+b+1)}{ (2n+a + b ) (2n+  a + b + 1)}
\hat{P}^{(a,b)}_{n-1}. \nonumber
\eea

Since these operator act at fixed $(a,b)$ labels, it looks like we could define the action of $Z$ and
$Z(1-Z) \partial_{Z}$ on the space of holomorphic intertwiners at fixed representation labels.
Thus, taking into account the normalization coefficients of the normalised states we get an action
of, say, $Z$ of the type
\be
Z C_{\vj}^{k} = \alpha_{\vj}(k) C_{\vj}^{k+1}  + \beta_{\vj}(k) C_{\vj}^{k}  + \gamma_{\vj}(k) C_{\vj}^{k-1}
\ee
A closer look at these coefficients shows however that $\alpha_{\vj}(j_{1}+j_{2}) $
(or $\alpha_{\vj}(j_{3}+j_{4}) $) is infinite! This means that the operator of multiplication by $Z$ is
not defined on the entire Hilbert space. It is an operator with a ``domain of definition'' restricted to
the states $C_{\vj}^{k}$ with $k<\mathrm{max}(j_{1}+j_{2},j_{3}+j_{4})$.
The same conclusion applies to the operator $Z(1-Z) \partial_{Z}$.

\end{appendix}


\end{document}